\documentclass[preprint,5p,twocolumn,status=final]{elsarticle} 

\usepackage{ifthen}
\usepackage{graphicx}
\usepackage[svgnames,table]{xcolor}
  \definecolor{lightgray}{gray}{0.9}
\usepackage{footnote} 
\makesavenoteenv{table}
\makesavenoteenv{table*}
\makesavenoteenv{tabular}
\usepackage{url}
\usepackage{array}
\usepackage[colorlinks,allcolors=Blue]{hyperref} %
\usepackage{cleveref} 
  \Crefname{section}{Section}{Sections}
  \Crefname{paragraph}{Section}{Sections}
  \Crefname{table}{Table}{Tables}
  \Crefname{figure}{Figure}{Figures}
\usepackage{multirow}
\usepackage{tabularx}
\usepackage{wrapfig}
\usepackage[english]{babel}
\usepackage{balance} %
\usepackage{enumerate}
\usepackage{ragged2e}  %
  \newcolumntype{L}{>{\raggedright\arraybackslash}X}
  \newcolumntype{?}{!{\vrule width 1pt}}
\usepackage{amssymb}
\usepackage{colortbl}
\usepackage{paralist}
\usepackage{subfig}
\usepackage{tikz}
  \usetikzlibrary{arrows,arrows.meta,positioning,automata,
    shapes,decorations.pathreplacing,calc}
  
\usepackage[inline,nomargin]{fixme}
\usepackage{ctable}

\newcommand{\wrt}{with respect to\ }

\newcommand{\cf}{cf.\ } 		
\newcommand{\ies}{i.e.,\ } 		
\newcommand{\ie}{that is,\ }			
 		
\newcommand{\egs}{e.g.\ } 		
\newcommand{\Eg}{For example,\ }
\newcommand{\eg}{for example,\ }
\newcommand{\Egs}{E.g.\ }

\newcommand{\vs}{vs.\ }	

\usepackage{mdframed}
\newmdtheoremenv[%
hidealllines=true,
leftmargin=0,%
rightmargin=0,
backgroundcolor=gray!30,%
ntheorem]{finding}{Finding}

\makeatletter
\newcommand{\trackedcontentNamed}[3]{%
  \textbf{#3}%
  \def\@currentlabel{#3}%
  \label{#1#2}%
}
\newcommand{\Construct}[2]{\trackedcontentNamed{C-}{#1}{#2}}
\newcommand{\SubConstruct}[3]{%
  \trackedcontentNamed{SC-}{#2}{#3}}
\newcommand{\Hypothesis}[2]{h\trackedcontentNamed{H-}{#1}{#2}:}
\newcommand{\Question}[2]{q\trackedcontentNamed{Q-}{#1}{#2}:}
\newcommand{\Finding}[3]{
  \begin{finding}%
    \label{F-#1}
    #3
  \end{finding}
}

\makeatother

\newcommand{\ConstructRef}[1]{\ref{C-#1}}
\newcommand{\SubConstructRef}[1]{\ref{SC-#1}}
\newcommand{\SubConstructRefP}[1]{}
\newcommand{\HypothesisRef}[1]{h\ref{H-#1}}
\newcommand{\HypothesisRefP}[1]{}
\newcommand{\QuestionRef}[1]{q\ref{Q-#1}}
\newcommand{\QuestionRefM}[1]{\mathit{q\ref{Q-#1}}}
\newcommand{\FindingRef}[1]{\textbf{F\ref{F-#1}}}

\newcommand{\mycite}[2][]{#1~\cite{#2}}
\newcommand{\median}[2]{\mathsf{med}(#1)\in\{\mathit{#2}\}}

 \usepackage{charter} %
\usepackage{doi}

\usepackage{amssymb} 

\journal{Elsevier Information and Software Technology}

\begin{document}
\renewcommand*{\appendixname}{} 

\begin{frontmatter}
  \title{Safety Practice and its Practitioners: Exploring a Diverse
    Profession\tnoteref{preprintcr}}

  \tnotetext[preprintcr]{
    Funded by the Deutsche Forschungsgemeinschaft (DFG, German
    Research Foundation) -- GL 915/1-1.
    \newline
    \copyright\ 2018. This manuscript version is
    made available under the CC-BY-NC-ND 4.0 license
    \url{http://creativecommons.org/licenses/by-nc-nd/4.0/}.
    \newline
    \textbf{Reference Format:}
    Gleirscher, M., \& Nyokabi, A.. Private Communication (\today).
    Department of Computer Science, University of York, United
    Kingdom. eprint: arXiv reference
  }

  \author[york,tum]{Mario Gleirscher}
  \ead{mario.gleirscher@york.ac.uk}
  \author[siemens,tum]{Anne Nyokabi}
  \ead{anne.nyokabi@tum.de}
  \address[york]{Department of Computer Science, University of
    York, United Kingdom
    \\ Deramore Lane, Heslington, York YO10 5GH}
  \address[siemens]{Process Industries and Drives Division, Siemens AG, Germany
    \\ Schuhstra{\ss}e 60, 91052 Erlangen}
  \address[tum]{Formerly: Institut f\"ur Informatik, Technische Universit\"at 
    M\"unchen, Germany
    \\ Boltzmannstr. 3, 85748 Garching bei M\"unchen}

\begin{abstract}
  \small
  \begin{inparadesc}
  \item[\textbf{Context:}] %
    System safety refers to a diverse engineering discipline
    assessing and improving various aspects of
    safety %
    in socio-technical systems and their software-intensive
    sub-systems.  While system safety has been a vital area of applied
    research for many decades, its practice and practitioners seem
    empirically still not well studied.  Beyond anecdotal
    evidence---case reports, interviews, discussion forums, blogs, and
    ``war stories'' serving as interesting examples in textbooks---and
    surveys, we are missing open, large-scale, and long-term
    empirical investigations %
    that promote knowledge transfer and research validation.
  \item[\textbf{Objective:}] In this article, we explore means
    for work that safety practitioners rely on, factors influencing
    the performance of these professionals, and their perception of
    their role in the system life cycle.  Along with that we want to
    examine observations from previous research.
  \item[\textbf{Methods:}] We build a construct of safety practice,
    collect data for this construct using an on-line survey, summarise
    and interpret the collected data, and investigate several
    hypotheses based on the previous observations.
  \item[\textbf{Results:}] We analyse and present the responses of 124
    practitioners in safety-critical system and software projects.
    Aside from other findings, our data 
    \begin{inparaitem} 
    \item suggests that safety decision making mainly depends on expert
      opinion and project memory,
    \item lacks evidence that safety is typically
      a cost-benefit question, %
    \item does not exhibit the prejudice that formal methods are not
      beneficial, %
    \item leaves it unclear as to whether or not standards and methods
      have become inadequate, and %
    \item indicates that safety is not typically confused with
      reliability. %
    \end{inparaitem}
    Additionally, we contribute a research design directing towards
    explanatory empirical studies of safety practice.
  \item[\textbf{Conclusions:}] We believe that empirical research of
    safety practice is still in an early stage, bearing the risk of
    undesirable mismatches of the state of the art and the state of
    practice.  However, this situation offers great opportunities for
    research.
  \end{inparadesc}
\end{abstract}

\end{frontmatter}

\section{Introduction}
\label{sec:introduction}

\emph{System safety practice}~(safety practice for short) is a
remarkably diverse field spanning many disciplines involved in the
system life cycle, influenced by heterogeneous criticality-driven
\emph{safety cultures}~\cite{Perrow1984,Sorensen2002,Choudhry2007}
across various application domains, geographical regions, and
regulatory authorities.

Researchers have surveyed and investigated practised approaches to
\emph{accident prevention}, \eg in the chemical plant and nuclear
power plant sectors~\cite{Sorensen2002} and in the construction
industries~\cite{Choudhry2007}.  However, our literature search has
not uncovered a single officially published empirical investigation
(\ies a case or field study, a controlled field experiment, a survey
of practitioners) of the \emph{effectiveness of practised approaches}
to prevent or reduce software and (control) systems' contributions to
hazards.

In the following, we highlight the motivations for our study, describe
observations from previous research, outline our research objective,
and summarise the contributions of this work.

\subsection{Problem Statement}
\label{sec:motivation}

From exploratory content analysis of more than 200 selected
\emph{question and answer posts} on several safety practitioners'~(SP)
\emph{on-line channels} of a period of~4 years and one expert
interview~\cite{Hussein2016}, we observe that SPs
\begin{enumerate}
\item discuss various issues with the application of standards,
  calculation of failure rates,
  correct planning of safety tests, and
  completeness of hazard analyses;
\item are missing a standardised way of integrating safety with
  security activities;
\item are concerned about the adequacy of methods,
  a lack of safety education, and
  the misunderstanding of their role.
\end{enumerate}

From exploratory content analysis of more than 370 \emph{case reports}
(\ies on incidents and accidents) from the aviation, automotive, and
railway domains and 7 \emph{semi-structured interviews} with SPs from
these domains~\cite{Yang2016}, we observe that
\begin{enumerate}
\item human errors and specification errors were more often reported
  as accident root causes than software implementation errors---this
  is consistent with the findings in \cite[p.~30f]{HSE2003};
\item no IT security problems were reported;
\item reports in general, and comparably often in the automotive
  domain, were non-informative of subtle accident root causes~(\ies
  causes lying outside the possibilities, budgets, or obligations of
  accident analysts and investigators);
\item few of the selected reports at least suggest that accidental
  complexity~\cite{Brooks1995a}---particularly, missing or
  mistaken %
  maintenance, refactoring, evolution, or migration---negatively
  affects system safety;
\item interviewees report issues of unclear separation of
  system-level and software-level activities (\cf\cite{Knight2002});
\item interviewees state that available methods are currently just
  appropriate in their domains but can easily get insufficient for
  complex future applications.
\end{enumerate}

These observations fuel some almost negligently accepted
computer-related risks---as regularly archived by~\mycite[Neumann et
al.]{Neumann2018}---as well as occasional but recent worries about the
state of the practice and education in safety engineering in
particular~\cite{McDermid2014} and in software engineering~(SE) in
general~\cite{Osterweil2018}.

\subsection{Research Objectives}
\label{sec:research-objective}

These findings certainly ask for more evidence.  In line with the
research agenda in \cite{Martins2016},
safety engineering research might, hence, pose clarifying questions
such as:
\begin{enumerate}
\item %
  Which means are SPs familiar with and which do they currently use?
  How clear, unambiguous, consistent, up-to-date, and effective are
  those means?
\item %
  What are the SPs' current problems, challenges, needs, and
  expectations?
\item %
  How do SPs view their profession, role, and contribution in
  the life cycle?
\end{enumerate}

\subsection{Contributions}
\label{sec:contributions}

This work contributes to safety research in several ways:
\begin{itemize}
\item First, we present results of a cross-sectional self-administered
  on-line survey among SPs: Particularly, we sample some of their
  experiences, views, opinions, and their self-perception.

\item Then, we test several comparative
  hypotheses~(\Cref{sec:hypotheses-tests}) on safety practice and SPs
  and interpret our test results~(\Cref{sec:interpr-results}) \wrt
  findings and experience from previous
  work~\cite{Yang2016,Hussein2016}. This way, we also elaborate on
  results in~\cite{Nyokabi2017}.

\item Furthermore, we respond to the request from~\mycite[Alexander et
  al.]{Alexander2010} and \mycite[Rae et al.]{Rae2010} for applying
  improved methodology in empirical research of safety practice, as
  well as the desire of a stronger involvement of SPs in research
  evaluation such as stated by \mycite[Martins and
  Gorschek]{Martins2016}.

\item Moreover, we contribute a research
  design~(\Cref{sec:research-design}) for similar empirical
  assessments.  This setting might as well be applicable to other SE
  domains~(see, \egs\cite{Valerdi2009}).
\end{itemize}

\subsection{Overview}
\label{sec:overview}

\Cref{fig:overview} provides an overview of the research procedures
for this article.  After discussing terminology and related work in
\Cref{sec:background} and describing our research method in
\Cref{sec:research-design}, we present our results in
\Cref{sec:results}.  Particularly, we describe our sample in
\Cref{sec:sample} and summarise the results of all valid responses in
\Cref{sec:summ-answ-quest}.  \Cref{sec:hypotheses-tests} highlights
the results of several hypotheses tests.  Our discussion follows in
\Cref{sec:discussion}, with the interpretation of our test results in
\Cref{sec:interpr-results} and the examination of threats to the
validity of our study in \Cref{sec:threats-validity}.  We summarise
our findings in \Cref{sec:conclusions}.  \Cref{tab:datasummary}
contains a detailed summary of the response
data.

\begin{figure}
  \centering
  \footnotesize
\begin{tikzpicture}
  [dim/.style={align=flush left},
  tit/.style={align=left,anchor=south west},
  itm/.style={align=center,minimum
    height=3em,minimum width=2cm,fill=SkyBlue!40},
  lab/.style={align=center}]

  \node[itm] (prst) at (0,0) {Observations\\\& Problem\\Statement\\(\Cref{sec:motivation})};
  \node[itm] (reob) at ($(prst)+(22em,0)$) {Research\\Objectives\\(\Cref{sec:research-objective})}; 

  \node[dim,minimum height=23.5em,minimum width=21.5em,fill=gray!10] (vali)
  at ($(reob)+(-6em,-15.75em)$) {}; 
  \node[tit,anchor=south east,align=right]
  at (vali.south east) {Validity Procedures\\(\Cref{sec:validity-procedure,sec:threats-validity})};

  \node[dim,minimum height=12em,minimum width=20.5em,fill=white] (cons)
  at ($(reob)+(-6em,-10.5em)$) {};
  \node[tit,anchor=north west,align=left] at (cons.north west) {Construct\\(\Cref{sec:construct})};
  
  \node[itm] (rego) at ($(reob)+(0,-7em)$) {Research Goal\\\& Questions\\(\Cref{sec:hypoth-rese-quest})}; 
  \node[itm] (suin) at ($(rego)+(0,-7em)$) {Survey\\Instrument\\(\Cref{sec:questionnaire})}; 
  \node[itm] (wohy) at ($(suin)+(-12em,0)$) {Working\\Hypotheses\\(\Cref{sec:hypotheses})}; 
  \node[itm] (hyan) at ($(suin)+(-12em,-7em)$) {Hypotheses\\Analysis\\(\Cref{sec:hypotheses-tests})}; 
  \node[itm] (resp) at ($(suin)+(0,-7em)$) {Responses\\(\Cref{sec:summ-answ-quest})}; 
  \node[itm] (find) at ($(hyan)+(-10em,0)$) {Findings\\(\Cref{sec:interpr-results})}; 
  \node[itm] (rewo) at ($(find)+(0,10em)$) {Related Work\\(\Cref{sec:relwork})}; 

  \draw[arrows={-latex},thick,tips=proper] 
  (prst) edge[above] node[lab] {motivate} (reob)
  (rego) edge[right] node[lab] {refine} (reob)
  (suin) edge node[lab,right] {measures} (rego)
  (suin) edge node[lab,above] {measures} (wohy)
  (wohy) edge[sloped,above] node[lab] {derived from} (prst)
  (rewo) edge[sloped,above] node[lab] {justifies} (wohy)
  (rewo) edge node[lab,fill=white] {shares,\\opposes} (prst)
  (resp) edge node[lab,below] {provide\\data for} (hyan)
  (hyan) edge[left] node[lab] {checks,\\extends} (wohy)
  (resp) edge[bend left,sloped,above] node[lab] {yield} (find)
  (hyan) edge[above] node[lab] {yields} (find)
  (find) edge node[lab,fill=white] {compared to\\(\Cref{sec:relat-exist-find})} (rewo)
  (suin) edge[left] node[lab] {used to sample\\(\Cref{sec:sample})} (resp)
  ;
\end{tikzpicture}

   \caption{Overview of the research method for this article
    \label{fig:overview}
  }
\end{figure}

\section{Background}
\label{sec:background}

We introduce important terms as well as related work we will revisit
in our discussions below.

\subsection{Terminology and Definitions}
\label{sec:terminology}

The \emph{life cycle} of an \emph{engineered system} typically refers
to the phases of design, implementation, release, maintenance,
operation, and disposal.
\emph{Dependability} then encompasses the handling of reliability,
availability, maintainability, and safety in the life cycle, \eg by
improving fault-tolerance.

In this work, we focus the discipline of \emph{system
  safety},\footnote{From software, electrical, electronics, control,
  and systems engineering.} including \emph{functional safety}.
System safety is usually situated in the context of safety of
machinery,\footnote{From mechanical engineering.} process
safety,\footnote{From automation and plant engineering.} structural
safety,\footnote{From construction or civil engineering.} or
occupational health and safety.  These disciplines have in common the
identification, assessment, and management of operational risk.  This
procedure includes the prevention or handling of \emph{undesired
  events} at any stage~(\egs hazards or safety risks, incidents, and
accidents) and of any type~(\egs human error, software faults, and
system failures).
In addition, \emph{security of information technology}~(IT security or
\emph{security} for short) is the discipline of protecting
computer-based systems and data against malicious attacks and
unauthorised access.

Then, \emph{safety practice} denotes the practical aspects of system
safety in both industrial settings and applied research.
Based on this, we consider a \emph{safety practitioner} as a
person who supports or performs \emph{safety decision making},
particularly, by identifying \emph{hazards} and assessing their causes
and consequences, the design of hazard countermeasures~(also known as
hazard controls), the assurance of safety, or by performing research
and consultancy for these \emph{safety activities}.  Importantly,
there are many \emph{means}---\ie best practices, methods,
techniques, and standards---to apply in these activities.

\subsection{Related Work}
\label{sec:relwork}

As indicated in \Cref{sec:introduction}, there are only few
cross-disciplinary exploratory inquiries of safety practice and its
practitioners.  The following studies demonstrate the importance of
empirical methods (interviews and related survey methods such as focus
groups and questionnaires) in further examining safety practice.

\mycite[Dwyer]{Dwyer1992}, for multiple disciplines, and
\mycite[Knight]{Knight2002}, for software engineering, characterise
safety practice from their experience, forecasting the ongoing trend
of increased automation, the increasingly critical interplay between
the involved engineering domains, and the corresponding challenges for
future safety research.

\paragraph{Adequacy of Means of Work in Safety Practice}

Safety-critical systems are subjected to \emph{automation} (\ies the
use of qualified and verified tool chains) for their development,
testing, and overall assurance.  \mycite[Graaf et al.]{Graaf2003}
and \mycite[Kasurinen et al.]{Kasurinen2010}
investigate challenges and obstacles to adoption of new methods,
languages, and tools in embedded system RE, architecture design, and
software testing.  Our study explores this direction within safety
practice.

\mycite[Hatcliff et al.]{Hatcliff2014} summarise particular challenges
in the certification of software-dependent systems and suggest
improvements, stressing the concept of ``designed-in
safety/security.''
These works inspired and underpin our hypotheses but are different
from our survey approach to examining safety practice and its
practitioners.

\mycite[Chen et al.]{Chen2018a} report on
the challenges and best practices of using assurance cases.  
Our questionnaire about safety practice includes more general
questions about methods, training, and interaction, backed by a larger
number of data points.

For organisations that engineer safety-critical systems,
\mycite[Ceccarelli and Silva]{Ceccarelli2015} provide a framework for
compliance checking during and after the introduction of new safety
standards (\egs DO-178B) into an organisation.  In our study, we are
asking SPs whether safety standards known and used by our
respondents, actually improve the organisation's safety practice.

\mycite[McDermid and Rae]{McDermid2014} report on their cross-domain
insights into the practice of engineering safety-critical systems,
discussing the question: ``How did systems get so safe despite
inadequate and inadequately applied techniques?''  Without presuming
that modern systems are acceptably safe, we interrogate SPs about
their means of work.

\mycite[Wang and Wagner]{Wang2018} investigate decision processes in
safety activities.  For complex and highly critical systems such
processes are usually committee- or group-driven to reduce
organisational single points of failure.  The authors examine whether
such decision making is prone to a number of pitfalls known as
``groupthink'' and studied in group psychology.  Being more
exploratory in nature, our study design differs from the
psychology-based construct used in~\cite{Wang2018} and yet addresses a
fraction of it.

\paragraph{Process Factors influencing Safety Practice}

Requirements engineering~(RE) and, particularly, requirements
specification, are critical points of failure in every safety-critical
system project.  Examining research on the communication and
validation of safety requirements in industrial projects,
\mycite[Martins and Gorschek]{Martins2016} %
conclude that there is a lack of evidence for the usefulness and
usability of recent safety research.  We want to contrast their
finding with how practitioners currently perceive the adequacy of
their means of work.

\mycite[Nair et al.]{Nair2015} present results from a survey of 52~SPs
on how they \emph{manage the variety of safety evidence} for critical
computer-based systems.  Good evidence management implies many
technical challenges in safety practice.
Particularly, traceability is crucial for change impact analysis
(CIA), \ie the analysis of how changes of safety-critical artefacts
(\egs specifications, issue databases, designs) are propagated and
whether these changes have negative safety impact.
\mycite[Borg et al.]{Borg2016} report on 14~interviews
with SPs about their CIA activities, finding that SPs 
have difficulties in understanding the motivation of CIA, are
overwhelmed by the information they have to process when conducting
CIA, and struggle with trusting and updating former CIAs.
From a %
cross-sectional %
survey of 97 practitioners, \mycite[De la Vara et al.]{Vara2016}
observe insufficient CIA tool support.
Our study examines such means of work from a more general viewpoint.

In the \Cref{sec:disc-motiv-behind,sec:acceptance}, we establish
relations between these works and our study.  In
\Cref{tab:relat-exist-find} in \Cref{sec:relat-exist-find}, we compare
their findings with our results.

\section{Survey Planning}
\label{sec:research-design}

This section describes our survey
design~(\Cref{sec:hypoth-rese-quest}),
the survey instrument~(\Cref{sec:questionnaire}),
our working hypotheses~(\Cref{sec:hypotheses}),
the procedure for data collection~(\Cref{sec:daten-collection}) and
analysis~(\Cref{sec:data-analysis}), and instrument
evaluation~(\Cref{sec:validity-procedure}).  We follow the guidelines in
\cite{Fink2016-HowConductSurveys} for planning and conducting our
survey and
\cite{Jedlitschka2008-ReportingExperimentsSoftware,Kitchenham2008,Kitchenham2007-Evaluatingguidelinesreporting}
for the reporting.

\subsection{Research Goal and Questions}
\label{sec:hypoth-rese-quest}

From our project experience summarised in~\Cref{sec:motivation} and
previous research~\cite{Nyokabi2017,Yang2016,Hussein2016}, we have
learned about typical issues in \emph{safety practice}.  This
cross-sectional survey design aims at resuming these issues.  The
\emph{objective} of this exploratory study is
\begin{quote}
  to investigate \emph{safety practice and its practitioners} and to
  examine observations we made during our preliminary research.
\end{quote}
For this, we explore three \emph{research questions}:
\begin{enumerate}[RQ1]
\item Which \emph{means} do SPs typically rely on in their activities?
  How helpful %
  are those means to them?
\item Which typical \emph{process factors} have influence on SPs'
  decisions and performance?
\item How do SPs perceive and understand their role %
  in the \emph{process} or life cycle?
\end{enumerate}

\subsubsection{Construct}
\label{sec:construct}

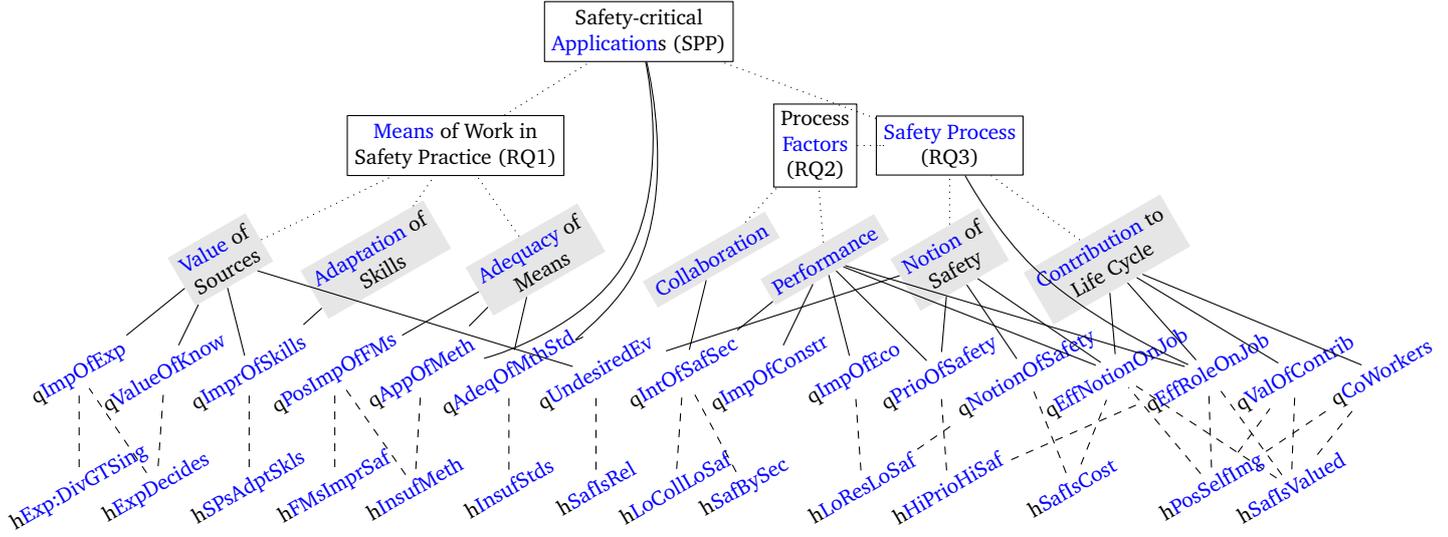
\begin{figure*}
  \centering

\begin{tikzpicture}[>=latex,line join=bevel,scale=0.17, every node/.style={scale=.8}]
\begin{scope}
  \pgfsetstrokecolor{black}
  \definecolor{strokecol}{rgb}{1.0,1.0,1.0};
  \pgfsetstrokecolor{strokecol}
  \definecolor{fillcol}{rgb}{1.0,1.0,1.0};
  \pgfsetfillcolor{fillcol}
\end{scope}
\begin{scope}
  \pgfsetstrokecolor{black}
  \definecolor{strokecol}{rgb}{1.0,1.0,1.0};
  \pgfsetstrokecolor{strokecol}
  \definecolor{fillcol}{rgb}{1.0,1.0,1.0};
  \pgfsetfillcolor{fillcol}
\end{scope}
\begin{scope}
  \pgfsetstrokecolor{black}
  \definecolor{strokecol}{rgb}{1.0,1.0,1.0};
  \pgfsetstrokecolor{strokecol}
  \definecolor{fillcol}{rgb}{1.0,1.0,1.0};
  \pgfsetfillcolor{fillcol}
\end{scope}
\begin{scope}
  \pgfsetstrokecolor{black}
  \definecolor{strokecol}{rgb}{1.0,1.0,1.0};
  \pgfsetstrokecolor{strokecol}
  \definecolor{fillcol}{rgb}{1.0,1.0,1.0};
  \pgfsetfillcolor{fillcol}
\end{scope}
\begin{scope}
  \pgfsetstrokecolor{black}
  \definecolor{strokecol}{rgb}{1.0,1.0,1.0};
  \pgfsetstrokecolor{strokecol}
  \definecolor{fillcol}{rgb}{1.0,1.0,1.0};
  \pgfsetfillcolor{fillcol}
\end{scope}
  \node (H11_RQ_B2) at (838.5bp,270.0bp) [draw,draw=none,rotate=30] {\QuestionRef{H11_RQ_B2}};
  \node (H11_RQ_B6) at (647.5bp,270.0bp) [draw,draw=none,rotate=30] {\QuestionRef{H11_RQ_B6}};
  \node (ValueOfSPs_Self) at (2341.5bp,522.0bp) [draw,draw=none,align=center,fill=gray!20,rotate=30] {\SubConstructRef{ValueOfSPs-Self} to\\ Life Cycle};
  \node (H24) at (1992.5bp,18.0bp) [draw,draw=none,rotate=30] {\HypothesisRef{H24}};
  \node (H26) at (1029.5bp,18.0bp) [draw,draw=none,rotate=30] {\HypothesisRef{H26}};
  \node (H20) at (2572.5bp,18.0bp) [draw,draw=none,rotate=30] {\HypothesisRef{H20}};
  \node (H22) at (1394.5bp,18.0bp) [draw,draw=none,rotate=30] {\HypothesisRef{H22}};
  \node (ValueOfKnowledgeSource) at (396.5bp,522.0bp) [draw,draw=none,align=center,fill=gray!20,rotate=30] {\SubConstructRef{ValueOfKnowledgeSource} of\\ Sources};
  \node (H16_RQ_F4) at (2945.5bp,270.0bp) [draw,draw=none,rotate=30] {\QuestionRef{H16_RQ_F4}};
  \node (H18_RQ_F5_ab) at (2363.5bp,270.0bp) [draw,draw=none,rotate=30] {\QuestionRef{H18_RQ_F5.ab}};
  \node (H3_RQ_E1) at (274.5bp,270.0bp) [draw,draw=none,rotate=30] {\QuestionRef{H3_RQ_E1}};
  \node (H17_RQ_E2) at (1220.5bp,270.0bp) [draw,draw=none,rotate=30] {\QuestionRef{H17_RQ_E2}};
  \node (Means) at (912.5bp,774.0bp) [draw,rectangle,align=center] {\SubConstructRef{Means} of Work in\\Safety Practice (RQ1)};
  \node (SafSecInteraction) at (1473.5bp,522.0bp) [draw,draw=none,align=center,fill=gray!20,rotate=30] {\SubConstructRef{SafSecInteraction}};
  \node (H1_RQ_G3) at (1786.5bp,270.0bp) [draw,draw=none,rotate=30] {\QuestionRef{H1_RQ_G3}};
  \node (H11b) at (647.5bp,18.0bp) [draw,draw=none,rotate=30] {\HypothesisRef{H11b}};
  \node (H18) at (2264.5bp,18.0bp) [draw,draw=none,rotate=30] {\HypothesisRef{H18}};
  \node (H6_RQ_C1) at (459.5bp,270.0bp) [draw,draw=none,rotate=30] {\QuestionRef{H6_RQ_C1}};
  \node (H11) at (822.5bp,18.0bp) [draw,draw=none,rotate=30] {\HypothesisRef{H11}};
  \node (H14) at (86.5bp,18.0bp) [draw,draw=none,rotate=30] {\HypothesisRef{H14}};
  \node (H16) at (2745.5bp,18.0bp) [draw,draw=none,rotate=30] {\HypothesisRef{H16}};
  \node (H17) at (1220.5bp,18.0bp) [draw,draw=none,rotate=30] {\HypothesisRef{H17}};
  \node (H7_RQ_F1_a) at (2563.5bp,270.0bp) [draw,draw=none,rotate=30] {\QuestionRef{H7_RQ_F1.a}};
  \node (ProcessFactor) at (1701.5bp,774.0bp) [draw,rectangle,align=center] {Process\\ \SubConstructRef{ProcessFactor}\\(RQ2)};
  \node (SafetyActivities) at (1996.5bp,774.0bp) [draw,rectangle,align=center] {\SubConstructRef{SafetyActivities}\\(RQ3)};
  \node (H22_RQ_D1) at (1412.5bp,270.0bp) [draw,draw=none,rotate=30] {\QuestionRef{H22_RQ_D1}};
  \node (ProcessPerformance) at (1722.5bp,522.0bp) [draw,draw=none,align=center,fill=gray!20,rotate=30] {\SubConstructRef{ProcessPerformance}};
  \node (ImprovementOfSkills) at (737.5bp,522.0bp) [draw,draw=none,align=center,fill=gray!20,rotate=30] {\SubConstructRef{ImprovementOfSkills} of\\ Skills};
  \node (H8) at (1546.5bp,18.0bp) [draw,draw=none,rotate=30] {\HypothesisRef{H8}};
  \node (NotionOfSafety) at (1996.5bp,522.0bp) [draw,draw=none,align=center,fill=gray!20,rotate=30] {\SubConstructRef{NotionOfSafety} of\\ Safety};
  \node (H3) at (256.5bp,18.0bp) [draw,draw=none,rotate=30] {\HypothesisRef{H3}};
  \node (H14_RQ_C2) at (86.5bp,270.0bp) [draw,draw=none,rotate=30] {\QuestionRef{H14_RQ_C2}};
  \node (H1) at (1804.5bp,18.0bp) [draw,draw=none,rotate=30] {\HypothesisRef{H1}};
  \node (H6) at (459.5bp,18.0bp) [draw,draw=none,rotate=30] {\HypothesisRef{H6}};
  \node (H1_RQ_G2) at (1601.5bp,270.0bp) [draw,draw=none,rotate=30] {\QuestionRef{H1_RQ_G2}};
  \node (AdequacyOfStandardsMethods) at (1087.5bp,522.0bp) [draw,draw=none,align=center,fill=gray!20,rotate=30] {\SubConstructRef{AdequacyOfStandardsMethods} of\\ Means};
  \node (H24_RQ_E3) at (1974.5bp,270.0bp) [draw,draw=none,rotate=30] {\QuestionRef{H24_RQ_E3}};
  \node (H20_RQ_F2) at (2755.5bp,270.0bp) [draw,draw=none,rotate=30] {\QuestionRef{H20_RQ_F2}};
  \node (CurrentApplication) at (1314.5bp,1026.0bp) [draw,rectangle,align=center] {Safety-critical\\ \SubConstructRef{CurrentApplication}s (SPP)};
  \node (H26_RQ_B4) at (1029.5bp,270.0bp) [draw,draw=none,rotate=30] {\QuestionRef{H26_RQ_B4}};
  \node (H18_RQ_F5) at (2165.5bp,270.0bp) [draw,draw=none,rotate=30] {\QuestionRef{H18_RQ_F5}};
  \draw [->,dashed,-] (H17) ..controls (1220.5bp,79.426bp) and (1220.5bp,189.18bp)  .. (H17_RQ_E2);
  \draw [->,dashed,-] (H16) ..controls (2700.7bp,79.97bp) and (2619.9bp,191.95bp)  .. (H7_RQ_F1_a);
  \draw [->,dotted,-] (Means) ..controls (1012.7bp,836.84bp) and (1197.1bp,952.41bp)  .. (CurrentApplication);
  \draw [->,solid,-] (H6_RQ_C1) ..controls (444.12bp,331.52bp) and (416.59bp,441.64bp)  .. (ValueOfKnowledgeSource);
  \draw [->,dotted,-] (ImprovementOfSkills) ..controls (780.53bp,583.97bp) and (858.3bp,695.95bp)  .. (Means);
  \draw [->,dashed,-] (H16) ..controls (2650.4bp,80.749bp) and (2475.8bp,195.95bp)  .. (H18_RQ_F5_ab);
  \draw [->,dotted,-] (ProcessPerformance) ..controls (1717.4bp,583.43bp) and (1708.2bp,693.18bp)  .. (ProcessFactor);
  \draw [->,solid,-] (H7_RQ_F1_a) ..controls (2448.0bp,322.2bp) and (2288.9bp,401.95bp)  .. (2181.5bp,504.0bp) .. controls (2102.0bp,579.54bp) and (2037.2bp,694.63bp)  .. (SafetyActivities);
  \draw [->,solid,-] (H20_RQ_F2) ..controls (2652.3bp,332.84bp) and (2462.4bp,448.41bp)  .. (ValueOfSPs_Self);
  \draw [->,dashed,-] (H26) ..controls (1029.5bp,79.426bp) and (1029.5bp,189.18bp)  .. (H26_RQ_B4);
  \draw [->,dashed,-] (H16) ..controls (2794.8bp,80.06bp) and (2883.9bp,192.41bp)  .. (H16_RQ_F4);
  \draw [->,solid,-] (H22_RQ_D1) ..controls (1489.4bp,332.51bp) and (1629.9bp,446.73bp)  .. (ProcessPerformance);
  \draw [->,solid,-] (H6_RQ_C1) ..controls (528.36bp,332.42bp) and (653.95bp,446.27bp)  .. (ImprovementOfSkills);
  \draw [->,dashed,-] (H20) ..controls (2665.4bp,80.749bp) and (2835.9bp,195.95bp)  .. (H16_RQ_F4);
  \draw [->,dashed,-] (H3) ..controls (260.89bp,79.426bp) and (268.73bp,189.18bp)  .. (H3_RQ_E1);
  \draw [->,dashed,-] (H1) ..controls (1800.1bp,79.426bp) and (1792.3bp,189.18bp)  .. (H1_RQ_G3);
  \draw [->,solid,-] (H24_RQ_E3) ..controls (1979.9bp,331.43bp) and (1989.4bp,441.18bp)  .. (NotionOfSafety);
  \draw [->,dashed,-] (H18) ..controls (2288.7bp,79.607bp) and (2332.1bp,190.1bp)  .. (H18_RQ_F5_ab);
  \draw [->,dashed,-] (H11b) ..controls (647.5bp,79.426bp) and (647.5bp,189.18bp)  .. (H11_RQ_B6);
  \draw [->,dashed,-] (H6) ..controls (459.5bp,79.426bp) and (459.5bp,189.18bp)  .. (H6_RQ_C1);
  \draw [->,dashed,-] (H24) ..controls (1988.1bp,79.426bp) and (1980.3bp,189.18bp)  .. (H24_RQ_E3);
  \draw [->,dashed,-] (H18) ..controls (2240.3bp,79.607bp) and (2196.9bp,190.1bp)  .. (H18_RQ_F5);
  \draw [->,solid,-] (H7_RQ_F1_a) ..controls (2351.7bp,333.45bp) and (1954.5bp,452.49bp)  .. (ProcessPerformance);
  \draw [->,dotted,-] (SafSecInteraction) ..controls (1529.7bp,584.15bp) and (1631.7bp,696.87bp)  .. (ProcessFactor);
  \draw [->,dotted,-] (SafetyActivities) ..controls (1825.3bp,837.25bp) and (1505.6bp,955.4bp)  .. (CurrentApplication);
  \draw [->,solid,-] (H26_RQ_B4) ..controls (1131.8bp,320.82bp) and (1258.6bp,396.49bp)  .. (1309.5bp,504.0bp) .. controls (1390.8bp,675.69bp) and (1342.3bp,916.14bp)  .. (CurrentApplication);
  \draw [->,dotted,-] (ProcessFactor) ..controls (1852.9bp,774.0bp) and (1855.7bp,774.0bp)  .. (SafetyActivities);
  \draw [->,dashed,-] (H22) ..controls (1398.9bp,79.426bp) and (1406.7bp,189.18bp)  .. (H22_RQ_D1);
  \draw [->,solid,-] (H11_RQ_B2) ..controls (1022.3bp,315.5bp) and (1194.9bp,378.41bp)  .. (1276.5bp,504.0bp) .. controls (1380.9bp,664.57bp) and (1339.1bp,914.56bp)  .. (CurrentApplication);
  \draw [->,solid,-] (H16_RQ_F4) ..controls (2793.8bp,333.29bp) and (2512.2bp,450.77bp)  .. (ValueOfSPs_Self);
  \draw [->,solid,-] (H18_RQ_F5_ab) ..controls (2358.1bp,331.43bp) and (2348.6bp,441.18bp)  .. (ValueOfSPs_Self);
  \draw [->,solid,-] (H24_RQ_E3) ..controls (1912.3bp,332.24bp) and (1799.2bp,445.34bp)  .. (ProcessPerformance);
  \draw [->,dashed,-] (H11) ..controls (826.4bp,79.426bp) and (833.37bp,189.18bp)  .. (H11_RQ_B2);
  \draw [->,dashed,-] (H3) ..controls (214.7bp,79.97bp) and (139.16bp,191.95bp)  .. (H14_RQ_C2);
  \draw [->,dashed,-] (H20) ..controls (2570.3bp,79.426bp) and (2566.4bp,189.18bp)  .. (H7_RQ_F1_a);
  \draw [->,solid,-] (H22_RQ_D1) ..controls (1427.4bp,331.52bp) and (1454.0bp,441.64bp)  .. (SafSecInteraction);
  \draw [->,dashed,-] (H20) ..controls (2521.0bp,80.06bp) and (2427.9bp,192.41bp)  .. (H18_RQ_F5_ab);
  \draw [->,solid,-] (H1_RQ_G2) ..controls (1631.1bp,331.7bp) and (1684.4bp,442.56bp)  .. (ProcessPerformance);
  \draw [->,dashed,-] (H11) ..controls (779.47bp,79.97bp) and (701.7bp,191.95bp)  .. (H11_RQ_B6);
  \draw [->,solid,-] (H17_RQ_E2) ..controls (1415.8bp,333.41bp) and (1781.7bp,452.25bp)  .. (NotionOfSafety);
  \draw [->,dashed,-] (H14) ..controls (86.5bp,79.426bp) and (86.5bp,189.18bp)  .. (H14_RQ_C2);
  \draw [->,dashed,-] (H8) ..controls (1513.6bp,79.789bp) and (1454.5bp,191.02bp)  .. (H22_RQ_D1);
  \draw [->,dashed,-] (H16) ..controls (2747.9bp,79.426bp) and (2752.3bp,189.18bp)  .. (H20_RQ_F2);
  \draw [->,dotted,-] (NotionOfSafety) ..controls (1996.5bp,583.43bp) and (1996.5bp,693.18bp)  .. (SafetyActivities);
  \draw [->,solid,-] (H18_RQ_F5) ..controls (2124.0bp,331.88bp) and (2049.2bp,443.48bp)  .. (NotionOfSafety);
  \draw [->,dashed,-] (H1) ..controls (1894.3bp,80.694bp) and (2059.0bp,195.67bp)  .. (H18_RQ_F5);
  \draw [->,solid,-] (H7_RQ_F1_a) ..controls (2508.7bp,332.15bp) and (2409.4bp,444.87bp)  .. (ValueOfSPs_Self);
  \draw [->,solid,-] (H26_RQ_B4) ..controls (1043.7bp,331.52bp) and (1069.0bp,441.64bp)  .. (AdequacyOfStandardsMethods);
  \draw [->,solid,-] (H14_RQ_C2) ..controls (163.4bp,332.51bp) and (303.91bp,446.73bp)  .. (ValueOfKnowledgeSource);
  \draw [->,dotted,-] (ValueOfKnowledgeSource) ..controls (525.82bp,585.16bp) and (765.19bp,702.06bp)  .. (Means);
  \draw [->,dotted,-] (AdequacyOfStandardsMethods) ..controls (1044.5bp,583.97bp) and (966.7bp,695.95bp)  .. (Means);
  \draw [->,solid,-] (H11_RQ_B6) ..controls (757.46bp,332.98bp) and (960.25bp,449.12bp)  .. (AdequacyOfStandardsMethods);
  \draw [->,dotted,-] (ValueOfSPs_Self) ..controls (2255.8bp,584.6bp) and (2098.9bp,699.2bp)  .. (SafetyActivities);
  \draw [->,solid,-] (H17_RQ_E2) ..controls (1013.0bp,333.45bp) and (623.78bp,452.49bp)  .. (ValueOfKnowledgeSource);
  \draw [->,dashed,-] (H24) ..controls (2135.8bp,81.248bp) and (2401.6bp,198.53bp)  .. (H7_RQ_F1_a);
  \draw [->,dashed,-] (H20) ..controls (2617.5bp,79.97bp) and (2698.8bp,191.95bp)  .. (H20_RQ_F2);
  \draw [->,solid,-] (H11_RQ_B2) ..controls (900.0bp,332.24bp) and (1011.7bp,445.34bp)  .. (AdequacyOfStandardsMethods);
  \draw [->,solid,-] (H1_RQ_G3) ..controls (1770.9bp,331.52bp) and (1742.9bp,441.64bp)  .. (ProcessPerformance);
  \draw [->,solid,-] (H18_RQ_F5_ab) ..controls (2202.7bp,333.2bp) and (1902.7bp,451.16bp)  .. (ProcessPerformance);
  \draw [->,solid,-] (H18_RQ_F5_ab) ..controls (2272.2bp,332.69bp) and (2104.8bp,447.67bp)  .. (NotionOfSafety);
  \draw [->,solid,-] (H3_RQ_E1) ..controls (304.37bp,331.7bp) and (358.04bp,442.56bp)  .. (ValueOfKnowledgeSource);
\end{tikzpicture}

   \caption{Research design for our main construct ``safety practice
    and its practitioners'' (SPP, \Cref{sec:construct}).
    The base (h)ypotheses layer \emph{is backed by data of} 
    the (q)uestionnaire layer~(dashed edges).  The latter layer contains questions
    \emph{providing data about} (solid edges) 
    expectations and challenges (boxes in grey).  These expectations
    and challenges are \emph{formulated
     over}~(dotted edges) the \ConstructRef{StudyObjects} of safety practice~(framed boxes).
    For sake of brevity,
    classification criteria~(\Cref{tab:classification}) are omitted.
    \label{fig:researchdesign}
  }
\end{figure*}

For this objective and these research questions, we introduce the
construct \emph{safety practice and its practitioners~(SPP)}.  This
construct incorporates SPs' processes, tasks, roles, methods, tools,
and infrastructures and, by interrogating them via a questionnaire,
their views and opinions of safety practice.  SPP is divided into
three sub-constructs: \ConstructRef{class} of SPs,
\ConstructRef{StudyObjects} of safety practice, and
\ConstructRef{Challenges} \& challenges in safety practice.  The 
construct is visualised in \Cref{fig:researchdesign}.

\begin{table}[t]
  \caption{Classification criteria for characterising the population and
    for sample assessment.
    \textbf{Legend:} MC\dots multiple-choice, (N)ominal or (O)rdinal scale
    \label{tab:classification}}
  \footnotesize
  \begin{tabularx}{\columnwidth}{Xl}
    \toprule
    \textbf{\Construct{class}{Classification} Criterion} & \textbf{Scale} \\
    \midrule
    \SubConstruct{class}{CQ_1_edu}{Educational Background} & N / MC \\
    \SubConstruct{class}{CQ_2_dom}{Application Domains} & N / MC \\
    \SubConstruct{class}{CQ_3_exp}{Level of Experience} & O / duration in years \\
    \SubConstruct{class}{CQ_5_std}{Familiarity with Standards} & N / MC \\
    \SubConstruct{class}{CQ_5_meth}{Familiarity with Methods} & N / MC \\ %
    \SubConstruct{class}{CQ_4_geo}{Geographical Regions} & Open / MC \\
    \SubConstruct{class}{CQ_6_lang}{Native Languages} & N / MC \\ %
    \SubConstruct{class}{CQ_7_buslang}{Working Languages} & N / MC \\ %
    \SubConstruct{class}{CQ_8_role}{Safety-related Roles} & N / MC \\ %
    \bottomrule
  \end{tabularx}
\end{table}

\begin{table}[t]
  \caption{Constituents of safety practice and
    practitioner's expectations and challenges.
    \textbf{Legend:} (N)ominal or (O)rdinal scale, (T)ruth
    values as nominal scale, *~\dots half-open or open.
  }
  \label{tab:constructs}
  \footnotesize
  \begin{tabularx}{\columnwidth}{LL}
    \toprule
    \textbf{Construct}
    & \textbf{Scales}
    \\\midrule
    \multicolumn{2}{l}{\Construct{StudyObjects}{Constituents} \textbf{of Safety Practice}}
    \\\midrule
    \SubConstruct{StudyObjects}{SafetyActivities}{Safety Process} (activities, roles, and practitioners)
    & N / \egs decisions, hazard identification, resources
    \\ %
    \SubConstruct{StudyObjects}{ProcessFactor}{Factors} (constraints and issues)
    & T / \egs lack of resources, high schedule pressure
    \\
    \SubConstruct{StudyObjects}{Means}{Means} (conventional techniques; formal methods; tools; norms; %
      skills; knowledge sources)
    & N* / \egs FMEA, ISO26262, FMEA expertise, expert opinions
    \\
    \SubConstruct{StudyObjects}{CurrentApplication}{Application} domains (current, new, complex)
    & N* / \egs systems based on adaptive control, machine learning
    \\
    \midrule
    \multicolumn{2}{l}{\Construct{Challenges}{Expectations} \textbf{ \& Challenges in Safety Practice} (as perceived by SPs)}
    \\\midrule
    \SubConstruct{Challenges}{ProcessPerformance}{Performance} of safety activities
    & O / high \dots low performance
    \\ %
    \SubConstruct{Challenges}{AdequacyOfStandardsMethods}{Adequacy} of means
    & O / high \dots low adequacy
    \\
    \SubConstruct{Challenges}{SafSecInteraction}{Collaboration} between safety and security engineers
    & O / effective \dots ineffective collaboration
    \\ %
    \SubConstruct{Challenges}{ValueOfKnowledgeSource}{Value} of knowledge sources to SPs
    & O / high \dots low, per class of methods or standards 
    \\
    \SubConstruct{Challenges}{ImprovementOfSkills}{Adaptation} and improvement of SPs' skills
    & O / high \dots low self-improvement/adaptation
    \\
    \SubConstruct{Challenges}{NotionOfSafety}{Notion}, perception, and priority of safety activities
    & N*
    \\
    \SubConstruct{Challenges}{ValueOfSPs-Self}{Contribution} of SPs to system life cycle
    & O / high \dots low contribution
    \\\bottomrule	
  \end{tabularx}
\end{table}

The criteria for classification in \Cref{tab:classification} and the
break-down in \Cref{tab:constructs} are results from the first
author's experience from research in system safety, from
collaborations with industry, from expert interview transcripts, and
from the supervision of the three thesis projects documented
in~\cite{Nyokabi2017,Yang2016,Hussein2016}.  The bottom-up creation of
the SPP construct took place along the lines of grounded
theory~\cite{Corbin2015-BasicsQualitativeResearch} %
based on these materials and further experience gained during the
survey execution.

Below, we use the following prefixes when referencing important
content items:
\begin{inparaitem}
\item[RQ] for research questions,
\item[h] for working hypotheses, 
\item[q] for questions in the questionnaire, and
\item[F] for findings.
\end{inparaitem}
References will have the shape
$\langle X \rangle\langle Label \rangle[-\langle o \rangle]$ where
$X \in \{RQ,h,q,F\}$ and $o$ can refer to an answer option in the
questionnaire.  Additionally, we provide legends along with the
corresponding figures.

\subsection{Survey Participants and Population}
\label{sec:population}

Safety practitioners are our direct study subjects, our target group.
A \emph{safety practitioner} is a person whose professional activities
as a practitioner or researcher in industry or academia are tightly
related to the engineering of safety-critical systems.
\Cref{tab:classification} lists criteria we use to characterise and
identify members of the \emph{population} of SPs.
\emph{Safety practice}, as described in \Cref{sec:terminology}, is our
indirect study object.
SPs participating in our study are also called \emph{study or survey 
  participants} or \emph{respondents}.

\subsection{Data Collection Instrument: On-line Questionnaire}
\label{sec:questionnaire}

\Cref{tab:questions} provides details on the (q)uestions we discuss in
this work.
For sake of conciseness, concept traceability, and compact
presentation in this article, we consolidated the questions stated in
our questionnaire, of course taking care of maintaining their original
meanings.  For verification of this transformation, the whole original
questionnaire and its code book are documented in~\cite{Nyokabi2017}.

\begin{table}
  \caption{Scales used in the questionnaire
    \label{tab:scales}}
  \footnotesize
  \begin{tabularx}{1.0\linewidth}{lL}
    \toprule
    \textbf{Type}
    & \textbf{Values} \\
    \midrule
    \emph{value}
    & very high (\emph{vh}), (\emph{h})igh,
    (\emph{m})edium, (\emph{l})ow, very low (\emph{vl})
    \\
    \emph{agreement}
    & strongly agree (\emph{sa}), (\emph{a})gree,
    neither agree nor disagree (\emph{nand}), (\emph{d})isagree,
    strongly disagree (\emph{sd})
    \\
    \emph{impact}
    & (\emph{h})igh, %
    (\emph{m})edium, (\emph{l})ow, (\emph{n})o impact %
    \\
    \emph{adequacy}
    & very adequate (\emph{va}), (\emph{a})dequate,
    slightly adequate (\emph{sa}), not adequate (\emph{na})
    \\
    \emph{frequency}
    & often, rarely/occasionally, never; or all,
    many, few, none
    \\
    \emph{choice}
    & single/multiple: (\emph{ch})ecked, (\emph{un})checked; or yes, no
    \\
    \bottomrule
  \end{tabularx}
\end{table}

\begin{table*}[t]
  \caption{Transcription and summary of selected questions from the
    questionnaire.  \textbf{Legend:} Nominal, (O)rdinal,
    \textsc{(L)ikert}-type scale, (T)ruth values as nominal scale, MC\dots multiple-choice, * \dots
    half-open or open.  
    \Crefrange{fig:H3_RQ_E1}{fig:H14_RQ_C2} show details on the
    options; Sec./Fig. serves the navigation.
  }
  \label{tab:questions}
  \footnotesize
  \begin{tabularx}{\textwidth}{Llllr}
    \toprule
    \textbf{Question}
    & \textbf{Scale} (see \Cref{tab:scales})
    & \textbf{Sec.}
    & \textbf{Fig.}
    & \textbf{N}
    \\\midrule
    \Question{H3_RQ_E1}{ValueOfKnow} %
    Of how much \emph{value}\SubConstructRefP{ValueOfKnowledgeSource} are specific \emph{knowledge sources}\SubConstructRefP{Means} for safety decision making?
    & L* / value per source
    & \ref{sec:sum-H3_RQ_E1}
    & \ref{fig:H3_RQ_E1}
    & 97
    \\
    \Question{H1_RQ_G2}{ImpOfConstr} %
    To which extent do specific \emph{process constraints and issues}\SubConstructRefP{ProcessFactor} negatively \emph{impact safety activities}\SubConstructRefP{ProcessPerformance}?
    & O* / impact per factor
    & \ref{sec:sum-H1_RQ_G2}
    & \ref{fig:H1_RQ_G2}
    & 93
    \\
    \Question{H1_RQ_G3}{ImpOfEco} %
    How often do \emph{economic factors}\SubConstructRefP{ProcessFactor} have a strong influence on the \emph{handling of hazards}\SubConstructRefP{ProcessPerformance}?
    & O / frequency 
    & \ref{sec:sum-H1_RQ_G3}
    & -- %
    & 93
    \\
    \Question{H26_RQ_B4}{AdeqOfMthStd} %
    Regarding a specific \emph{application domain}\SubConstructRefP{CurrentApplication}, how \emph{adequate}\SubConstructRefP{AdequacyOfStandardsMethods} are applicable safety \emph{standards and methods}\SubConstructRefP{Means} in ensuring safety?
    & O / adequacy per domain
    & \ref{sec:sum-H26_RQ_B4}
    & \ref{fig:H26_RQ_B4}
    & 102
    \\
    \Question{H11_RQ_B2}{AppOfMeth}
    The \emph{application of conventional techniques}~(\egs FMEA and FTA)\SubConstructRefP{Means} has become too difficult\SubConstructRefP{AdequacyOfStandardsMethods} for
\emph{complex applications of recent technologies}\SubConstructRefP{CurrentApplication}.
    & L / agreement
    & \ref{sec:sum-H11_RQ_B2}
    & \ref{fig:H11_RQ_B2}
    & 97
    \\
    \Question{H11_RQ_B6}{PosImpOfFMs} %
    Estimate the \emph{positive impact}\SubConstructRefP{AdequacyOfStandardsMethods} of formal methods\SubConstructRefP{Means} on safety activities and
system safety.
    & O / impact
    & \ref{sec:sum-H11_RQ_B6}
    & \ref{fig:H11_RQ_B6}
    & 58
    \\
    \Question{H6_RQ_C1}{ImprOfSkills}
    Specify your level of agreement with 4 statements about
    \emph{factors\SubConstructRefP{Means}\SubConstructRefP{ValueOfKnowledgeSource}
    improving a SP's skills}\SubConstructRefP{ImprovementOfSkills}.
    & L / agreement per statement
    & \ref{sec:sum-H6_RQ_C1}
    & \ref{fig:H6_RQ_C1}
    & 96
    \\
    \Question{H22_RQ_D1}{IntOfSafSec} %
    Specify your level of agreement with 10 statements\SubConstructRefP{ProcessPerformance} about the \emph{interaction of safety and security}\SubConstructRefP{SafSecInteraction} activities.
    & L / agreement per statement
    & \ref{sec:sum-H22_RQ_D1}
    & \ref{fig:H22_RQ_D1}
    & 95
    \\
    \Question{H18_RQ_F5}{NotionOfSafety} %
    How is safety \emph{viewed}\SubConstructRefP{NotionOfSafety} in your field of practice?
    & Nominal* / MC
    & \ref{sec:sum-H18_RQ_F5}
    & \ref{fig:H18_RQ_F5}
    & 95
    \\
    \Question{H24_RQ_E3}{PrioOfSafety} %
    Specify your level of agreement with 4 statements about \emph{factors\SubConstructRefP{NotionOfSafety} increasing the efficiency in safety activities}\SubConstructRefP{ProcessPerformance}.
    & L / agreement per statement 
    & \ref{sec:sum-H24_RQ_E3}
    & \ref{fig:H24_RQ_E3}
    & 97
    \\
    \Question{H7_RQ_F1.a}{EffRoleOnJob} 
    Is your \emph{job
      affected}\SubConstructRefP{ProcessPerformance}\SubConstructRefP{ValueOfSPs-Self} by any predominant definition of
      your \emph{role}\SubConstructRefP{SafetyActivities}? In either case, we request
      for comment. 
    & T* / comment
    & \ref{sec:sum-H7_RQ_F1.a}
    & --
    & 91
    \\
    \Question{H18_RQ_F5.ab}{EffNotionOnJob} Is your \emph{job
      affected}\SubConstructRefP{ProcessPerformance}\SubConstructRefP{ValueOfSPs-Self}
    by any predominant \emph{view of
      safety}\SubConstructRefP{NotionOfSafety}? In either case, we
    request for comment.  & T* / comment & \ref{sec:sum-H18_RQ_F5.ab}
    & -- & 95
    \\
    \Question{H17_RQ_E2}{UndesiredEv} %
    Specify your level of agreement with 5 statements\SubConstructRefP{Means}\SubConstructRefP{ValueOfKnowledgeSource} about \emph{safety activities}\SubConstructRefP{NotionOfSafety}.
    & L / agreement per statement 
    & \ref{sec:sum-H17_RQ_E2}
    & \ref{fig:H17_RQ_E2}
    & 97
    \\
    \Question{H20_RQ_F2}{ValOfContrib} %
    Of how much \emph{value}\SubConstructRefP{ValueOfSPs-Self} is your role as a practitioner or researcher in safety-critical
system developments?
    & L / value
    & \ref{sec:sum-H20_RQ_F2}
    & \ref{fig:H20_RQ_F2}
    & 95
    \\
    \Question{H16_RQ_F4}{CoWorkers} %
    How much \emph{value}\SubConstructRefP{ValueOfSPs-Self} do non-safety co-workers attribute to the \emph{role of a safety practitioner}?
    & L / value 
    & \ref{sec:sum-H16_RQ_F4}
    & \ref{fig:H16_RQ_F4}
    & 95
    \\
    \Question{H14_RQ_C2}{ImpOfExp} %
    Specify your level of agreement with 2 statements about the
    \emph{role of
      experience\SubConstructRefP{Means}\SubConstructRefP{ValueOfKnowledgeSource}
      in safety activities}.
    & L / agreement per statement
    & \ref{sec:sum-H14_RQ_C2}
    & \ref{fig:H14_RQ_C2}
    & 96
    \\
    \bottomrule	
  \end{tabularx}
\end{table*}

\subsubsection{Motivations underlying the Questions}
\label{sec:disc-motiv-behind}

In the following, we establish links between the questions and other
research summarised in \Cref{sec:relwork}.

\paragraph{\QuestionRef{H26_RQ_B4}}

\mycite[Bloomfield and Bishop]{Bloomfield2009} contrast prescriptive
regulation with goal-based regulation, reviewing current
practice, highlighting potential benefits of safety cases along with
the challenge of gaining sufficient confidence.
Starting from a general position, question~\QuestionRef{H26_RQ_B4} is
about \emph{norms adequacy} in general.

For maturity measurement, \mycite[Ceccarelli and
Silva]{Ceccarelli2015} work with a construct similar to the one in
\Cref{tab:constructs}.  By asking question \QuestionRef{H26_RQ_B4}, we
cover practitioners' views (and opinions) independent of a specific
norm.

The questions about \emph{adequacy of means}
(particularly, \QuestionRef{H26_RQ_B4}, \QuestionRef{H11_RQ_B2},
\QuestionRef{H11_RQ_B6}),
aim at the re-examination of known challenges as, \eg discussed by
\mycite[Kasurinen et al.]{Kasurinen2010} and \mycite[Graaf et
al.]{Graaf2003}.

The answer categories for question~\QuestionRef{H26_RQ_B4} are based
on industry sectors with a relatively high pace of innovation and/or
new, complex, but not yet well-understood system applications (\egs
self-driving cars).

\paragraph{\QuestionRef{H3_RQ_E1}}

\mycite[Lethbridge et al.]{Lethbridge2003}
observe that test and quality documentation is the most likely
maintained kind of documentation.  
With question \QuestionRef{H3_RQ_E1}, we want to find out about how
project documentation is used in safety decision making. 

Moreover, \mycite[Rae and Alexander]{Rae2017a} examine how
confidence in safety expert judgements (\egs
individual versus group judgements) is justified and leads to actual
validity of the conclusions the further stages of the safety life
cycle are based on.  The authors argue that expert risk assessments
exhibit low effectiveness in measuring risk as an objective quantity
and propose ``risk assessment as a means of describing, rather than
quantifying risk.''  Their analysis extends the background of
\QuestionRef{H3_RQ_E1}.

\paragraph{\QuestionRef{H22_RQ_D1} and \QuestionRef{H24_RQ_E3}}

While \mycite[Chen et al.]{Chen2018a} focus on the
aspect of \emph{training and collaboration in safety assurance},
our study crosses these aspects generally with the questions
\QuestionRef{H22_RQ_D1} and \QuestionRef{H24_RQ_E3} about interaction
in and efficiency of safety activities.

The questions \QuestionRef{H22_RQ_D1}, \QuestionRef{H20_RQ_F2}, and
\QuestionRef{H16_RQ_F4} address the integration of safety activities
with the life cycle, similar to \mycite[Bjarnason et
al.]{Bjarnason2013} on the alignment of RE and verification and
validation.

In contrast to tool support for optimal auditing as investigated by
\mycite[Dodd and Habli]{Dodd2012}, our questions (\ies
\QuestionRef{H7_RQ_F1.a}, \QuestionRef{H20_RQ_F2},
\QuestionRef{H16_RQ_F4}, and \QuestionRef{H14_RQ_C2})
help to solicit personal views of SPs as external auditors and
consultants.

\paragraph{\QuestionRef{H1_RQ_G2},
\QuestionRef{H1_RQ_G3}, and \QuestionRef{H18_RQ_F5}}

As summarised in \Cref{sec:motivation} and as discussed in
\cite{Yang2016}, we presume negative consequences of ``accidental
complexity''~\cite{Brooks1995a} on system safety.  \mycite[Lim et
al.]{Lim2012} examine the perception of technical debt, highlighting
the inevitable trade-off between software quality and business value.
In an unfortunate case, an acceptance of technical debt can lead to an
acceptance of low software quality, and for some systems, to an
acceptance of accidental complexity.  Whenever this reasoning applies
to a safety-critical system, we should ask whether this system is
taken in by an unacceptable trade-off between safety and business
value?  Asking the questions \QuestionRef{H1_RQ_G2},
\QuestionRef{H1_RQ_G3}, and \QuestionRef{H18_RQ_F5}, we inversely
probe the demand for investigations of the safety impact of technical
debt.

Based on the SPP construct, we interrogate SPs about supportive
factors (\QuestionRef{H3_RQ_E1}, \QuestionRef{H24_RQ_E3}) and
obstacles (\QuestionRef{H1_RQ_G2}) in safety decision making, gathered
from our previous interviews in~\cite{Yang2016,Hussein2016}.

\subsubsection{Notes on the Questionnaire}
\label{sec:notes-questionnaire}

\label{sec:legend}
Some questions in \Cref{tab:questions} are half-open, \ie we allow
respondents to extend the list of given answer options.
The scales used for encoding the answers in the column ``Scale'' are
described in~\Cref{tab:scales}.  We treat value and agreement as a
5-level \textsc{Likert}-type scale.  Value, impact, adequacy, and
frequency scales are equipped with a ``do not know~(\emph{dnk})''
option.
Together with ``neither agree nor disagree~(\emph{nand})'' answers,
participants are given two ways to stay indecisive.  This way, we try
to reduce bias by forced responses.
From comparative analysis, we conclude that it is safe to
\textbf{discard} \emph{dnk}-answers and
missing answers from our analyses.

We expect survey participants to spend 20--30~minutes on the
questionnaire.  Although we do not collect personal data, they can
leave us their email address if they want to receive our results.

\subsection{Working Hypotheses}
\label{sec:hypotheses}

We derive working hypotheses from our
observations~(\Cref{sec:motivation}) from previous
research~\cite{Yang2016,Hussein2016,Nyokabi2017}.
\Cref{tab:hypotheses} contains two types of working
\emph{(h)ypotheses} we want to analyse and test with the data we
collect from the survey participants.  First, the \emph{base
  hypotheses} incorporate observations, assumptions, or prejudices,
either identified from our previous research or already made by other
researchers.  Additionally, we elaborate \emph{comparative hypotheses}
during \emph{exploratory analysis}~\cite{Streb2010} of the responses.

Some hypotheses in \Cref{tab:hypotheses} are directly measured by a
single compound question~(see, \egs\HypothesisRef{H14}
and~\QuestionRef{H14_RQ_C2}).  We do not collect data for each
individual construct referred to in such hypothesis-question pairs.

\Cref{fig:researchdesign} summarises the survey design presented in
\Crefrange{sec:hypoth-rese-quest}{sec:hypotheses} by showing important
interrelationships between the base hypotheses, the questions of the
questionnaire, and the parts of the SPP construct.

\begin{table*}[h]
  \caption{Overview of hypotheses (used as $H_1$ in the tests).
    \textbf{Legend}: See~\Cref{sec:questionnaire}.  The quantification
    ranges a--j refer to the answer options of the questions
    associated with the hypotheses, see
    \Crefrange{fig:H3_RQ_E1}{fig:H14_RQ_C2}.  The original
    questionnaire is documented in more detail in \cite{Nyokabi2017}.}
  \label{tab:hypotheses}
  \footnotesize
  \begin{tabularx}{\textwidth}{LL}
    \toprule
    \textbf{Hypothesis} %
    & \textbf{Supported if} \dots~(AC, \Cref{sec:acceptance-criteria}) %
    \\\midrule

    \multicolumn{2}{l}{\emph{Base Hypotheses}}
    \\\midrule
    
    \Hypothesis{H3}{ExpDecides}
    SPs' activities\SubConstructRefP{SafetyActivities} \emph{mainly depend on}
      (d) expert opinion and (g) experience from similar projects\SubConstructRefP{Means}.
    & $\forall o\in\{d,g\}\colon\median{\QuestionRefM{H3_RQ_E1}^o}{h,vh} \land o$ among 3 highest valued (of 7) knowledge sources $\land\;\median{\QuestionRefM{H14_RQ_C2}^a}{a,sa}$
    \\
    \Hypothesis{H1}{LoResLoSaf}
    There is a lack of resources\SubConstructRefP{ProcessFactor}
    \emph{that has a negative impact on} the performance of safety activities\SubConstructRefP{ProcessPerformance}. 
    & $\forall o\in\{a,d\}\colon \median{\QuestionRefM{H1_RQ_G2}^o}{m,h}
      \land
      \median{\QuestionRefM{H1_RQ_G3}}{often}
      \land
      \QuestionRefM{H18_RQ_F5}^c \leq 30$
    \\
    \Hypothesis{H26}{InsufStds}
    Safety activities for highly-automated applications\SubConstructRefP{CurrentApplication} \emph{lack support of}\SubConstructRefP{AdequacyOfStandardsMethods} appropriate standards and methods\SubConstructRefP{Means}.
    & For $\geq 5$ out of 7 domains $o$: $\median{\QuestionRefM{H26_RQ_B4}^o}{sa,na}$
    \\
    \Hypothesis{H11}{InsufMeth}
    Conventional methods (\egs FMEA, FTA)\SubConstructRefP{Means}
    \emph{are challenging to
    apply}\SubConstructRefP{AdequacyOfStandardsMethods} to complex modern applications\SubConstructRefP{CurrentApplication}.
    & $\median{\QuestionRefM{H11_RQ_B2}}{a,sa} 
      \land
      \QuestionRefM{H11_RQ_B6}_{\mathit{m+h}} > 25\%$
    \\
    \Hypothesis{H11b}{FMsImprSaf}
    The use of formal methods\SubConstructRefP{Means} \emph{has a positive impact on} the performance of safety activities\SubConstructRefP{AdequacyOfStandardsMethods}.
    & $\median{\QuestionRefM{H11_RQ_B6}}{m,h}$
    \\
    \Hypothesis{H6}{SPsAdptSkls}
    SPs improve their skills\SubConstructRefP{ImprovementOfSkills}
      \emph{towards} new applications\SubConstructRefP{CurrentApplication}, \egs \emph{by} studying recent
      results in safety research\SubConstructRefP{Means}.
    & $\forall o\in\{a,b\}\colon\median{\QuestionRefM{H6_RQ_C1}^o}{a,sa}$
    \\
    \Hypothesis{H8}{SafBySec}
    For current applications\SubConstructRefP{CurrentApplication}, the assurance of safety \emph{also depends strongly on}\SubConstructRefP{SafSecInteraction} the assurance of security.
    & $\forall o\in\{a,c,e,f\}\colon \median{\QuestionRefM{H22_RQ_D1}^o}{a,sa}$

    \\
    \Hypothesis{H18}{SafIsCost}
    Safety \emph{is more seen as}\SubConstructRefP{NotionOfSafety} a cost-increasing \emph{rather than} a cost-saving part in many application domains\SubConstructRefP{CurrentApplication}.
    & $\QuestionRefM{H18_RQ_F5}^a_{\mathit{ch}} > 60\%  %
      \land
      \forall o\in\{b,e\}\colon\QuestionRefM{H18_RQ_F5}^o_{\mathit{ch}} < 40\%$ %
    \\
    \Hypothesis{H22}{LoCollLoSaf}
    A lack of collaboration of safety and security engineers\SubConstructRefP{SafSecInteraction} \emph{has a negative impact on} safety activities\SubConstructRefP{ProcessPerformance}.
    & $\forall o\in\{h,i,j\}\colon \median{\QuestionRefM{H22_RQ_D1}^o}{a,sa}$
    \\
    \Hypothesis{H24}{HiPrioHiSaf}
    Prioritisation of safety in management decisions\SubConstructRefP{NotionOfSafety} enables SPs to \emph{perform their tasks more efficiently}\SubConstructRefP{ProcessPerformance}.
    & $\forall o\in\{a,b\}\colon \median{\QuestionRefM{H24_RQ_E3}^o}{a,sa}$
    \\
    \Hypothesis{H17}{SafIsRel}
    SPs understand safety as \emph{a special case of}\SubConstructRefP{NotionOfSafety} reliability. %
    & $\forall o\in\{a,e\}\colon \median{\QuestionRefM{H17_RQ_E2}^o}{a,sa}
      \land
      \forall o\in\{b,c\}\colon \median{\QuestionRefM{H17_RQ_E2}^o}{d,sd}$
    \\
    \Hypothesis{H16}{SafIsValued}
    SPs believe that their non-safety co-workers \emph{attribute high
value}~\SubConstructRefP{ValueOfSPs-Self} to SPs' contributions.
    & $\median{\QuestionRefM{H16_RQ_F4}}{h,vh}
      \land
      \median{\QuestionRefM{H20_RQ_F2}}{m,h,vh}$
    \\
    \Hypothesis{H20}{PosSelfImg}
    SPs perceive their \emph{contribution as highly valuable}\SubConstructRefP{ValueOfSPs-Self}.
    & $\median{\QuestionRefM{H20_RQ_F2}}{h,vh}
      \land
      \median{\QuestionRefM{H16_RQ_F4}}{m,h,vh}$
    \\\midrule
    
    \multicolumn{2}{l}{\emph{Comparative Hypotheses}}
    \\\midrule
    
    \Hypothesis{H14}{Exp:DivGTSing} %
    SPs with high diverse expertise\SubConstructRefP{Means}
    \emph{better perform in safety
      activities}\SubConstructRefP{ProcessPerformance}\SubConstructRefP{ValueOfSPs-Self})
    than SPs with low singular expertise\SubConstructRefP{Means}.
    &
    $\forall o\in\{a,b\}\colon
    \median{\QuestionRefM{H14_RQ_C2}^o}{a,sa} \land
    \median{\QuestionRefM{H14_RQ_C2}^c}{nand,a,sa}$
    \\
    \Hypothesis{C1}{Value:SenLTJun}
    \emph{Senior SPs}\SubConstructRefP{CQ_3_exp} attribute lower
      value to their role\SubConstructRefP{ValueOfSPs-Self} in the
      system life-cycle than \emph{junior
        SPs}\SubConstructRefP{CQ_3_exp} (\cf\HypothesisRef{H16}, \HypothesisRef{H20}).
    & One-sided $\mathbf{U}$ succeeds with $p < 0.05$
    \\
    \Hypothesis{C2}{Adapt:SenGTJun}
    \emph{Senior SPs} agree more than \emph{junior SPs}\SubConstructRefP{CQ_3_exp} that skill adaptation~(\egs
      learning) is required and takes
      place\SubConstructRefP{ImprovementOfSkills} (\cf\HypothesisRef{H6}).
    & One-sided $\mathbf{U}$ succeeds with $p < 0.05$
    \\
    \Hypothesis{C7}{Adapt:AutoGTAero}
    SPs using \emph{automotive standards} agree more than SPs using
      \emph{aerospace standards}\SubConstructRefP{CQ_5_std} that
      skill adaptation~(\egs learning) is required and takes
      place\SubConstructRefP{ImprovementOfSkills} (\cf\HypothesisRef{H6}).
    & One-sided $\mathbf{U}$ succeeds with $p < 0.05$
    \\
    \Hypothesis{C6}{InsufMeth:EngDifSci}
    \emph{Engineering-focused SPs} agree different from \emph{research-focused SPs}\SubConstructRefP{CQ_8_role} with \HypothesisRef{H11}.
    & Two-sided $\mathbf{U}$ succeeds with $p < 0.05$
    \\
    \Hypothesis{C8}{InsufMeth:AutoGTAero}
    SPs using \emph{automotive standards} agree more than SPs using \emph{aerospace standards}\SubConstructRefP{CQ_5_std} with \HypothesisRef{H11}.
    & One-sided $\mathbf{U}$ succeeds with $p < 0.05$
    \\\bottomrule	
  \end{tabularx}
\end{table*}

\subsubsection{Motivations underlying the Hypotheses}
\label{sec:acceptance}

In the following, we justify our working hypotheses through
establishing links to other research (\Cref{sec:relwork}).

\paragraph{\HypothesisRef{H3}: SPs' activities mainly depend on expert
  opinion and experience from similar projects} %

It is well-known that experts are fallible (see, \egs recent
investigations in \cite{Rae2017,Rae2017a,Wang2018}) and, thus, relying
on experts in organisational (and engineering) decision making can
contribute to critical single points of failures in such
organisations.  Moreover, it is well-known that reusing (\egs cloning)
repositories from finished projects in similar new projects bears many
risks of errors in reuse or update of these data.  Our previous
interviews suggest that both these knowledge sources are used in
safety practice.

\paragraph{\HypothesisRef{H1}: A lack of resources has a negative
  impact on the performance of safety activities} %

The observations in \Cref{sec:motivation} motivate the collection of
evidence on whether or not \emph{a lack of resources might have a
  negative impact on safety activities}.  For this hypothesis,
``negative impact'' refers to, \eg deferred safety decisions,
hindered hazard identification and implementation of hazard controls,
or limited SPs' abilities to fill their role.
The conjecture that \emph{budgets constrain safety activities} is
further inspired by ``the willingness to accept some technical risks
to achieve business goals'' as concluded by Lim et
al.~\cite[p.~26]{Lim2012}.

\paragraph{\HypothesisRef{H26}: Safety activities for highly-automated
  applications lack support of appropriate standards and methods} %

The belief that \emph{safety practice is missing adequate standards
  and methods} has been discussed by
\mycite[Cant]{Cant2013} and \mycite[Knauss]{Knauss2017}.
Questions about the \emph{appropriateness of methods and standards}
have also been raised by \mycite[McDermid and Rae]{McDermid2014}.  The
idea behind \HypothesisRef{H26} is to understand the situation of SPs
in new, not yet matured industry sectors.  SPs would have the
opportunity to adapt their skills and to gain further
expertise~(\HypothesisRef{H6}).

\paragraph{\HypothesisRef{H6}: SPs improve their skills towards new
  applications, \egs by studying recent results in safety research}

\mycite[Hatcliff et al.]{Hatcliff2014} observe that ``industry's
capability to verify and validate these systems has not kept up''~(we
inquire \emph{willingness to improve skills} with \HypothesisRef{H6})
and that ``the gap between practice and capability is increasing''
because of \emph{more integrated and more complex software
  technologies}.  In contrast to the compliance framework presented by
\mycite[Ceccarelli and Silva]{Ceccarelli2015}, Hatcliff et
al.~highlight that showing compliance with existing norms cannot
guarantee safety.  Our study touches \emph{norms adequacy} with
\HypothesisRef{H26}.

\paragraph{\HypothesisRef{H11}: Conventional methods (\egs FMEA, FTA)
  are challenging to apply to complex modern applications} %

The observation that \emph{conventional methods have become
  inadequate} is broached by
\mycite[Knight]{Knight2002,Knight2012}.
Likewise, \mycite[McDermid and Rae]{McDermid2014} and \mycite[Hatcliff
et al.]{Hatcliff2014} underpin \HypothesisRef{H26} and
\HypothesisRef{H11}, though not the
long-standing~\cite{Bloomfield1991} and frequent expectation that
\emph{formal methods~(FM) have a positive impact on safety
  practice}~(\HypothesisRef{H11b}).

\paragraph{\HypothesisRef{H11b}: The use of formal methods has a
  positive impact on the performance of safety activities} %

The \emph{efficacy of FMs in practice} has been an only moderately
researched subject for many years, investigated, \eg by
\mycite[Barroca and McDermid]{Barroca1992} and \mycite[Woodcock et
al.]{Woodcock2009}.  One intention underlying~\HypothesisRef{H11b} is
to determine whether we have to further examine FM effectiveness to
cross-validate reported experiences~(\egs\cite{Lockhart2014}).

\paragraph{\HypothesisRef{H8}: For current applications, the assurance
  of safety also depends strongly on the assurance of security} %

Safety-critical applications of networked or connected (software)
systems have recently revived the question of \emph{how safety and IT
  security influence each other?}  Along these lines, the
justification of~\HypothesisRef{H8} is based on manifold
\emph{anecdotal evidence}~(see, \egs\cite{Checkoway2011}) that
security problems can cause safety violations and, possibly, vice
versa.

\paragraph{\HypothesisRef{H18}: Safety is more seen as a
  cost-increasing rather than a cost-saving part in many application
  domains} %

How are the practical achievements and implications of \emph{system
  safety and the effort spent therefor} related?  How relevant are
such utilitarian and controversial questions to SPs and their
organisations?  Touching this subject, \HypothesisRef{H18} is
formulated in the context of ``total cost of safety,'' \ie the cost
of accident prevention and accident consequences borne by
organisations that engineer and operate safety-critical systems.
\HypothesisRef{H18}'s truth might contribute negatively to the role of
SPs in an (engineering) organisation.

\paragraph{\HypothesisRef{H22}: A lack of collaboration of safety and
security engineers has a negative impact on safety activities}

According to \mycite[Conway]{Conway1968-HowdoCommitees}, the structure
of an engineered system converges towards the (communication)
structure of its engineering organisation.  \Eg in a safety-critical
distributed embedded system (\egs avionics, process automation, and
automotive architectures), team collaboration would determine the
architectural decomposition and direct communication links in the
architecture.  However, team collaboration not necessarily implies
keeping track of the impact of critical changes across all critical
relationships.  It is also known that critical relationships in a
complex architecture are far from obvious.  Sadly, such relationships
are sometimes only indirectly perceived as an undesired \emph{emergent
  property}.  Hence, we ask SPs about the collaborations between
so-called ``property engineers,'' \egs safety and security
engineers~(\QuestionRef{H22_RQ_D1}).

\paragraph{\HypothesisRef{H17}: SPs understand safety as a special
  case of reliability} %

\mycite[Leveson]{Leveson2012} stresses an observed misconception about
system safety, namely that the responsibility to make systems
\emph{safe enough} is reduced to the responsibility to make their
\emph{critical parts just reliable enough}.  Her claim stimulates the
question to which extent SPs are solely driven by reliability concerns
and which negative implications this might have.  Moreover,
\HypothesisRef{H17} is also motivated by
examinations~\cite{Napolano2015} of how findings from previous
accidents can be included in safety arguments.

\paragraph{\HypothesisRef{C7}: SPs using automotive standards agree
  more than SPs using aerospace standards that skill adaptation is
  required and takes place}

During our interviews we heard several times that system safety
practice in the automotive domain is for several reasons less
developed than in other domains, such as aerospace.  Hence, we assume
that automotive SPs are currently more strongly involved in or aware
of skill development in their domain than SPs in aerospace.

\subsection{Data Collection Procedure: Sampling}
\label{sec:daten-collection}
\label{sec:sampling}

To draw a diverse sample of safety practitioners, we
\begin{enumerate}
\item advertise our survey on safety-related on-line discussion channels,
\item invite practitioners and researchers in safety-related domains
  from our social networks, and
\item ask these people to disseminate information about our survey.
\end{enumerate}
Our sampling procedure can best be described as a mixture of
opportunity, volunteer, and cluster-based sampling.  The cluster is
formed by survey participants from several of these channels.  We
expect to get a sample stronger than non-probabilistic but,
because of a lack of control of the sampling process, weaker than
uniformly random.

\paragraph{Sample Representativeness}
\label{sec:representation}

To check how well our final sample \emph{appropriately represents
  safety practice and its practitioners}, %
the questionnaire measures the classification criteria
in~\Cref{tab:classification}.  See~\cite{Nyokabi2017} for the question
used for this.

\subsection{Analysis Procedure}
\label{sec:data-analysis}

This section describes the analysis of the responses, the checking of
the working hypotheses, and our tooling.

\subsubsection{Analysis of Responses}
\label{sec:analysis-responses}

We use instruments of descriptive statistics~\cite{Haslam2009} such as
median~($\mathsf{med}$), %
mean~($\mu$), variance~($\mathsf{var}$), and frequency histograms to
summarise the responses per question.

\paragraph{Half-Open and Open Questions}
\label{sec:openquestions}

Some questions of our questionnaire are \emph{half-open}, \ie we allow
to add another answer option by providing an extra scale and a text
field, and some questions are \emph{open}, \ie we only provide a text
field.

Particularly, most demographic questions are half-open
multiple-choice~(MC) questions, \ie they have an extra text field
``Other''.  We use the answers from this text field to extend and
revise the classifications imposed by the given answer options.
See \Cref{sec:sample} for the results.

Furthermore, we close some of the main questions using qualitative content
analysis and coding~\cite{Neuendorf2016}.  For some half-open
questions, we extend the statement lists and nominal scales
accordingly.  The results of this step are shown in
\Cref{sec:summ-answ-quest} when discussing the questions in the
\Cref{sec:sum-H3_RQ_E1,sec:sum-H1_RQ_G2,sec:sum-H18_RQ_F5,sec:sum-H7_RQ_F1.a,sec:sum-H18_RQ_F5.ab}.

\subsubsection{Hypothesis Analysis and Statistical Tests}
\label{sec:tests}

We use \emph{non-statistical analysis} for all base hypotheses for
which we directly\footnote{\Eg our construct envisages
  hypothesis \HypothesisRef{H11} to refer to
  \emph{\SubConstructRef{AdequacyOfStandardsMethods} of means}.
  However, to keep our questionnaire lean, with question
  \QuestionRef{H11_RQ_B2}, we directly measure \emph{agreement} for
  one \emph{instance} of this hypothesis.} collect
data~(\Cref{tab:hypotheses}).

For most comparative hypotheses, we apply the
\textsc{Mann-Whitney}~$\mathbf{U}$ 
test~\cite{Haslam2009}~($\mathbf{U}$ for short) to check for
difference.
We use $\mathbf{U}$ if the following
assumptions hold:
\begin{itemize}
\item exactly one \textsc{Likert}-type or ordered-categorical dependent
  variable~(DV), %
\item random division into two
  groups, %
\item group members are not
  paired, %
\item treatments via independent variables~(IV) are already 
  applied, %
\item group sizes may differ and be small~($<30$),
\item per-group distributions of the DV may be
  dissimilar %
  and non-\textsc{Gaussian}.
\end{itemize}

Let $H$ be a hypothesis and $\alpha$ be the maximum chance of a
Type~I error, \ie incorrect rejection of the null
hypothesis~$H_0$. %
$\mathbf{U}$ tries to reject $H_0$ with a confidence 
of~$1-\alpha$.  We require $p<\alpha$ for the Type I error $p$ of
\emph{incorrectly distinguishing two groups of respondents} \wrt
$H$. %
If $\mathbf{U}$ succeeds to reject $H_0$ then the \textbf{support of
  the desired alternative hypothesis} $H_1$ is increased.  Failure of
$\mathbf{U}$ in rejecting $H_0$~(\ies $p \geq\alpha$) denies any
conclusion on $H$ from the given data set~\cite[p.~168]{Shull2008}.
The medium maturity %
and criticality %
of our hypotheses~(for an exploratory study) and the medium accuracy
of our data~(from a survey method) %
make it reasonable to stick with the typical choice of
$\alpha = 0.05$.

\paragraph{Acceptance Criteria (AC)}
\label{sec:acceptance-criteria}

The criteria in~\Cref{tab:hypotheses} describe the
aggregation of the question scales in~\Cref{tab:scales} to match the
hypotheses.  These criteria are built from symbols of the kind
$q\langle\mathit{id}\rangle^{\langle\mathit{answer\;
    option}\rangle}_{\langle\mathit{scale\; value(s)}\rangle}$
referring to the questions in~\Cref{tab:questions}.
We require $\mathsf{med}$ to be non-central to express a large
supportive majority.  Alternatively, percentage thresholds (\egs
$>25\%$) express the desired variance or shape of the distribution.
In hypothesis tests, we mainly use classification
criteria~(\Cref{tab:classification}) as IVs.

\subsubsection{Tooling}
\label{sec:tooling}

We use Unipark\footnote{See \url{http://www.unipark.de}.} as a
platform for implementing on-line surveys and for data
collection~(\Cref{sec:daten-collection}) and temporary storage.  For
statistical analysis and data visualisation~(\Cref{sec:tests})
we use GNU R\footnote{See \url{https://www.r-project.org}.} and
Unipark.  Content analysis and coding takes place in typical
spreadsheet applications.

\subsection{Validity Procedure after Survey Planning}
\label{sec:validity-procedure}

In the following, we evaluate the face and content validity of our
instrument, and the internal and construct validity of our study.
Although, we did not perform an independent pilot study according to
\cite{Kitchenham2008}, we took several measures to assess the validity
of our study.

\subsubsection{Instrument Evaluation: Face and Content Validity}
\label{sec:face-cont-valid}

First, both authors performed several internal walk-throughs to
improve the survey design and the data analysis procedure.

Second, along the lines of a \emph{focus group}, we asked independent
persons to complete the questionnaire and to provide feedback via an
extra form field in the questionnaire and via email.  This dry run
took place between 13 and 27 June 2017.  We gathered 7 independent
responses, from
2 postgraduate research assistants with experience in the survey method,
and with experience in safety-critical software,
systems, and requirements engineering,
1 master student with industrial work experience in safety-critical
systems engineering,
1 IT practitioner and English native speaker,
1 person with a health and safety background,
2 persons with a software engineering background.

The feedback from the these respondents resulted in
\begin{itemize}
\item an extension and balancing of answer options, 
\item the alignment of answer scales throughout the whole
  questionnaire,
\item improvement of the nomenclature (terms are now described on the
  questionnaire page they first appear).
\item an extension of open answer fields, and
\item linguistic improvements. 
\end{itemize}
These steps helped us to improve questionnaire completeness,
consistency, and comprehensibility and reduce researcher
bias~\cite[Sec.~3.3.4]{Nyokabi2017}.

\subsubsection{Internal Validity of the Analysis Procedure} %
\label{sec:validity-internal}

Why would the procedure in \Cref{sec:research-design} lead to
reasonable and justified results?

$\mathbf{U}$ is applicable only if groups are independent \wrt the
considered IV. %
This circumstance can cause problems with MC-questions.  \Eg with
\QuestionRef{H18_RQ_F5}, the same respondents might be in both groups
``all data points with choice (c)'' and ``all data points with choice
(a).''  Hence, these groups contain data points that can be dependent
in a certain but unknown way.  For comparisons with $\mathbf{U}$, we
reduce this issue by converting the responses to single-choice
questions using discriminating features such as the time of the first
answer option chosen.

The 7 test data points allowed us to validate our tooling~(\egs R
scripts, see \Cref{sec:tooling}).  Test data points are not included
in the final data set.

\subsubsection{Construct Validity} %
\label{sec:validity-construct}

Why would the construct~(\Cref{sec:construct}) appropriately
represent the phenomenon to investigate?

Because of the exploratory nature of our study, the sub-constructs and
their scales in \Cref{tab:constructs} represent the study object as
reconstructed from our analyses.  The working hypotheses and the
questionnaire represent an \emph{approximation and a selection} of
what needs to be measured and tested if we were to investigate this
study object~(\cf\Cref{fig:researchdesign}) in an explanatory study.
\Eg we assume that the 10 statements in \Cref{fig:H22_RQ_D1} for
question \QuestionRef{H22_RQ_D1} satisfactorily approximate the
``interaction of safety and security activities''~(\ies construct
\SubConstructRef{SafSecInteraction}) and its criticality.
Consequently, the scales in \Cref{tab:constructs} serve as a reference
to the internal validation of our study.

Not being unusual for an exploratory study, several of the hypotheses
are relatively weak and, therefore, even if accepted from our
collected data, only allow the derivation of \emph{restricted
  conclusions}. \Eg an accepted \HypothesisRef{H11b}~(\ies FMs have a
positive impact) reflects very much the personal experience,
perception, or opinion of our survey participants.  Their view has to
be distinguished from the question of actual FM effectiveness.  To
pursue such a question, we have to refine our research design using
the \emph{technology acceptance model}~\cite{Lee2003} and %
\emph{controlled field experiments}.

Our inquiry of SPs about supportive factors and obstacles in
safety decision making does not include safety evidence traceability
and management.  A future version of our construct and questionnaire
should therefore include the criteria examined by \mycite[Nair et
al.]{Nair2015}, \mycite[De la Vara et al.]{Vara2016} and \mycite[Borg
et al.]{Borg2016}.

Our construct and instrument are essentially new.  However, the
\ConstructRef{StudyObjects} overlap with the construct used
in~\cite{Ceccarelli2015} for safety process maturity assessment.
Furthermore, \mycite[Manotas et al.]{Manotas2016} employs a research
design analogous to ours, underpinning the appropriateness of our
approach to survey engineering practitioners.
Despite the drawbacks discussed before, we believe our design is
appropriate \wrt the expressive power of the working hypotheses.  In
summary, this construct can be a helpful guidance in the design of
successive studies.

\subsubsection{Reliability}
\label{sec:reliability}

A check for test-retest reliability (\egs changing attitudes of
respondents) and alternate form reliability are out of scope of this
exploratory study.  Hence, we do not plan to ask respondents to answer
the questionnaire more than once and
we run only one variant of the
questionnaire.

\section{Survey Results}
\label{sec:results}

In this section, we characterise our sample~(\Cref{sec:sample}),
summarise the responses~(\Cref{sec:summ-answ-quest}), and analyse our
hypotheses~(\Cref{sec:hypotheses-tests}).

\subsection{Survey Execution: Sample Size and Response Rate}
\label{sec:descr-data-points}

For the collection of data from the survey participants, we
\begin{enumerate}
\item advertised our survey over the channels in \Cref{tab:adplatforms}
  and
\item personally invited $>20$ persons.
\end{enumerate}
The sampling period lasted \emph{from 1 July 2017 til 25 September
  2017}.  In this period, we \emph{repeated step 1 up to three times}
to increase the number of participants.  The Unipark tracking data
shows that LinkedIn groups, ResearchGate, Twitter, and mailing lists
were effective in soliciting respondents, however, it is incomplete
and, hence, does not disclose which channels were most effective.

\begin{savenotes}
\begin{table}
  \caption{Safety-related channels we advertised our survey on (sorted
    alphabetically by category, full list in~\cite[pp.~92f]{Nyokabi2017})
    \label{tab:adplatforms}}
  \footnotesize
  \begin{tabular}{p{2cm}l}
    \toprule
    \textbf{Channel Type} & \textbf{Example/References}
    \\\midrule
    Facebook sites & \Egs Int.~Society of SPs \\ %
    General panels & SurveyCircle, \url{www.surveycircle.com} \\ 
    LinkedIn groups & \Egs on ARP 4754, DO-178, ISO 26262\\ %
    Mailing lists & \Egs system safety %
    (U Bielefeld,\footnote{See~\url{http://www.systemsafetylist.org}.}
    formerly U York) \\
    Newsletters & GI requirements engineering \\
    Personal websites & \Egs profiles on Twitter, LinkedIn, Xing \\ %
    ResearchGate & Q\&A forums on \url{www.researchgate.net} \\ %
    Xing groups & \Egs safety engineering \\ %
    Other channels & \Egs board of certified safety professionals \\
    \bottomrule
  \end{tabular}
\end{table}
\end{savenotes}

\begin{figure}
  \centering
  \includegraphics[width=.8\linewidth]{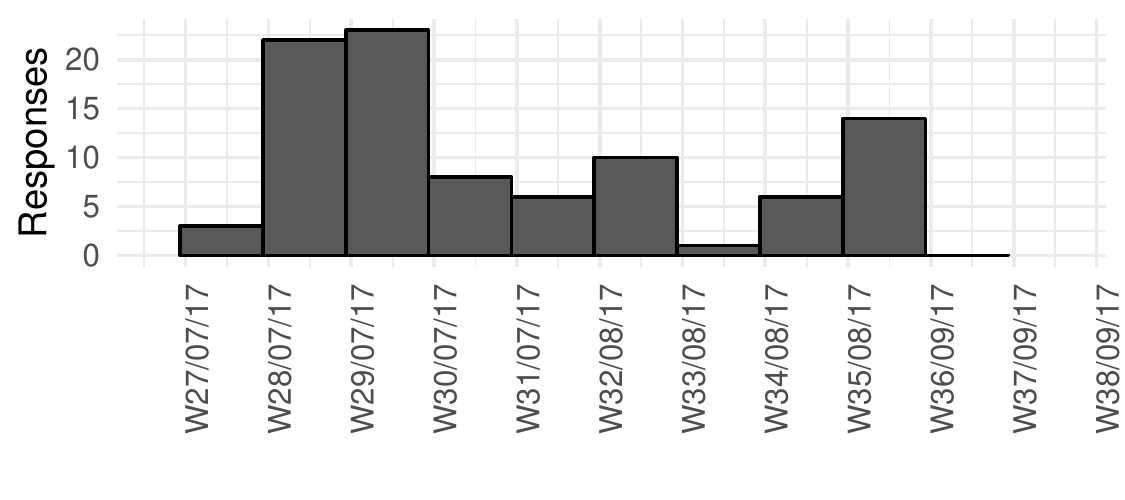}
  \caption{History of responses
    \label{fig:responsehist}}
\end{figure}

After 565 views of the questionnaire, our final sample contains
$N'=124$ (partial) responses with $N=93$ completed questionnaires and
$N=91$ (73\%) complete\footnote{Apart from two options of the
  classification question \SubConstructRef{CQ_8_role} (66, 76) and the
  question \QuestionRef{H11_RQ_B6} (62), we had at least 91 up to 124
  responses for each question.} data points.
\Cref{fig:responsehist} depicts the distribution of responses over
time.  According to our questionnaire tool, respondents spent
20~minutes on average to provide complete data points, 50\% spent
within 14 and 24 minutes time.

Given the numbers of members for some channels we used~(\egs for
LinkedIn groups), we estimate the return rates of responses per
channel to range from $~0.1$ to $~5\%$.

\label{sec:summary-sample}
From the sub-groups, we can build from the sample according to our
classification criteria, the smallest we like to reason about
below are of the size of around 15.

\Cref{tab:datasummary} provides a detailed enumeration of data
summaries (\ies number of answers per option) for all questions.

\subsection{Description of the Sample}
\label{sec:sample}

We describe our sample in the following and estimate the extent to
which it represents~(\Cref{sec:sampling}) the population of SPs.  For
each classification criterion according to~\Cref{tab:classification},
we provide a chart or we name the up to 10 most frequently occurring
answers, ordered by frequency.  Percentages (\%) right of the bars
indicate the fraction of the 93 completed questionnaires, shown in
parentheses the (N)umber of respondents who chose the corresponding
option.
Note that most of the classification questions allow MC
answers~(\cf\Cref{tab:classification} and
\cite[pp.~56ff]{Nyokabi2017}).

\paragraph{\SubConstructRef{CQ_1_edu}} %

\begin{figure}[t]
  \raggedleft
  \includegraphics[width=.97\columnwidth]{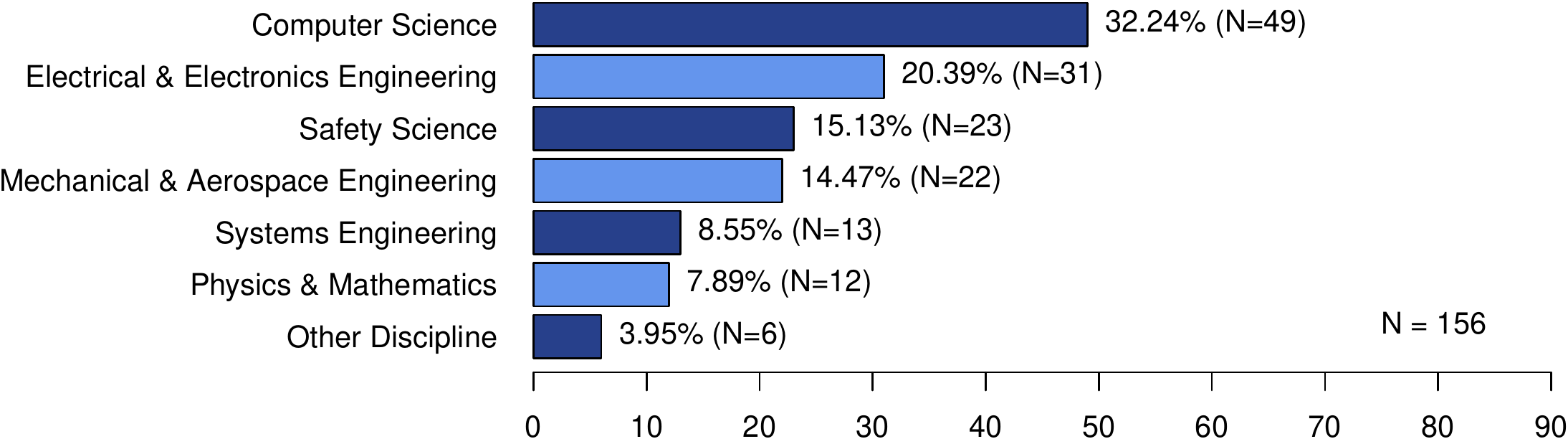}
  \caption{
    \SubConstructRef{CQ_1_edu} (frequency, MC)
    \label{fig:CQ_1_edu}
  }
\end{figure}

\Cref{fig:CQ_1_edu} summarises the \emph{educational background} of all
respondents:
\begin{itemize}
\item \emph{Computer scientists} include software engineers and computer
  engineers
\item \emph{Electrical and electronics engineers}
\item \emph{Safety scientists} include safety engineers, occupational safety
  practitioners, health and safety practitioners, human factors
  engineers, ergonomics engineers
\item \emph{Mechanical and aerospace engineers}
\item \emph{Systems engineers} include poly-technical systems engineers,
  information systems engineers, business technologists, engineering
  business administrators, engineering project managers
\item \emph{Physicists and mathematicians}
\item \emph{Other discipline} includes chemists, biochemists, civil
  engineers, language scientists
\end{itemize}

\paragraph{\SubConstructRef{CQ_2_dom}}

\begin{figure}[t]
  \centering
  \includegraphics[width=\columnwidth]{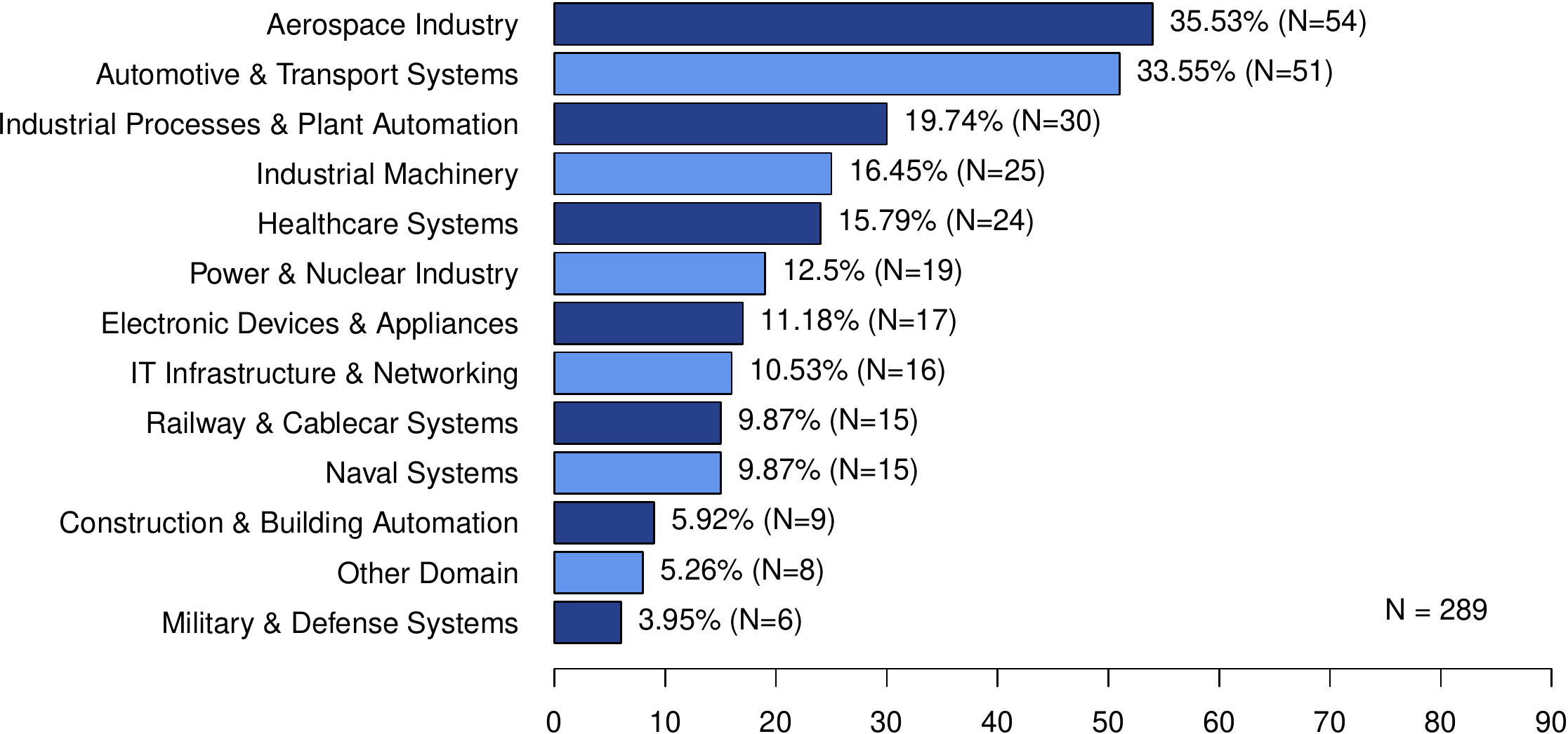}
  \caption{
    \SubConstructRef{CQ_2_dom} (frequency, MC)
    \label{fig:CQ_2_dom}
  }
\end{figure}

\Cref{fig:CQ_2_dom} summarises the \emph{application domain} of all
respondents where
``aerospace'' includes space telescopes;
``industrial processes and plant automation'' includes manufacturing,
chemical processes, oil and gas, energy infrastructure, and small
power plants;
``railway systems'' includes railway signalling;
``construction and building automation'' includes civil engineering
applications; and
``other domains'' includes food safety, biological safety, research and
development, and environment, health, and safety preparations.

\paragraph{\SubConstructRef{CQ_3_exp}}

\begin{figure}[t]
  \centering
  \includegraphics[width=.7\columnwidth]{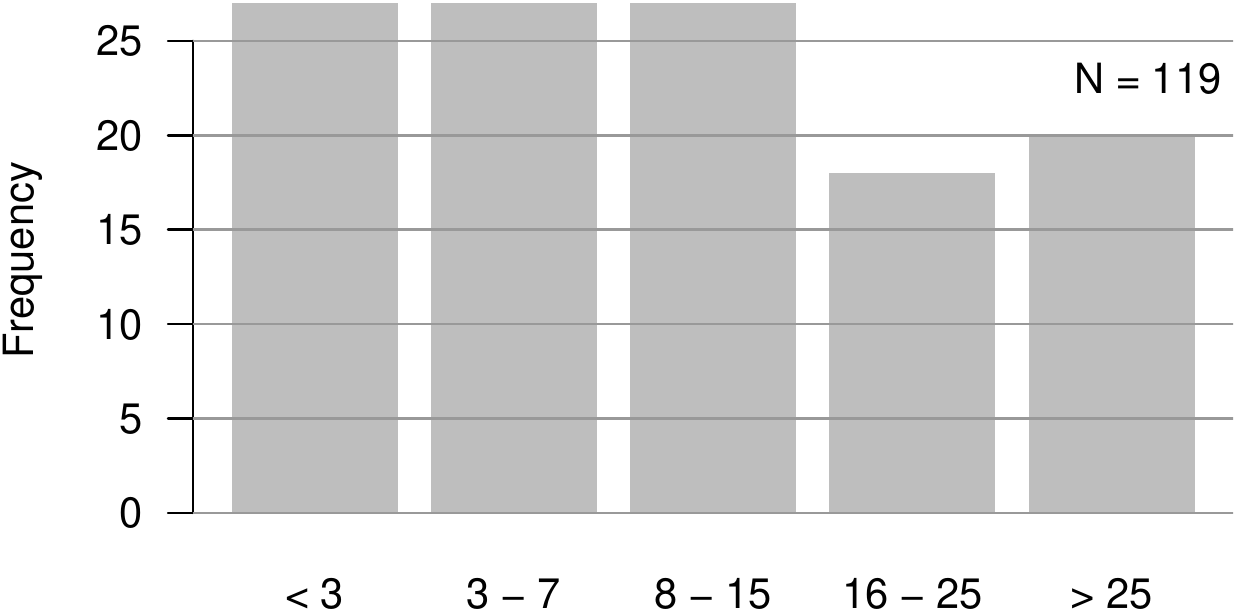}
  \caption{
    \SubConstructRef{CQ_3_exp} (time intervals in years)
    \label{fig:CQ_3_exp}
  }
\end{figure}

\Cref{fig:CQ_3_exp} indicates that our sample of SPs is moderately
balanced across all experience levels.

\paragraph{\SubConstructRef{CQ_5_std}} %

\begin{figure}[t]
  \raggedleft
  \includegraphics[width=.92\columnwidth]{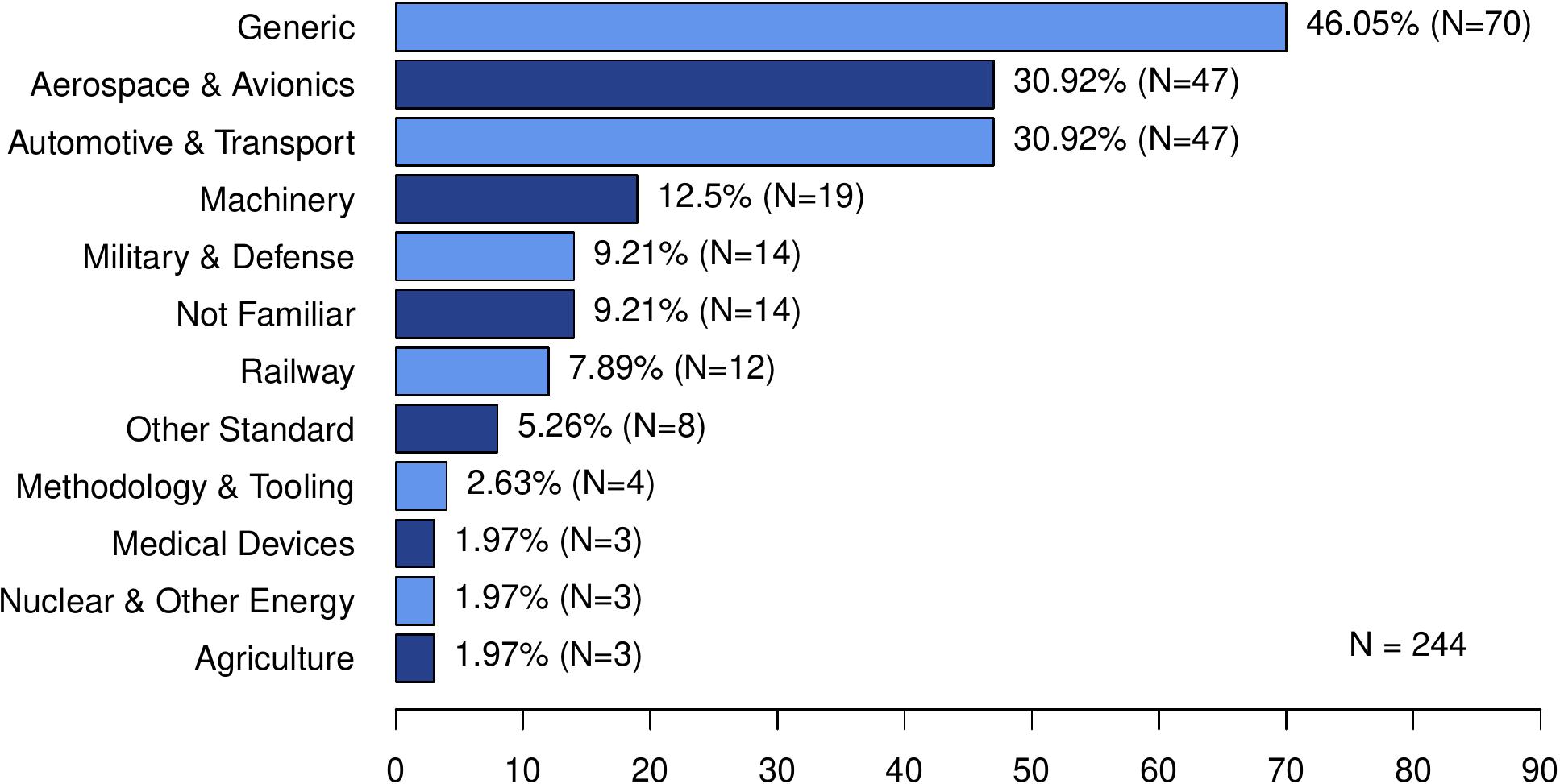}
  \caption{
    \SubConstructRef{CQ_5_std} and
    use by domain~(frequency, MC)
    \label{fig:CQ_5_std}
  }
\end{figure}

\Cref{fig:CQ_5_std} provides an overview of \emph{safety-related}
standards our respondents are familiar with (distinguished by
generality or by application domain):
Standards from aerospace~(\egs ARP 4761, DO-178, DO-254),
generic standards~(\egs ISO 61508, DIN VDE 0801)
automotive~(\egs ISO 26262),
machinery~(\egs ISO 13849, 25199, DIN EN 62061, MRL 2006/42/EG),
military~(\egs MIL-STD 882, UK Def Std 00-55),
railway~(\egs CENELEC EN 50126, 50128, 51029, 62061),
power plants~(\egs IEC 60880, 61513, 62138, 60987, 62340, IEC 800), and
medical devices~(\egs IEC 80001, ISO 14971, AAMI/UL 2800).
14 participants were neither familiar with any of the given 
standards nor did they specify other standards.

\paragraph{\SubConstructRef{CQ_5_meth}} %

\begin{figure}[t]
  \raggedleft
  \includegraphics[width=.88\columnwidth]{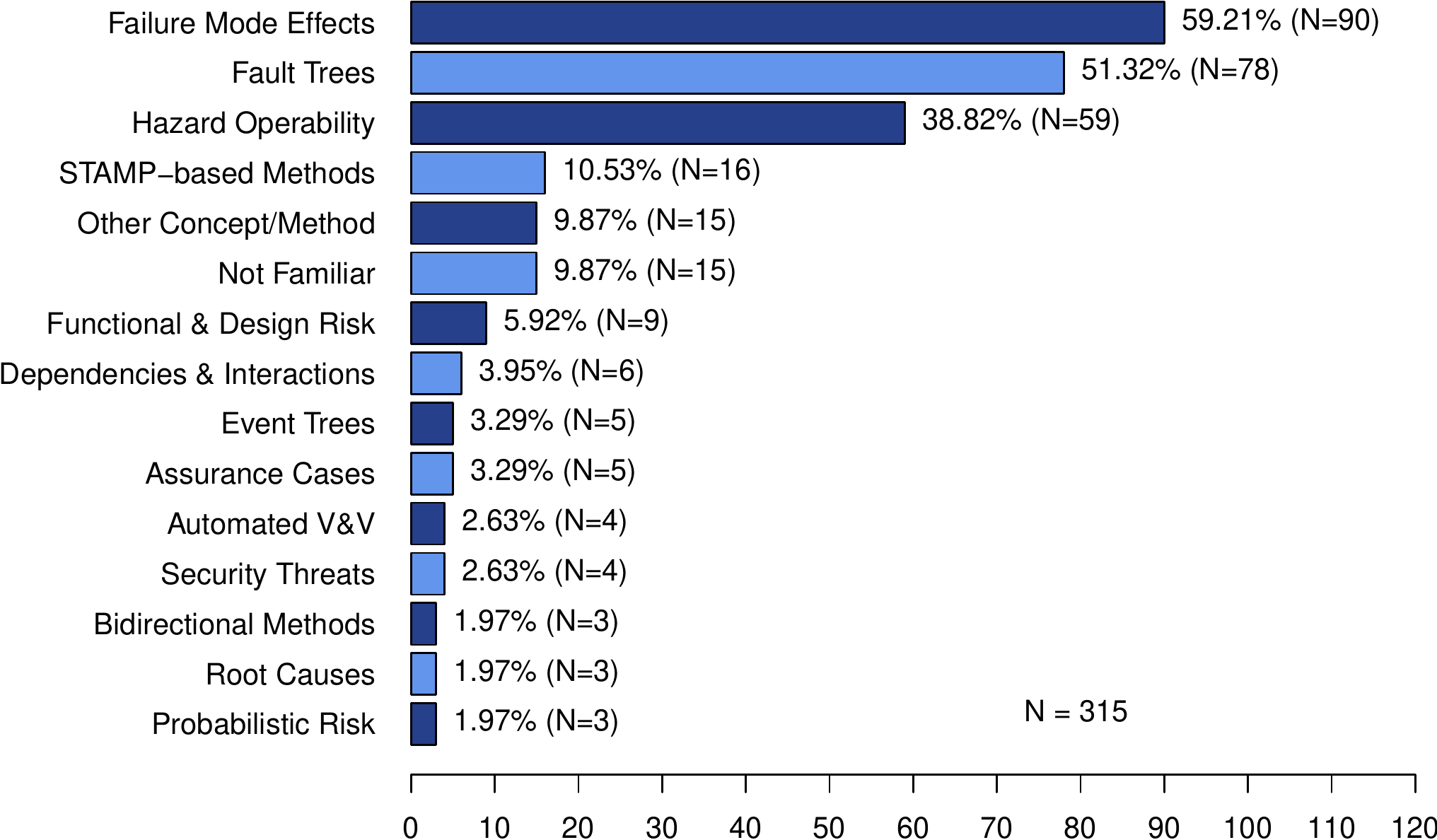}
  \caption{
    \SubConstructRef{CQ_5_meth} and concepts~(frequency, MC)
    \label{fig:CQ_5_meth}
  }
\end{figure}

\Cref{fig:CQ_5_meth} shows the familiarity of our respondents with
prevalent \emph{concepts of safety analysis} and the corresponding
\emph{classes of methods}, techniques, or
notations:\footnote{Abbreviations are described in \Cref{tab:abbrev}
  in \Cref{sec:list-abbreviations}.} \Eg
FMEA, FMECA, or FMEDA to assess \emph{failure mode effects};
HazOp studies, ergonomic work analysis and intervention methodology to
assess \emph{hazard operability};
\emph{STAMP-based methods} for hazard~(STPA) and accident~(CAST) analysis;
FHA, FFA, PHA, or PHL to assess risk at a \emph{functional or
abstract design level};
common cause~(CCA) or common mode~(CMA) analysis to include \emph{dependencies and interactions};
fault injection and property checking as techniques of \emph{automated}
validation and verification~(V\&V);
STRIDE or CORAS to assess and handle \emph{security threats};
\emph{bidirectional} methods such as Bowties or cause-consequence analysis;
\textsc{Markov} chains for \emph{probabilistic} risk analysis, and
GSN and SACM to build \emph{assurance cases}.
For ``Other'', our participants mentioned a
variety of approaches~(no more than twice):
5S, %
5W, %
CASS, %
coexistence analysis,
FRAM, %
HazRAC,
HEART,
HRA, %
MTA,
(O)SHA,
SAR,
SCRA,
SHARD,
SSHA, 
Poka Yoke, %
prognostic analysis, %
WBA,
ZHA, ZSA.

Only 4 respondents state familiarity with methods to assess and handle
\emph{security threats}.  15 respondents \emph{neither checked any of
  the given methods nor did they specify other methods} that are
relevant in their safety activities.

\paragraph{\SubConstructRef{CQ_4_geo}} %

DE (24.3\%), UK (16.4\%), US (15.3\%), AU (6.2\%), FR (5.1\%), IT
(3.4\%), CA (3.4\%), CN (2.8\%), and CH (2.8\%).

\paragraph{\SubConstructRef{CQ_6_lang}} %

\begin{figure}[t]
  \raggedleft
  \includegraphics[width=.8\columnwidth]{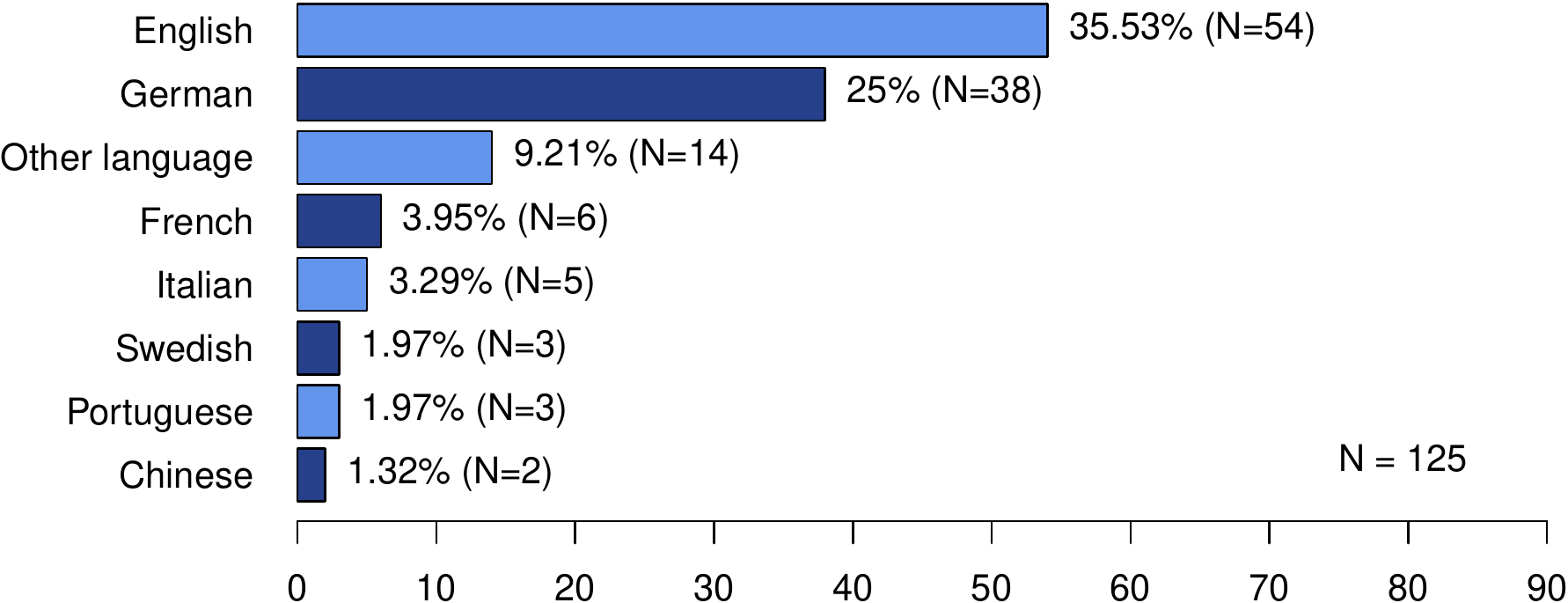}
  \caption{
    \SubConstructRef{CQ_6_lang} and concepts~(frequency, MC)
    \label{fig:CQ_6_lang}
  }
\end{figure}

\Cref{fig:CQ_6_lang} provides an overview of the \emph{languages
  spoken} by the respondents.

\paragraph{\SubConstructRef{CQ_7_buslang}}

\begin{figure}[t]
  \raggedleft
  \includegraphics[width=.82\columnwidth]{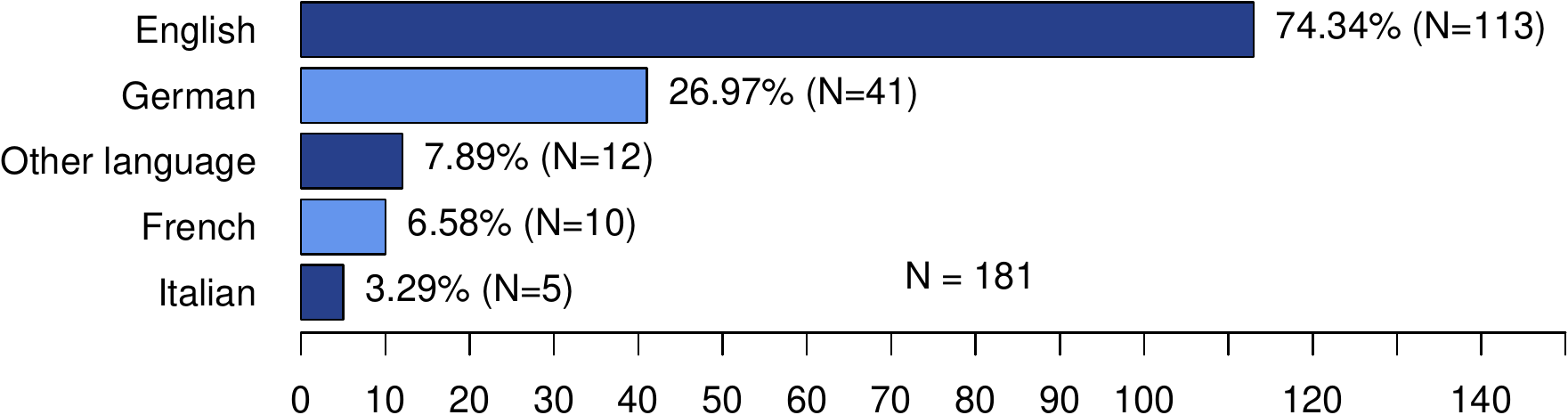}
  \caption{
    \SubConstructRef{CQ_7_buslang} and concepts~(frequency, MC)
    \label{fig:CQ_7_buslang}
  }
\end{figure}

\Cref{fig:CQ_7_buslang} provides an overview of the \emph{languages
  used at work} by the respondents.

\paragraph{\SubConstructRef{CQ_8_role}} %

\begin{figure}[t]
  \raggedleft
  \includegraphics[width=\columnwidth]{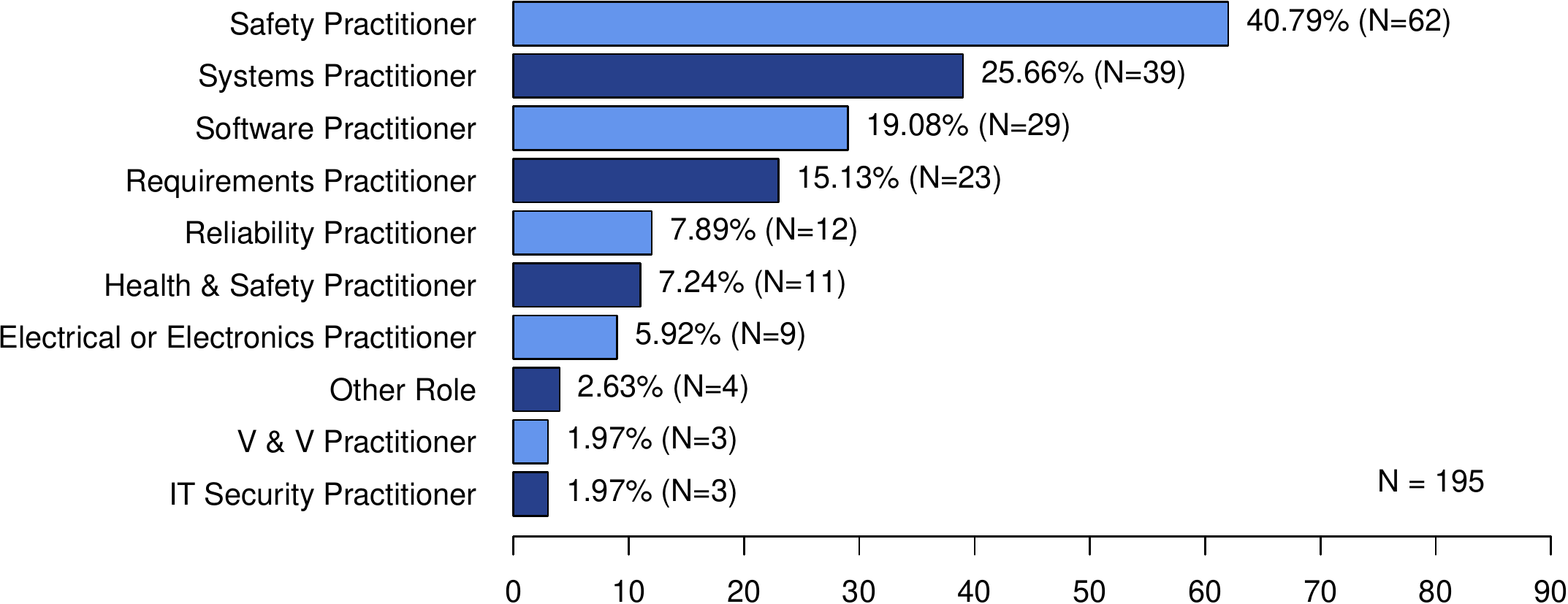}
  \includegraphics[width=.95\columnwidth]{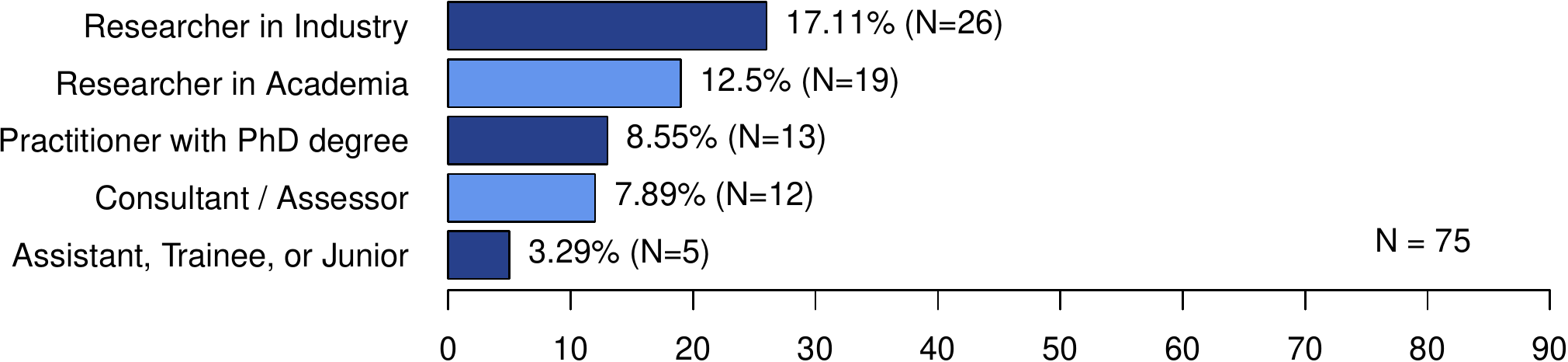}
  \caption{
    \SubConstructRef{CQ_8_role} (frequency, MC)
    split into 
    disciplines~(top) and responsibility profiles~(bottom)
    \label{fig:CQ_8_role}
  }
\end{figure}

In \Cref{fig:CQ_8_role}, the term \emph{practitioner} includes
the profile of an engineer and a manager.
Regarding engineering \emph{disciplines} and domains, ``safety practitioner''
includes engineers or managers in system safety, functional safety, or
in other safety domains as well as technology risk managers in
general; ``software practitioner'' includes developers, architects,
and tool developers; ``systems practitioners'' includes system
analysts and system architects; ``health \& safety practitioner''
includes occupational safety practitioners, human factors engineers,
and ergonomists; and ``V \& V practitioner'' includes test and
assurance practitioners.
For ``Other'', our respondents include a civil engineer, a project
manager, a method engineer, and a maintainability engineer.

Regarding \emph{responsibility profiles}, the category ``Consultant /
Assessor'' includes independent evaluators, auditors, regulators, and
inspectors dealing with safety certification.

\subsection{Summary of Responses}
\label{sec:summ-answ-quest}

In this section, we summarise the responses to the questions in
\Cref{tab:questions}.

\paragraph{Guide to the Figures}

The following text and figures complement each other.  For
\textsc{Likert}-type scales, we use centred diverging stacked bar
charts as recommended by~\cite{Robbins2011}.
$\mathsf{med}$ denotes the median and ``ex'' indicates the number of
excluded data points per answer option.

\subsubsection{\QuestionRef{H3_RQ_E1}: Value of Knowledge Sources}
\label{sec:sum-H3_RQ_E1}

\begin{figure}[t]
  \includegraphics[width=\columnwidth]{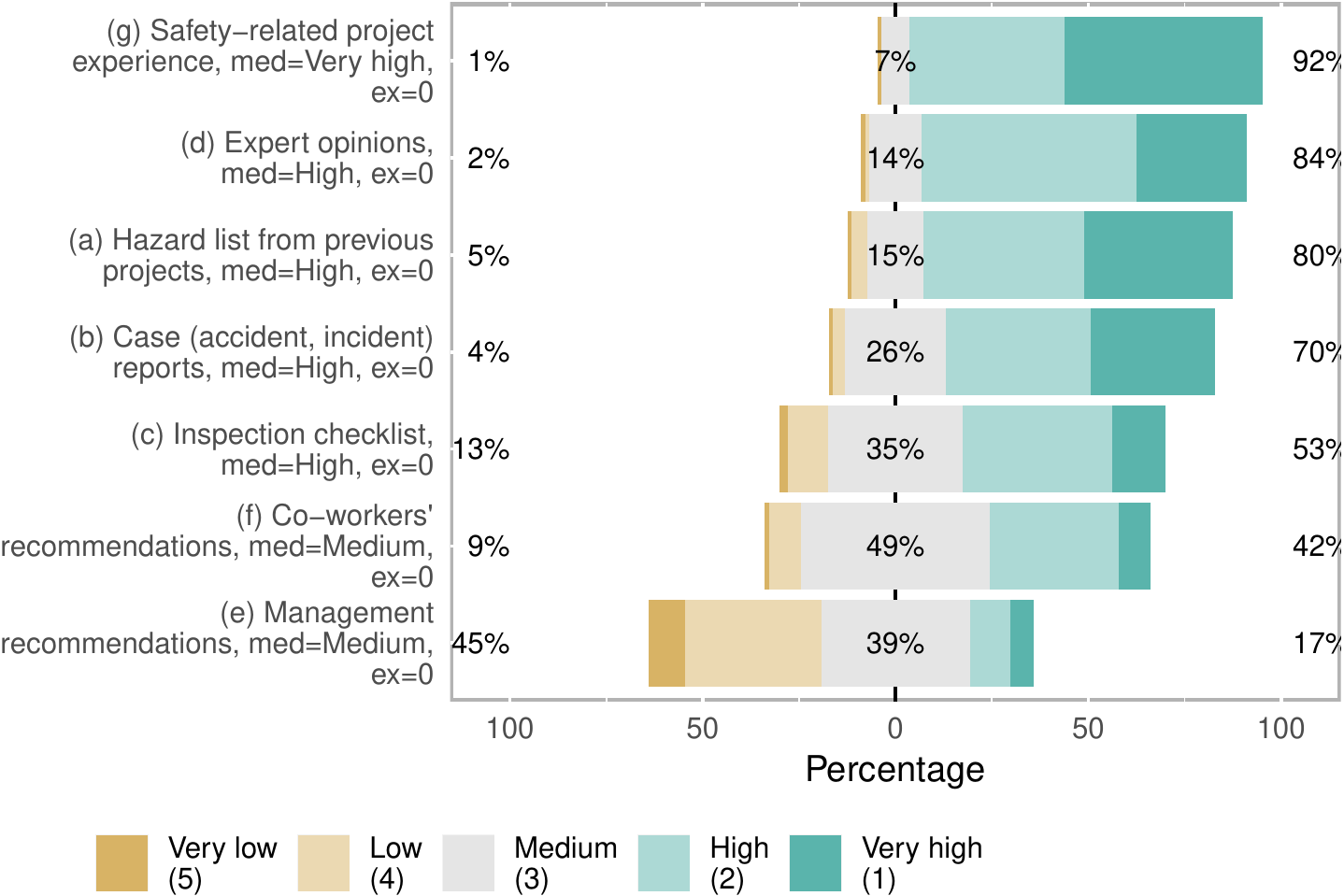}
  \caption{
    \QuestionRef{H3_RQ_E1}~($N=97$): Value of knowledge sources --
    Of how much \emph{value}\SubConstructRefP{ValueOfKnowledgeSource}
    are specific \emph{knowledge sources}\SubConstructRefP{Means}
    for safety decision making?
    \label{fig:H3_RQ_E1}}
\end{figure}

\Cref{fig:H3_RQ_E1} shows that, among the knowledge sources we asked
our participants to rate, \emph{expert opinion}, \emph{previous
  experience in safety-related projects}, and \emph{case reports}
represent the three highest valued knowledge sources used in safety
activities and safety decision making.  \emph{Management
  recommendations} turn out to be the lowest valued knowledge source.

The following knowledge sources, or resources in more general, were
additionally mentioned to be of \emph{very high or high value}:

Four respondents referred to the concept of \emph{adversarial
  thinking}, mentioning ``creative mind'', ``imagination'', ``analysis
capability,'' and ``acceptance of human fallibility.''  Three
respondents pointed to the concept of \emph{domain expertise and
  experience}, mentioning ``gut feel'', ``subject matter knowledge of
the application,'' and ``...real work and related
problems in reference situations...''
Furthermore, they mentioned
\emph{education},
\emph{specification documents and tools}~(\egs ``use of SPARK''),
\emph{independent assessment},
\emph{in-service monitoring logs}, and
\emph{previously certified similar systems}.

\subsubsection{\QuestionRef{H1_RQ_G2}: Constraints on Safety Activities}
\label{sec:sum-H1_RQ_G2}

\begin{figure}
  \includegraphics[width=\columnwidth]{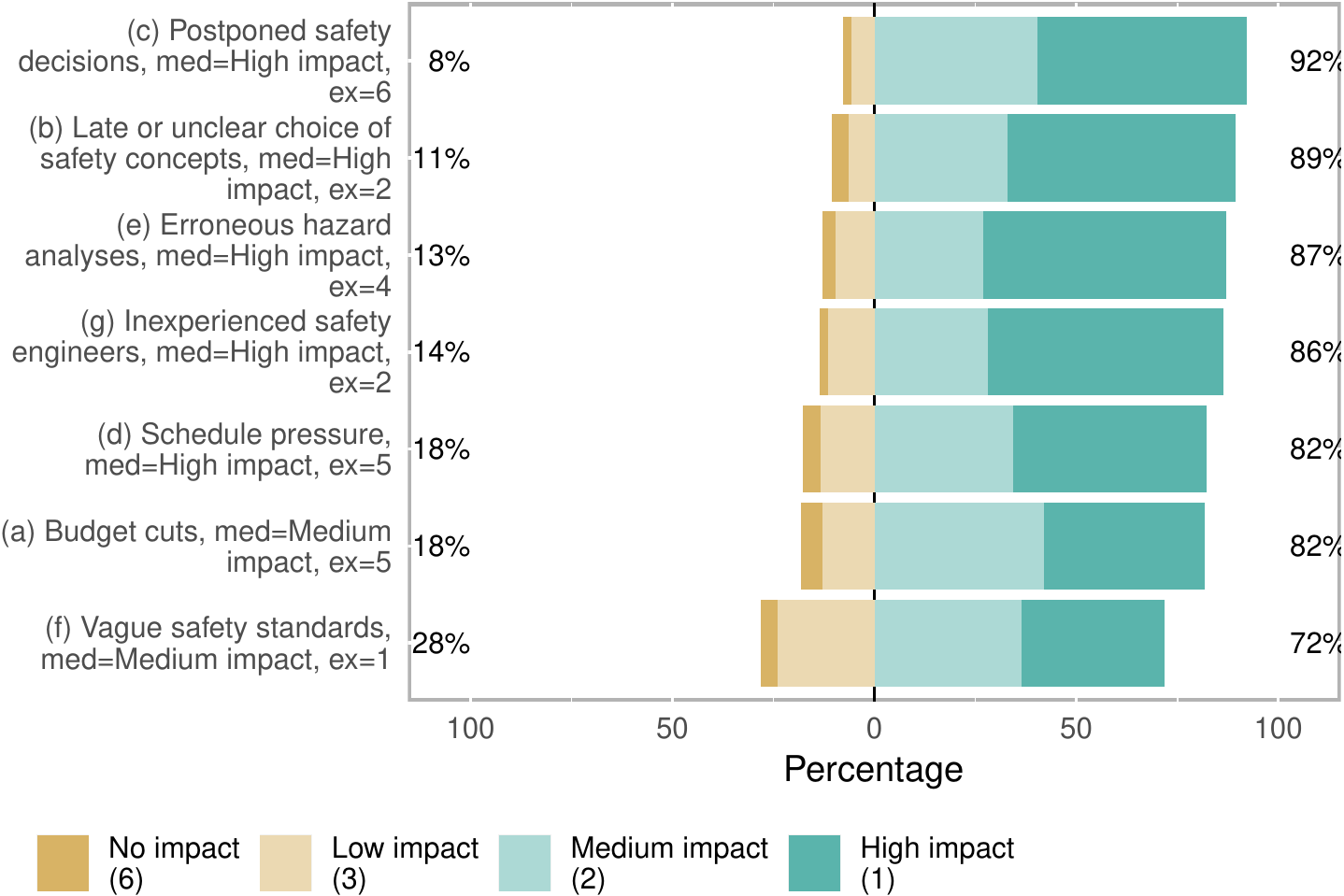}
  \caption{ \QuestionRef{H1_RQ_G2}~($N=100$): Negative impact on
    safety activities -- To which extent do specific \emph{process
      constraints and issues}\SubConstructRefP{ProcessFactor}
    negatively \emph{impact safety activities}\SubConstructRefP{ProcessPerformance}?
    \label{fig:H1_RQ_G2}}
\end{figure}

According to \Cref{fig:H1_RQ_G2}, \emph{inexperienced safety
  engineers~(g)} and \emph{erroneous hazard analyses~(e)} gained the
most ratings in the category ``significant negative impact on safety
activities.''  \emph{Postponed safety decisions~(c)} achieved
the largest consensus.  \emph{Vague safety standards~(f)} constitutes 
the bottom of this ranking but is still rated with medium or high
negative impact by the majority of respondents.

The following factors (\ies process constraints and issues) were
additionally mentioned to have \emph{high negative impact} on safety
activities:

Eight respondents broach the issue of \emph{missing management
  expertise and support}: ``Lack of education of managers in need for
safety'' identifies one respondent from the oil and gas industry.
Another one states that there is a ``general perception that safety is
only paper work'' and perceives a ``lack of safety knowledge within
management.''  One practitioner was even pointing to a
``lack of general safety culture.''

Three participants criticise that \emph{the degree of collaboration is
  too low}: They perceive a ``lack of system level engineering
experience'' as well as ``soloed working practices without a clear
view of [an] integrated safety concept'' and that the organisation is
``minimising [the] involvement of safety process/engineers into [the]
development process.''

Regarding \emph{incomplete or inadequate hazard lists}, respondents
mention ``unidentified hazard domains'' and ``imagined safety cases
not based on real workers experience.''  Along with that, one
practitioner mentions the issue of ``poorly defined requirements'':
Such requirements, when coming from upstream, are known to have a
negative effect on many downstream engineering activities.  Conversely,
inadequate hazard lists resulting from such activities can again have a
negative impact on downstream sub-system requirements specification.

Regarding \emph{compliance with norms}, one respondent was criticising
the ``transfer of concern from assessment to compliance,'' in other
words, \emph{compliance bias}.  Two others are broaching the opposite
phenomenon of \emph{compliance ignorance}, mentioning ``general ISO
26262 standard ignorance'' and a ``lack of understanding of regulatory
framework.''

Furthermore, according to another participant's experience there is
``too much faith in testing'' and ``reluctance to use formal
methods.''

\subsubsection{\QuestionRef{H1_RQ_G3}: Influence of Economic Factors}
\label{sec:sum-H1_RQ_G3}

More than a third~(36\%) %
of the survey participants share the view
that economic factors \emph{often strongly influence} the way how
hazards are handled, about half of them~(48\%) %
think that such
influence happens \emph{rarely or occasionally}~(median), and for
9\% %
such influences are \emph{not recognisable}.

\subsubsection{\QuestionRef{H26_RQ_B4}: Adequacy of Methods and Standards}
\label{sec:sum-H26_RQ_B4}

\begin{figure}[t]
  \includegraphics[width=\columnwidth]{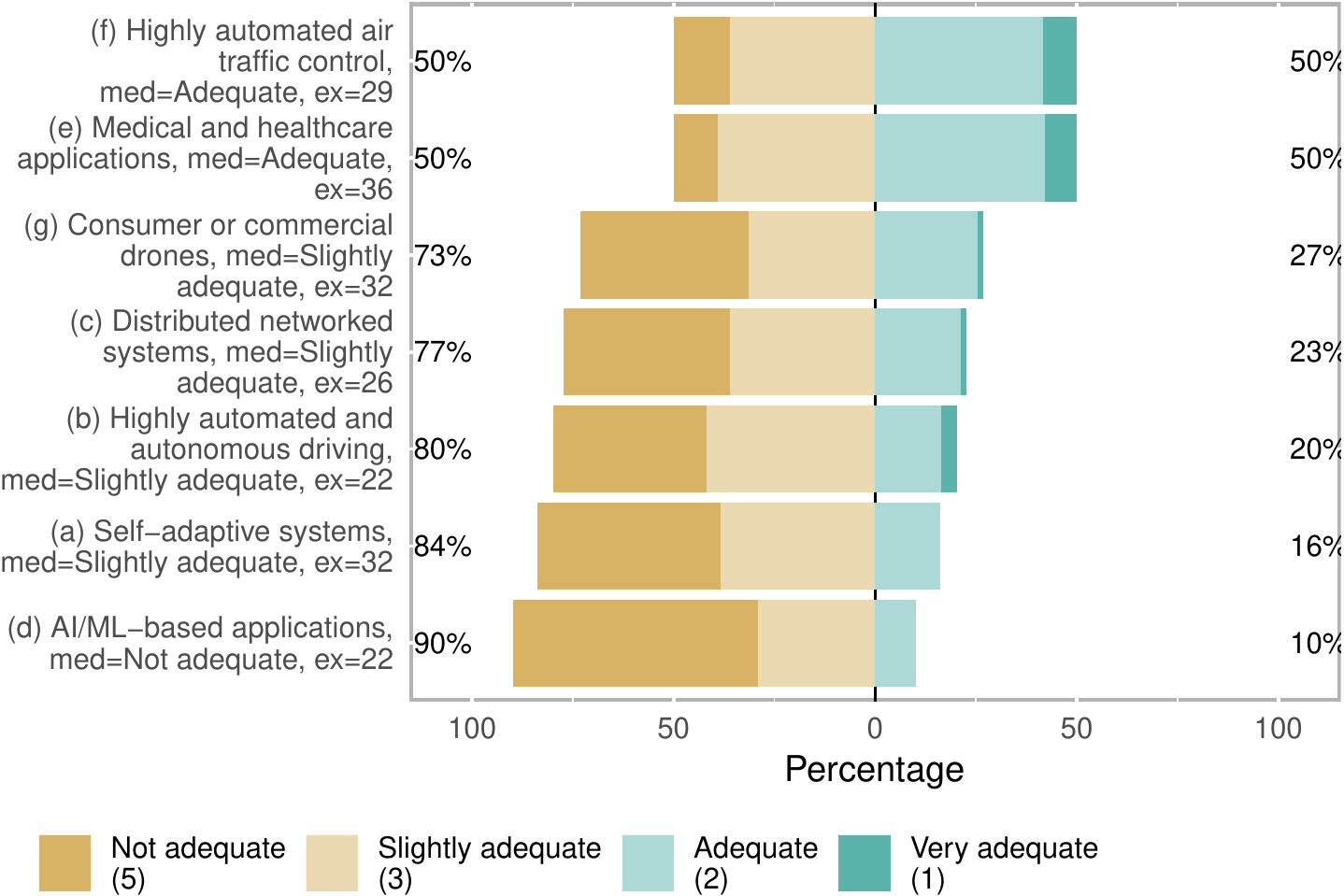}
  \caption{
    \QuestionRef{H26_RQ_B4}~($N=102$): Adequacy of methods and
    standards --
    Regarding a specific \emph{application domain}\SubConstructRefP{CurrentApplication},
    how \emph{adequate}\SubConstructRefP{AdequacyOfStandardsMethods}
    are applicable
    safety \emph{standards and methods}\SubConstructRefP{Means} in ensuring safety?
    \label{fig:H26_RQ_B4}}
\end{figure}

According to \Cref{fig:H26_RQ_B4}, traffic control~(f) and medical and
healthcare applications~(e) are most often believed to be supported by
\emph{adequate} methods and standards.  However, for all domains,
at least $50\%$ of the respondents think that the available means are
only \emph{slightly} or \emph{not at all adequate} for safety
assurance.  This question exhibits a relatively large number of
\emph{dnk}-answers.

\subsubsection{\QuestionRef{H11_RQ_B2}: Applicability of Methods}
\label{sec:sum-H11_RQ_B2}

\begin{figure}[t]
  \centering
  \includegraphics[width=\columnwidth]{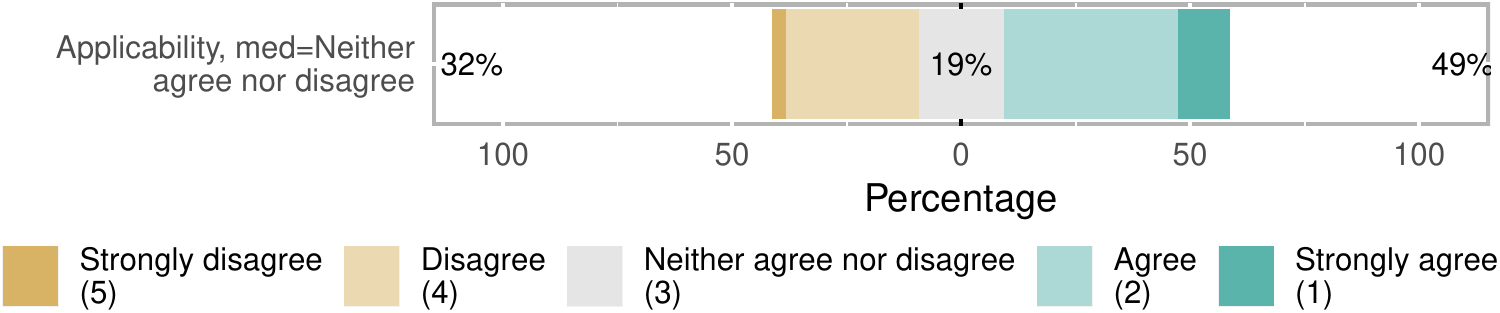}
  \caption{\QuestionRef{H11_RQ_B2}~($N=97$): Applicability of methods
    -- The \emph{application of conventional
      techniques}~(\egs FMEA and FTA)\SubConstructRefP{Means} has
    become too difficult\SubConstructRefP{AdequacyOfStandardsMethods}
    for \emph{complex applications of recent
      technologies}\SubConstructRefP{CurrentApplication}.
    \label{fig:H11_RQ_B2}}
\end{figure}

The \emph{nand}-median in \Cref{fig:H11_RQ_B2} shows that there is no
tendency or no clear consensus among respondents on whether or not
conventional methods have become too difficult to apply in current
applications.

\subsubsection{\QuestionRef{H11_RQ_B6}: Positive Impact of Formal Methods}
\label{sec:sum-H11_RQ_B6}

\begin{figure}[t]
  \centering
  \includegraphics[width=\columnwidth]{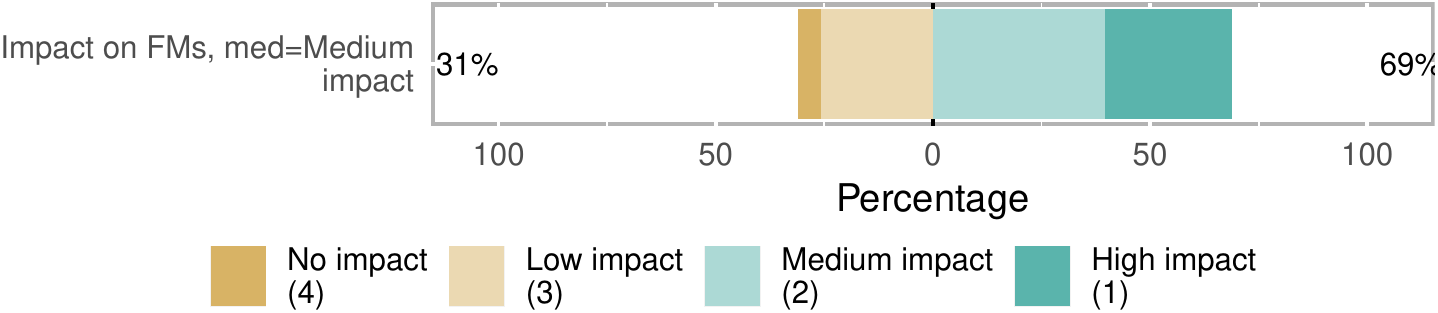}
  \caption{
    \QuestionRef{H11_RQ_B6}~($N=58$): Positive impact of formal
    methods --
    Estimate the \emph{positive impact}\SubConstructRefP{AdequacyOfStandardsMethods}
    of formal methods\SubConstructRefP{Means} on safety activities
    and system safety.
    \label{fig:H11_RQ_B6}}
\end{figure}

The median of ``medium impact'' in~\Cref{fig:H11_RQ_B6} indicates a
consensus among the participants on that the use of FMs might have a
\emph{positive} impact on the effectiveness of safety activities.
However, we only have a low number of responses resulting from missing
answers and we excluded \emph{dnk}-answers.

\subsubsection{\QuestionRef{H6_RQ_C1}: Improvement of Skills}
\label{sec:sum-H6_RQ_C1}

\begin{figure}
  \centering
  \includegraphics[width=\columnwidth]{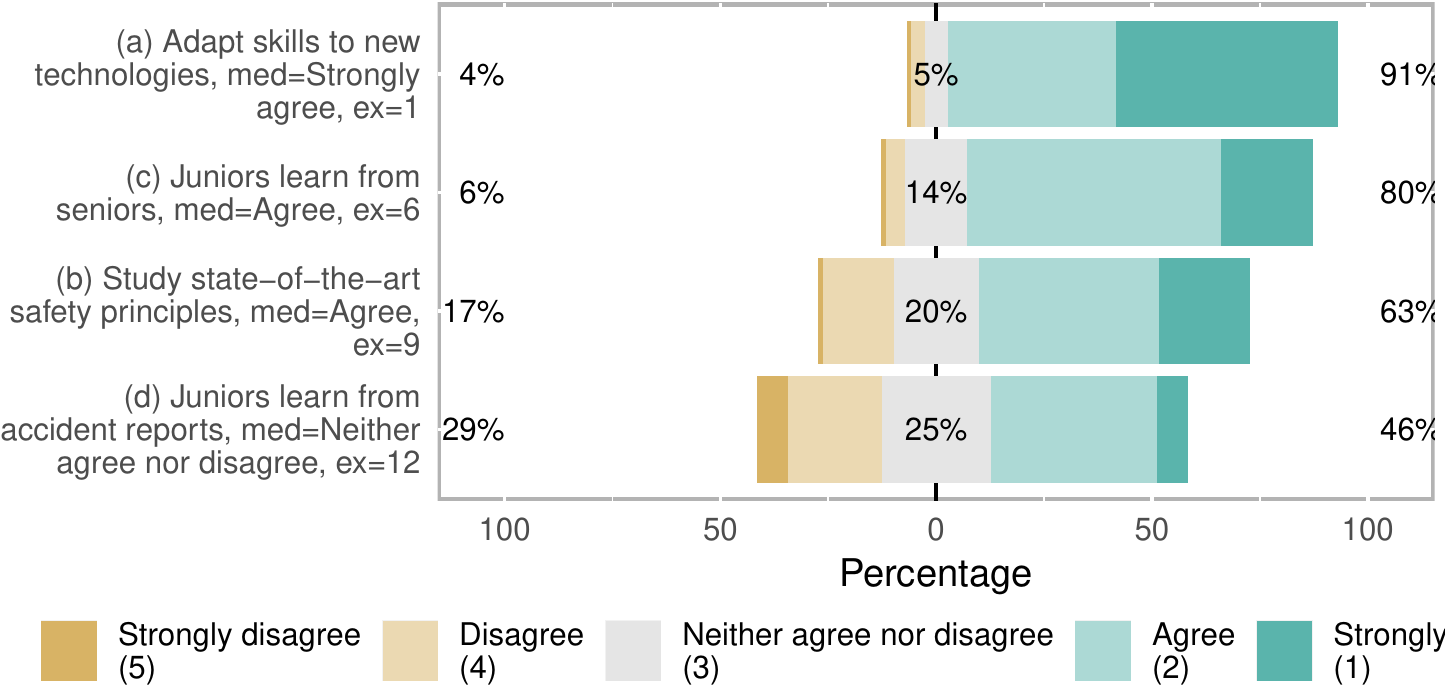}
  \caption{\QuestionRef{H6_RQ_C1}~($N=96$): Improvement of skills --
    Specify your level of agreement with 4 statements about
    \emph{factors\SubConstructRefP{Means}\SubConstructRefP{ValueOfKnowledgeSource}
      improving a SP's
      skills}\SubConstructRefP{ImprovementOfSkills}.
    \label{fig:H6_RQ_C1}}
\end{figure}

According to \Cref{fig:H6_RQ_C1}, SPs agree moderately~(39\%) to
strongly~(54\%) with the requirement to adapt their professional
skills to new technologies.  However, significantly less consensus was
achieved among the respondents on whether junior SPs should learn from
accident reports.

\subsubsection{\QuestionRef{H22_RQ_D1}: Interaction of Safety and Security}
\label{sec:sum-H22_RQ_D1}

\begin{figure}[t]
  \centering
  \includegraphics[width=\columnwidth]{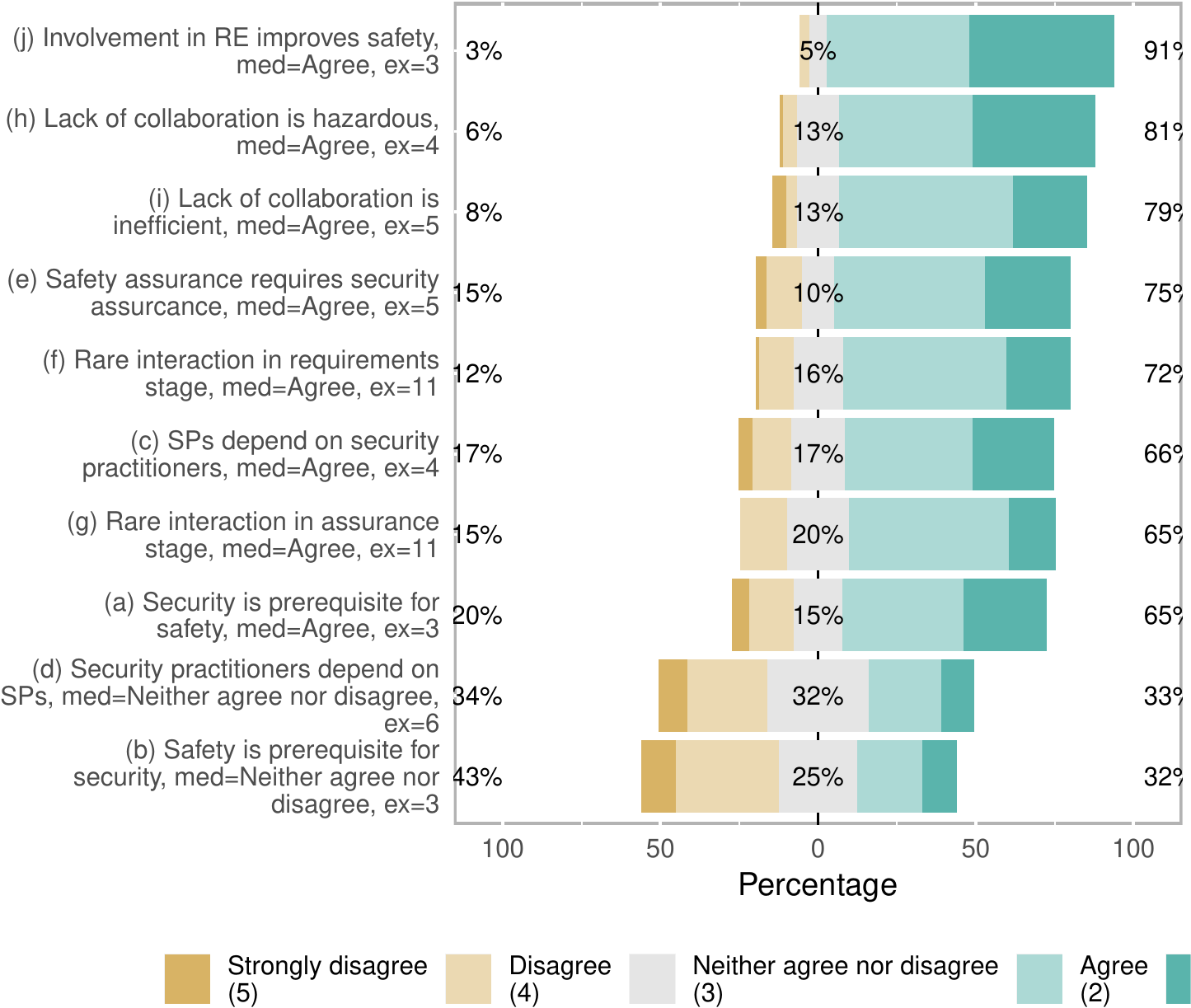}
  \caption{\QuestionRef{H22_RQ_D1}~($N=95$): Interaction of safety and
    security -- Specify your level of agreement with 10
    statements\SubConstructRefP{ProcessPerformance} about the
    \emph{interaction of safety and
      security}\SubConstructRefP{SafSecInteraction} activities.
    \label{fig:H22_RQ_D1}}
\end{figure}

The high moderate and strong agreement in \Cref{fig:H22_RQ_D1}
indicates that most of our participants perceive interactions between
safety and security as critical\SubConstructRefP{SafSecInteraction}.

SPs clearly agree on that interaction between safety and security
practitioners during requirements engineering and system assurance
rarely occurs (g,f).  Furthermore, clear agreement is achieved for the
``negative influence of a lack of collaboration~(between safety and
security engineers)''~(h,i) and for the ``positive influence of such a
collaboration''~(j) on safety activities.  However, we acknowledge 7\%
of disagreement with the ``requirement of ultimate IT security for
safety.''

No clear consensus is achieved regarding the \emph{dependence of
  security on safety}~(b,d).  As opposed to that, respondents agree on
the \emph{dependence of safety on security}~(a,c,e).

\subsubsection{\QuestionRef{H18_RQ_F5}: Notion of Safety}
\label{sec:sum-H18_RQ_F5}

\begin{figure}
  \centering
  \includegraphics[width=.8\columnwidth]{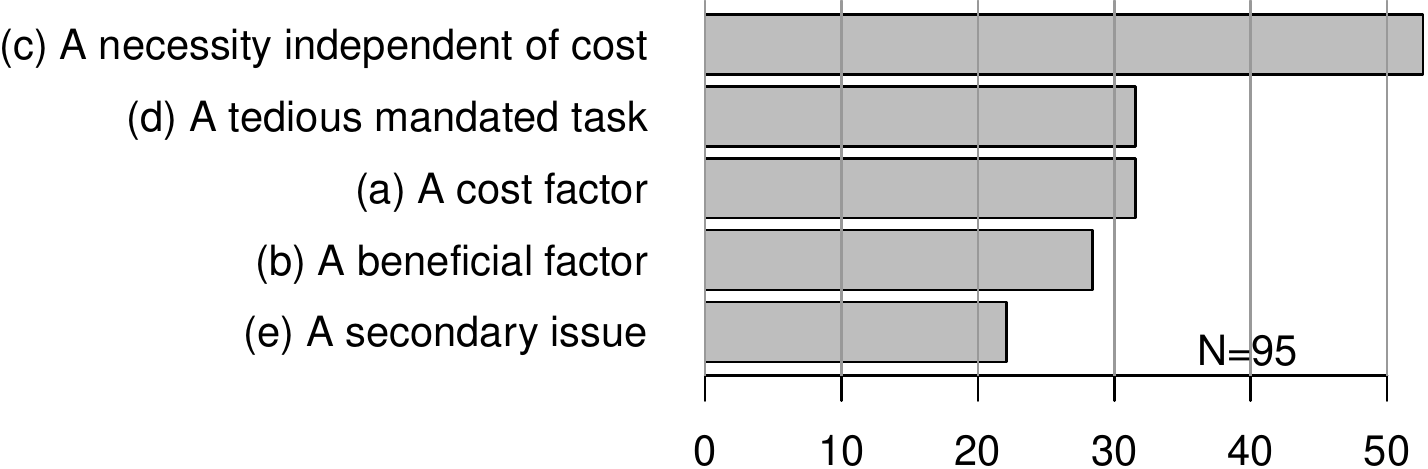}
  \caption{\QuestionRef{H18_RQ_F5}~($N=95$, MC): Frequency of safety
    notions --
    How is safety \emph{viewed}\SubConstructRefP{NotionOfSafety} in
    your field of practice? It is viewed as \dots
    \label{fig:H18_RQ_F5}}
\end{figure}

The multiple-choice answers in \Cref{fig:H18_RQ_F5} show that many
participants seem to be reluctant to associating cost/benefit schemes
with management decision making in system safety~(a,b).  Accordingly,
many responses indicate that safety is treated as a cost-independent
necessity~(c).
However, 51~(32\%) responses were given to the view of safety as an
``important, yet secondary, and tedious mandated issue''~(d,e).

Beyond the five given answer options, the notions of safety
additionally given by our respondents range from
a ``huge effort generating source'',
a ``marketing gadget'',
a ``high level product performance characteristic'',
a ``regulation'',
a ``general and common demand'',
a ``must have'' up 
to being ``essential.''

Importantly, two respondents add that \emph{it depends} ``on the
manager or the engineer'' or ``on the stakeholder and on the safety
professional.''  An ergonomist with 3 to 7 years of work experience
says that ``ergonomists usually are seen as added value to [the field]
because we try to work to improve performance and health at the same
time, safety is the natural outcome of this methodology.''

\subsubsection{\QuestionRef{H24_RQ_E3}: Priority of Safety}
\label{sec:sum-H24_RQ_E3}

\begin{figure}[t]
  \includegraphics[width=\columnwidth]{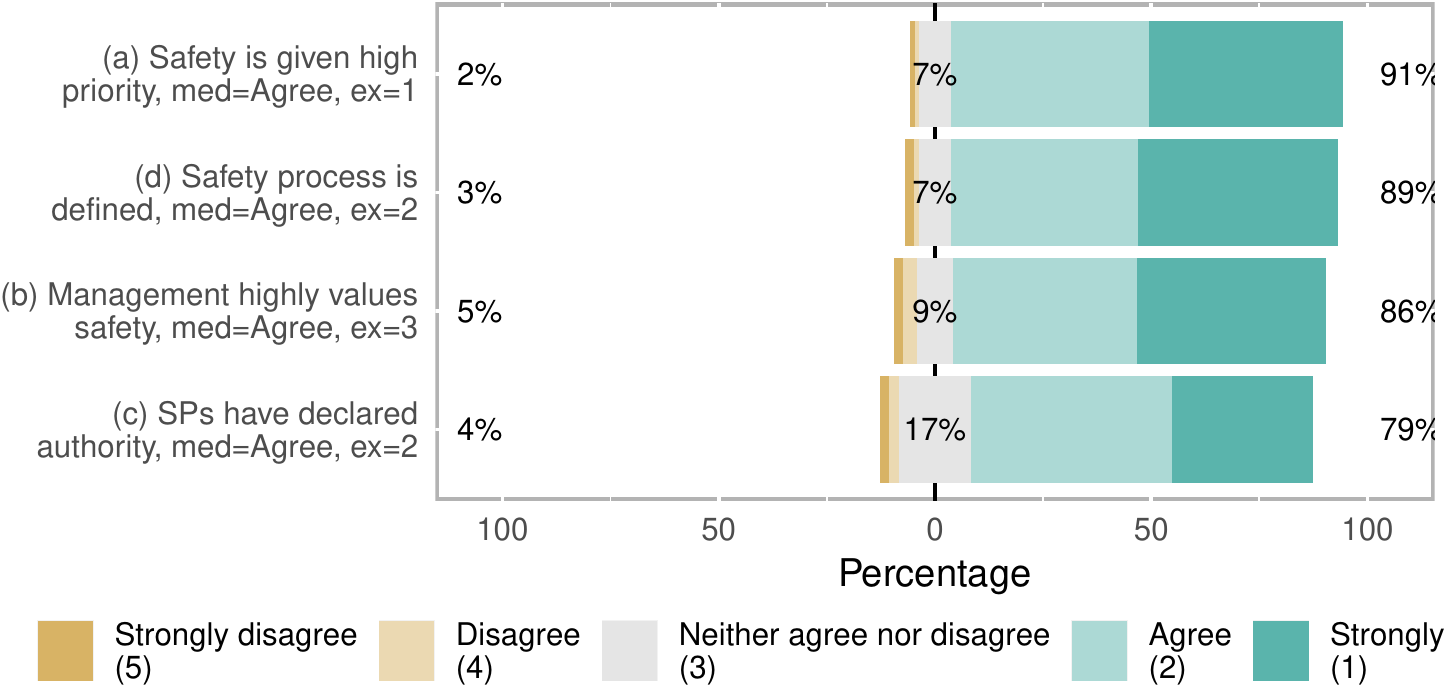}
  \caption{
    \QuestionRef{H24_RQ_E3}~($N=97$): Efficiency of safety activities
    --
    Specify your level of agreement with several statements about
    \emph{factors\SubConstructRefP{NotionOfSafety} increasing the efficiency
      in safety activities}\SubConstructRefP{ProcessPerformance}.
    \label{fig:H24_RQ_E3}}
\end{figure}

From \Cref{fig:H24_RQ_E3}, we can see a clear consensus of the
respondents for all given options~(a--d).  Particularly, increased
priority of safety decisions~(a) and defined safety processes~(d)
positively contribute to the efficiency of safety activities.  The
\Cref{sec:sum-H7_RQ_F1.a,sec:sum-H18_RQ_F5.ab} provide more details on
the factors believed to increase the efficiency and effectiveness of
safety activities \textbf{as well as dual factors} assumed to decrease
the efficiency and effectiveness thereof.  Along the way, comparably
many SPs~(17\%) do not offer any agreement on authority~(c).

\subsubsection{\QuestionRef{H7_RQ_F1.a}: Effect of Role Model
  on a SP's Job}
\label{sec:sum-H7_RQ_F1.a}
\label{sec:effect-safety-notion}

We asked our respondents to comment on \emph{whether and how their job
  is affected by a clear definition of their role}, if any, in their
organisations and application domains.

Apart from 5~\emph{dnk}-answers, we received 56 answers saying ``yes''
and, thus, stating that the role of a SP is \emph{clearly defined}.
These SPs perceive or expect the
following \emph{positive consequences} on their job~(frequency given in
parentheses, in descending order):  Clear role definitions \dots
\begin{itemize}
\item have a general positive impact on a SP's activities~(24),
\item lead to clear responsibilities, authority, and escalation
  routes~(13),
\item allow good integration of safety activities into the surrounding
  system life cycle processes~(6),
\item can make the achievement of compliance easier~(1), and
\item let SPs maintain autonomy or independence to carry through their
  most critical activities~(1).
\end{itemize}
However, our study participants report on the following
\emph{negative effects} on their job:  Clear role
definitions can \dots
\begin{itemize}
\item make engineers entirely push away safety-related
  responsibilities as a consequence of separating teams into safety
  and non-safety co-workers~(2),
\item lead to complex process definitions~(1),
\item get rather independent SPs exposed to company-wide resource and
  risk management~(1), and
\item impose a wrong focus or unnecessarily constrain a SP's
  tasks~(1).
\end{itemize}

Moreover, 30 participants responded with a ``no'' and, hence, state
that the role of a SP is \emph{not clearly defined}.  These SPs
consider or expect the following \emph{positive consequence} on their
job: Unclear role definitions \dots
\begin{itemize}
\item may promote more freedom to act, \eg to develop and employ
  new and more effective safety approaches~(3).
\end{itemize}
However, our respondents also perceive several \emph{negative
  effects} on their job:  Unclear role definitions \dots
\begin{itemize}
\item can entail unclear or wrong responsibilities as well as limited
  authority, autonomy, and space for discretionary activity~(9),
\item promote unclear, one-sided, or late decision making, in the worst case,
  rushed processing of checklists~(6),
\item have a general negative impact on a SP's tasks~(4),
\item can lead to disintegrated conceptions of safety, separated
  communities with a lack of communication and
  coordination, promoting unnecessarily confined decisions~(3),
\item can decrease the appreciation of a SP's analysis
  capabilities~(2), and
\item increase the risk of unqualified personnel assuming the role of
  a SP~(2).
\end{itemize}

\subsubsection{\QuestionRef{H18_RQ_F5.ab}: Effect of Safety Notion on
  a SP's Job}
\label{sec:sum-H18_RQ_F5.ab}

We asked our respondents to comment on \emph{whether and how their job
  is affected by a predominant notion of
  safety}~(\QuestionRef{H18_RQ_F5}), if any, in their organisations
and application domains.
  
Apart from 9 \emph{dnk}-answers, we received 74 answers indicating a
``yes'' and, hence, stating that the notion of safety \emph{has an
  effect} on their job:
10 respondents do not provide a specific comment.  The others argue
from several \emph{notions of safety} they have perceived in their
environments.  Below, we provide answer frequencies and cite a few
answers underpinning the summary statements.

\paragraph{Non-supportive Notions of Safety}

24 participants describe their experiences with a
\emph{non-supportive, misunderstood, or underrated safety culture}.
They report that \dots

\dots \emph{SPs have difficulties to argue their findings}~(9): ``Right
now there is no ability to have the safety requirement override
standard functional requirements.'' --
``1.~Our job always gets delayed and we are the last to get the
inputs. 2.~Non-safety engineers always try to justify or avoid the
suggestions/findings. 3.~It is difficult to sell safety culture to
non-safety engineers/managers.'' --
``I have to spend extra time explaining that safety is not about
compliance or implementing controls.'' --
``As for now safety has not the degree of importance to support
testing views and arguments against system designers and
management.''

\dots \emph{SPs suffer from late decision making}~(5): ``If I am not
allowed to do my job early in the process (requirements stage), safety
becomes more costly and I as a safety practitioner am viewed as a late
check in the box to get through a program rather than an integral part
of a design team.''

\dots \emph{SPs' activities have no lasting value}~(1): ``The safety
practitioner is neither equipped, nor capable of making the decisions
needed for a higher level of safety.  Being mostly policemen,
enforcers and rule designers, little if any of their contributions
have any meaningful or lasting value.''

\paragraph{Supportive Notions of Safety}

20 respondents describe their experiences with or their view of a
\emph{supportive or highly-valued safety culture}.  They report that
\dots

\dots \emph{SPs' findings are important and heard}~(6):
``My job is important because safety is valued and considered
necessary.'' --
``Most people in my organisation understand the importance of
safety. This is positive.'' --
``There are not many people who practice safety, since it is a
tedious job. So we are highly valued.''

\dots \emph{SPs are properly included in the process}~(1): ``Safety is
fundamental to the work we do and is ingrained into our processes in
such a way that its impossible to ignore. While it makes jobs harder
with much more analysis and review processes and every stage of the
product's development, we know its vital.''

\paragraph{Other Notions of Safety}

9 SPs describe an \emph{ambivalent picture}, saying that it depends on
individual projects whether their jobs are negatively or positively
affected: ``Safety at the last two places I worked is a check box
activity at best. Other places I've worked it was started early in the
pre-design phase. Starting early is more cost and schedule effective
with a better end product.''

5 SPs refer to a \emph{regulation-driven} notion of safety:
``Positively affected. In aerospace, safety is part of fundamental
engineering principles, so the process is embedded in systems
engineering and does not get left out.''
  
From a \emph{budget- or schedule-driven} perspective, respondents~(4)
observe that ``the budget for tools and training is never enough'' and
that ``resources, budget, support depend on the view/culture of
safety.''

Finally, 12 respondents claim, by saying ``no'', that the notion of
safety does \emph{not have any effect} on their jobs.

\subsubsection{\QuestionRef{H17_RQ_E2}: Role of Undesired Events for Safety}
\label{sec:sum-H17_RQ_E2}

\begin{figure}[t]
  \centering
  \includegraphics[width=\columnwidth]{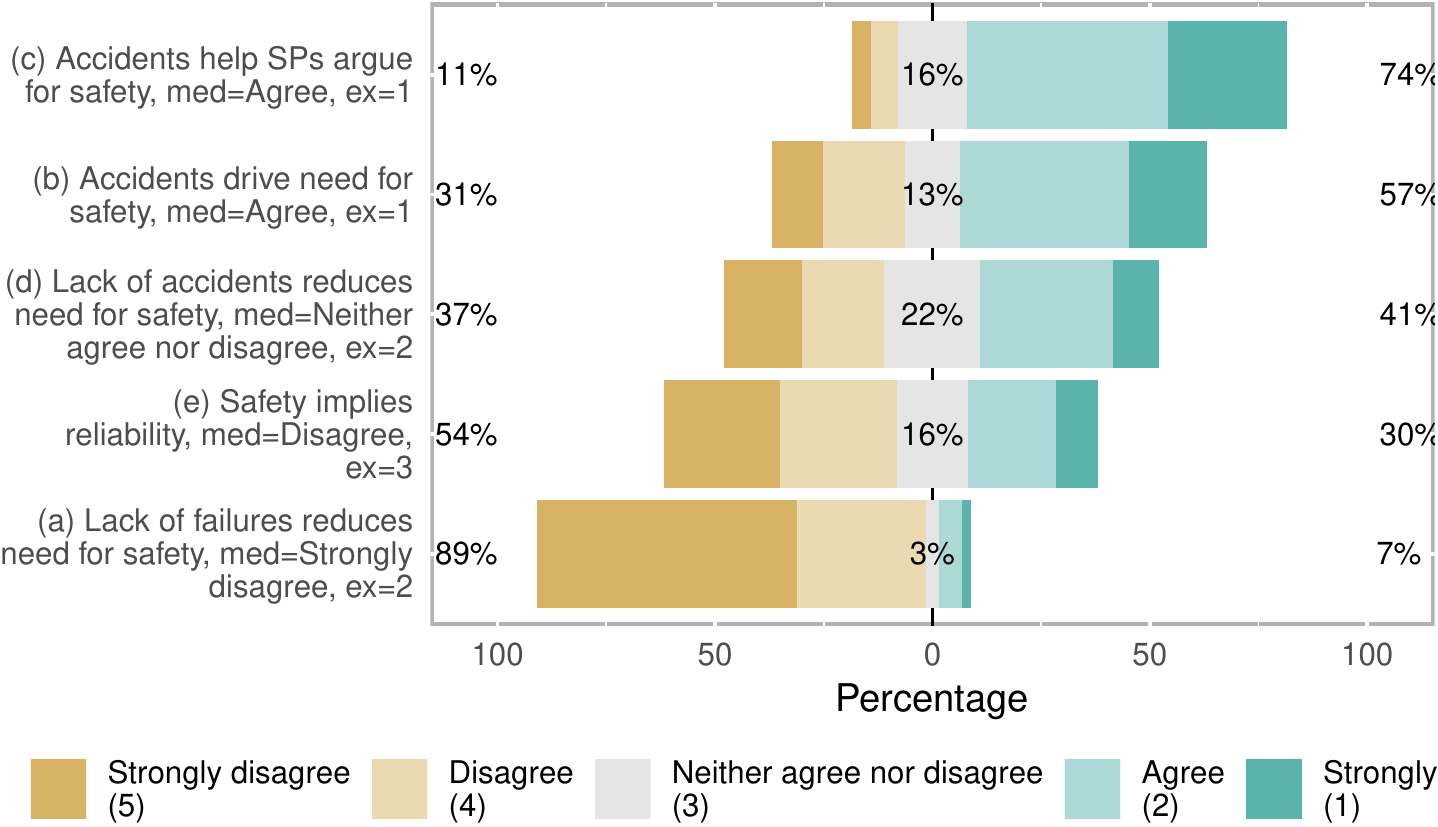}
  \caption{\QuestionRef{H17_RQ_E2}~($N=97$): Role of undesired
    events~(\ies failures, incidents, and accidents) for safety --
    Rate your level of agreement with 5
    statements\SubConstructRefP{Means}\SubConstructRefP{ValueOfKnowledgeSource}
    about \emph{safety
      activities}\SubConstructRefP{NotionOfSafety}.
    \label{fig:H17_RQ_E2}}
\end{figure}

\Cref{fig:H17_RQ_E2} shows a clearly disagreeing response on whether
lack of failures reduces the need for carrying through safety
activities~(a).
We have a more ambiguous agreement on whether \emph{safety implies
  reliability}~(e), \ie on whether having assured the safety of a
system \emph{usually includes} having also assured the reliability of
a system.
Moreover, known and reported accidents seem to be important
for the argumentation of the need for safety~(b,c).  However, the
agreement on whether a ``lack of accidents weakens arguments for the
need of safety''~(d) varies more.

\subsubsection{\QuestionRef{H20_RQ_F2}: Value of SPs' Contributions}
\label{sec:sum-H20_RQ_F2}

According to \Cref{fig:H20_RQ_F2}, the majority of respondents
perceives their role in the system life cycle as highly valuable or
better.  The analysis and comments in the
\Cref{sec:sum-H7_RQ_F1.a,sec:sum-H18_RQ_F5.ab} provide a more
differentiated picture of this answer.

\subsubsection{\QuestionRef{H16_RQ_F4}: Viewing SPs' Co-workers}
\label{sec:sum-H16_RQ_F4}

\begin{figure}[t]
  \centering
  \subfloat[\QuestionRef{H20_RQ_F2}~($N=95$): Value of SPs'
  contributions -- Of how much \emph{value}\SubConstructRefP{ValueOfSPs-Self}
  is your role as a practitioner or researcher in
  safety-critical system developments?]{
    \includegraphics[width=\columnwidth]{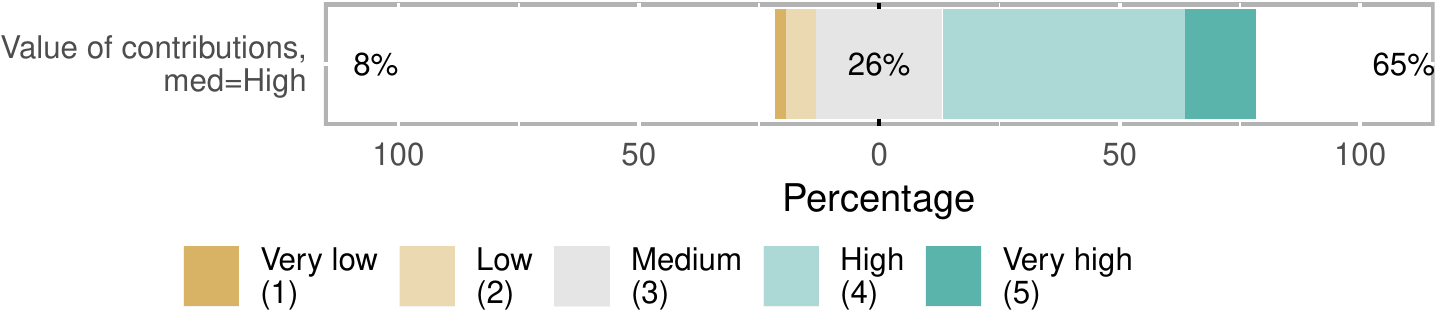}
    \label{fig:H20_RQ_F2}}

  \subfloat[\QuestionRef{H16_RQ_F4}~($N=95$): Viewing SPs' co-workers
  --
  How much \emph{value}\SubConstructRefP{ValueOfSPs-Self} do
  non-safety
  co-workers attribute to the \emph{role of a safety practitioner}?]{
    \includegraphics[width=\columnwidth]{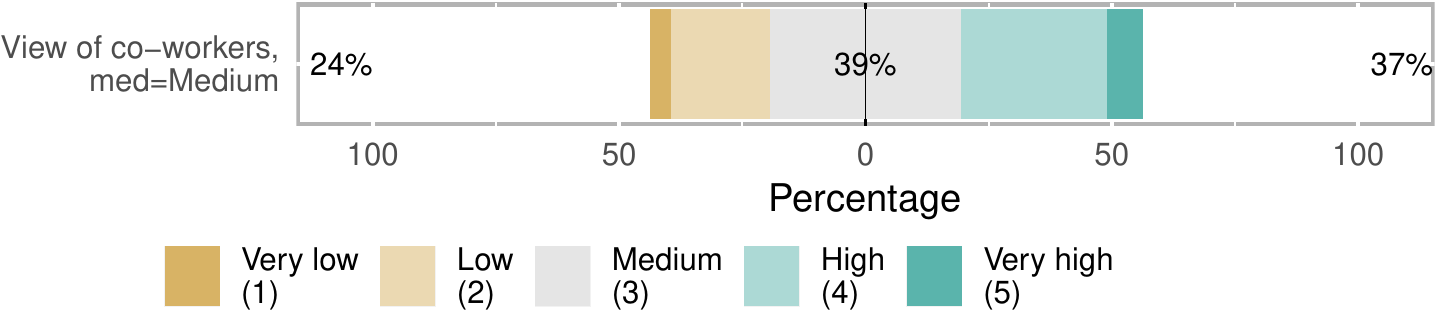}
    \label{fig:H16_RQ_F4}}
  \caption{Self-perception of SPs' role
    \label{fig:H20_H16}}
\end{figure}

\Cref{fig:H16_RQ_F4} suggests that the respondents vary strongly in
evaluating \emph{their contributions to the system life-cycle} when
trying to imagine their non-safety co-workers appreciation.

\subsubsection{\QuestionRef{H14_RQ_C2}: Influence of Experience}
\label{sec:sum-H14_RQ_C2}

\begin{figure}[t]
  \centering
  \includegraphics[width=\columnwidth]{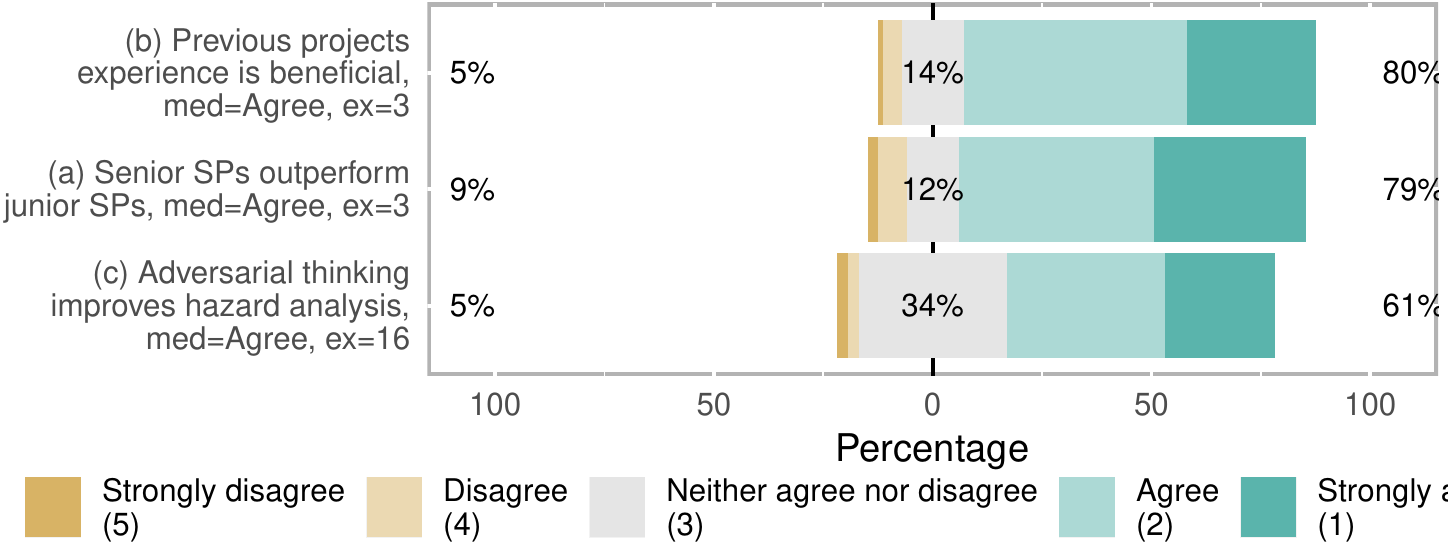}
  \caption{ \QuestionRef{H14_RQ_C2}~($N=96$): Role of experience --
    Specify your level of agreement with 3 statements about the
    \emph{role of
      experience\SubConstructRefP{Means}\SubConstructRefP{ValueOfKnowledgeSource}
     and adversarial thinking in safety activities}.
    \label{fig:H14_RQ_C2}}
\end{figure}

From the responses, \Cref{fig:H14_RQ_C2} shows that experience in
safety activities is believed to be positively associated with
\emph{improved hazard handling}~(a,b), particularly, experience from
similar previous projects~(b).  Adversarial thinking~(c) receives the
least agreement.

\subsection{Hypothesis Analysis and Test Results}
\label{sec:hypotheses-tests}
\label{sec:results-from-comp}

\Cref{tab:testresults} presents the test results for all hypotheses
listed in \Cref{tab:hypotheses} and based on the summary in
\Cref{sec:summ-answ-quest}.  Motivations for the acceptance criteria
given in \Cref{tab:hypotheses} are provided in \Cref{sec:acceptance}.
In summary, we were not able to find significant differences for the
\emph{pairs of groups (IVs) we compared} \wrt several DVs.

\begin{table*}[t]
  \caption{Results of hypotheses analysis for \HypothesisRef{H3} to \HypothesisRef{H20} and
    hypotheses tests for \HypothesisRef{H14} to \HypothesisRef{C8}.
    \textbf{Legend:}
    AC\dots acceptance criterion
    \label{tab:testresults}}
  \footnotesize
  \begin{tabularx}{1.0\linewidth}{LLL}
    \toprule
    \textbf{Hypothesis}:
    Construct-based proposition
    & \textbf{From the responses to} \dots
    & \dots \textbf{we conclude that} \dots
    \\\midrule

    \HypothesisRef{H3}:
    Dependence on expert opinion
    & \QuestionRef{H3_RQ_E1}~(\Cref{sec:sum-H3_RQ_E1}) and
      \QuestionRef{H14_RQ_C2}~(\Cref{sec:sum-H14_RQ_C2})
    & our AC is fulfilled.
    \\
    
    \HypothesisRef{H1}:
    Resources govern performance of SPs
    & \QuestionRef{H1_RQ_G2}~(\Cref{sec:sum-H1_RQ_G2})
      and \QuestionRef{H1_RQ_G3}~(\Cref{sec:sum-H1_RQ_G3})
    & the \QuestionRef{H1_RQ_G3}-part of our AC is not
      fulfilled. 
    \\
    
    \HypothesisRef{H26}:
    Inadequate means in high-automation
    & \QuestionRef{H26_RQ_B4}~(\Cref{sec:sum-H26_RQ_B4})
    & our AC is fulfilled.
    \\

    \HypothesisRef{H11}:
    Low method %
      adequacy %
    & \QuestionRef{H11_RQ_B2}~(\Cref{sec:sum-H11_RQ_B2}) and
      \QuestionRef{H11_RQ_B6}~(\Cref{sec:sum-H11_RQ_B6})
    & the \QuestionRef{H11_RQ_B2}-part of our AC is not
fulfilled by the \emph{nand}-median.
    \\
    
    \HypothesisRef{H11b}:
    Positive impact of formal methods
    & \QuestionRef{H11_RQ_B6}~(\Cref{sec:sum-H11_RQ_B6})
    & our AC is fulfilled.
    \\

    \HypothesisRef{H6}:
    Necessity of skill adaptation
    & \QuestionRef{H6_RQ_C1}~(\Cref{sec:sum-H6_RQ_C1})
    & our AC is fulfilled.
    \\

    \HypothesisRef{H8}:
    Dependence on IT security
    & \QuestionRef{H22_RQ_D1}~(\Cref{sec:sum-H22_RQ_D1})
    & our AC is fulfilled.
    \\

    \HypothesisRef{H18}:
    Safety is a cost-benefit question
    & \QuestionRef{H18_RQ_F5}~(\Cref{sec:sum-H18_RQ_F5})
    & the \QuestionRef{H18_RQ_F5}-a-part of
    our AC is not fulfilled.
    \\
    
    \HypothesisRef{H22}:
    Benefit of safety-security interaction
    & \QuestionRef{H22_RQ_D1}~(\Cref{sec:sum-H22_RQ_D1})
    & our AC is fulfilled.
    \\

    \HypothesisRef{H24}:
    Benefit of safety-as-a-priority
    & \QuestionRef{H24_RQ_E3}~(\Cref{sec:sum-H24_RQ_E3})
    & our AC is fulfilled. 
    \\

    \HypothesisRef{H17}:
    Safety is a special case of reliability 
    & \QuestionRef{H17_RQ_E2}~(\Cref{sec:sum-H17_RQ_E2})
    & none of the \QuestionRef{H17_RQ_E2}-parts of our AC are
      fulfilled.
    \\
    
    \HypothesisRef{H16}:
    High contribution to life cycle
    & \QuestionRef{H16_RQ_F4}~(\Cref{sec:sum-H16_RQ_F4}) and
      \QuestionRef{H20_RQ_F2}~(\Cref{sec:sum-H20_RQ_F2})
    & the \QuestionRef{H16_RQ_F4}-part of our AC is not
      fulfilled.
    \\

    \HypothesisRef{H20}:
    High contribution (self-image)
    & \QuestionRef{H20_RQ_F2}~(\Cref{sec:sum-H20_RQ_F2}) and
      \QuestionRef{H16_RQ_F4}~(\Cref{sec:sum-H16_RQ_F4})
    & our AC is fulfilled.
    \\

    \midrule
    & \textbf{From the comparison of} \dots
    & \dots \textbf{we conclude that} \dots
    \\\midrule
      
    \HypothesisRef{H14}:
    Benefit of diverse expertise
    & senior SPs with junior SPs (from responses to \QuestionRef{H14_RQ_C2}, \Cref{sec:sum-H14_RQ_C2})
    & our AC is fulfilled.
    \\

    \HypothesisRef{C1}:
    Assoc. of expertise \& value 
    & senior SPs with junior SPs
    & with $p = 0.15$, our AC is not fulfilled. %
    \\
    
    \HypothesisRef{C2}:
    Assoc. of expertise \& skill adaptation 
    & senior SPs with junior SPs
    & with $p = 0.052$, our AC is almost fulfilled. %
    \\

    \HypothesisRef{C7}:
    Assoc. of standards \& skill adaptation
    & SPs using automotive standards with SPs using aerospace standards
    & with $p = 0.09$, our AC is almost fulfilled.
    \\

    \HypothesisRef{C6}: %
    Assoc. of profession \& inadequate means
    & engineering-focused SPs with research-focused SPs
    & with $p = 0.12$, our AC is not fulfilled.
    \\

    \HypothesisRef{C8}:
    Assoc. of standards \& inadequate means
    & SPs using automotive standards with SPs using aerospace standards
    & with $p = 0.24$, our AC is not fulfilled.
    \\
    \bottomrule
  \end{tabularx}
\end{table*}

\section{Discussion}
\label{sec:discussion}

We interpret the responses~(\Cref{sec:interpr-results}), draw a
relationship to existing evidence~(\Cref{sec:relat-exist-find}), and
critically assess the validity of our
study~(\Cref{sec:threats-validity}).  From these discussions, we
derive our conclusions in~\Cref{sec:conclusions}.

\subsection{Interpretation of the Results and Findings}
\label{sec:interpr-results}

The following discussion takes into account the hypothesis analysis
and test results summarised in \Cref{tab:testresults}.  Details about
hypotheses and questions referred to in the text can be derived from
\Cref{tab:hypotheses,tab:questions}.

\subsubsection{Findings for RQ1: Means of Work in Safety Practice}

\paragraph{Hypothesis \HypothesisRef{H3}}

The support of \HypothesisRef{H3} should not be surprising as it
mirrors a rather typical situation in many engineering disciplines and
projects.  However, relying too much on knowledge of experts can in
the worst case go along with relying on a \emph{single point of
  failure of an organisation}.  Moreover, relying too much on
experience from similar projects can unfortunately go along with
\emph{wrongly transferring former conclusions~(\ies project memory)
  and not updating them} correspondingly.

\Finding{H3}{}{The responses suggest that safety
  mainly depends on expert opinion and project memory.}

\paragraph{Hypothesis \HypothesisRef{H26} is supported}

With regard to the offered application
domains~(\QuestionRef{H26_RQ_B4}, \Cref{sec:sum-H26_RQ_B4}), the
result for~\HypothesisRef{H26} is clearly negative: Our responses
indicate that \emph{inadequate methods or standards constitute a real
  issue in current high-automation safety practice}.
However, from~\QuestionRef{H1_RQ_G2} in \Cref{fig:H1_RQ_G2}, we know
SPs think that ``vague safety standards'' are problematic, though,
least problematic of all inquired process constraints and issues.
The 22 to 36 excluded \emph{dnk}-answers might stem from the fact that
most respondents can only make a statement for a small subset of the
inquired application domains.  We believe, the exclusion of these
responses does not weaken our observation.
Moreover, the observation of a lack of appropriate standards and
certification guidelines is anecdotally confirmed by \mycite[McDermid
and Rae]{McDermid2014} and empirically in the automated vehicle
testing domain by Knauss et al.~\cite[pp.~1878f]{Knauss2017}.

\Finding{H26}{}{Standards in the considered high automation domains
  seem to be inadequate.}

\paragraph{Hypothesis \HypothesisRef{H11}}

Because of overlapping \SubConstructRef{Means}, the rejection
of~\HypothesisRef{H11} stands in conflict with the support
of~\HypothesisRef{H26}.  On the one hand, we see a slight tendency
towards the first author's experience from interviews
\cite{Yang2016,Gleirscher2014a,Hussein2016}
suggesting~\HypothesisRef{H11} to be a justified hypothesis.
On the other hand, ambiguous agreement was given to ``have become too
difficult'' or, more generally, to ``have become inadequate.''  Asking
for agreement in question~\QuestionRef{H11_RQ_B2} should have been
substituted by asking for the \emph{level of}
\SubConstructRef{AdequacyOfStandardsMethods}.  However, we believe it
is safe to interpret the respondents' agreement that ``available
standards and methods have become too difficult'' as ``they are
challenging to apply.''  After all, we conclude that this construct
should better be measured by several questions to get more informative
and reliable results.

\Finding{H11}{}{From our data, we are not able to provide a clear
  general picture about the adequacy of means.}

The exploratory nature of our questionnaire made it necessary to
sacrifice the level of detail for certain questions,
\eg~\QuestionRef{H11_RQ_B2}, to keep the questionnaire short enough to
be feasible.  To get a more detailed response to this question, it has
to be repeated for each technique and standard and analysed for
sensitivity to, \eg industry-specific sub-groups of respondents.  A
more detailed questionnaire is subject of future
work~(\Cref{sec:futurework}).

\paragraph{Hypothesis \HypothesisRef{H11b} is supported}

The low number of valid responses to question~\QuestionRef{H11_RQ_B6}
certainly weakens the interpretation of the support of
\HypothesisRef{H11b}.  Both, the question~\QuestionRef{H11_RQ_B6} as
well as the notion of a formal method are very abstract.  Moreover,
the classification questions provide only little knowledge about our
respondents' experience with FMs.
\Finding{H11b}{}{Among informed respondents, formal methods are
  believed to be beneficial.}
Certainly, this finding requires another study with a more
specific research design. %

\subsubsection{Findings for RQ2: Impact of Process Factors}

\paragraph{Hypothesis \HypothesisRef{H1} is rejected}

Few respondents to question~\QuestionRef{H18_RQ_F5.ab} experience a
\emph{lack of resources} for safety activities.  This is consistent
with the data checked for the AC of \HypothesisRef{H1}.  Although the
responses suggest that the implication \emph{lack of resources has
  negative impact on safety} might hold, the antecedent of this
hypothesis is not broadly supported.

\Finding{H1}{}{Our data suggest that resources occasionally but
  not typically govern SPs' performance.}

However, by weakening~\HypothesisRef{H1}, we can acknowledge the
``often'' third of SPs showing a situation demanding for reaction
in the community.

\paragraph{Hypothesis \HypothesisRef{H18} is rejected}

We identify a weak positive association: Safety is most frequently
viewed as \emph{a cost-independent
  necessity}~(\QuestionRef{H18_RQ_F5}-c, \HypothesisRef{H18}) and the
median of \QuestionRef{H1_RQ_G3}~(\HypothesisRef{H1}) lies at
\emph{economic factors rarely or occasionally influence safety.}
So, for~\HypothesisRef{H18}, the many positive responses to the
options~(c,d,e) underpin the view of safety as a cost-independent
factor in management decision making.
We consider this to be positive but like to stress the need of an
in-depth explanatory study to confirm or refute this finding.

\Finding{H18}{}{Our responses suggest that safety is not typically a
  question of cost-benefit.}

\paragraph{Hypotheses \HypothesisRef{H22} is supported}

First of all, our data supports \HypothesisRef{H8} which states that
safety assurance strongly depends on security assurance.  
Interestingly, for~\HypothesisRef{H22}, SPs agree on both that \dots

\Finding{H22-1}{}{\dots a lack of collaboration or interaction
  downgrades the performance of safety
  activities~(\QuestionRef{H22_RQ_D1}-h,i), \textbf{and}}
\Finding{H22-2}{}{\dots interaction between safety and security
  practitioners \textbf{rarely occurs} in requirements and assurance
  activities~(\QuestionRef{H22_RQ_D1}-f,g).}

We consider this issue worthwhile to be monitored. %
Apart from desirable interactions at an organisational level,
potential \emph{dependence of security on
  safety}~(\QuestionRef{H22_RQ_D1}-b,d) is less obvious to our
respondents than potential \emph{dependence of safety on
  security}~(\QuestionRef{H22_RQ_D1}-a,c,e).  While the latter is
comparably well known, the former is more difficult to grasp.  Our
data shows this ambiguity but does not explain it.

\Finding{H22}{}{Overall, collaboration of safety and security experts
  is clearly viewed as beneficial.}

\paragraph{Hypothesis \HypothesisRef{H14} is supported}

Although the three propositions in \Cref{fig:H14_RQ_C2} seem obvious,
we included them in our questionnaire to confirm that such
occasionally important assumptions are actually made by
SPs~(\HypothesisRef{H14}).  For these assumptions to be formulated as
a hypothesis and tested accordingly, a further investigation would be
necessary. %
Hence, the support of~\HypothesisRef{H14} is not very informative
on its own but backs the support of~\HypothesisRef{H3}.

\Finding{H14}{}{Diverse expertise is perceived as beneficial for SPs.}

\paragraph{Hypothesis \HypothesisRef{C2} \vs \HypothesisRef{C7} and
  \HypothesisRef{H6}}

Among the comparative hypotheses,
only \HypothesisRef{C2} and \HypothesisRef{C7}
are close to being supported with $p = 0.052$ and $p = 0.09$.
The result for \HypothesisRef{H6} is unsurprising because senior
experts are professionals with longer experience and might have
witnessed training activities in their field more often than junior
SPs.  However, the small difference between both groups gives rise to
the conjecture that senior experts would avoid outdated skills as much
as junior professionals would.  An almost supported \HypothesisRef{C7}
gives rise to the conjecture that in automotive, currently in demand
of improvement of their safety practices, SPs spend corresponding
effort on skill improvement.

\Finding{C2}{}{The improvement of skills towards new
  technologies is generally agreed among respondents.}

\subsubsection{Findings for RQ3: Perception of Safety Practice}

\paragraph{Hypothesis \HypothesisRef{H17} is rejected}

Similarly, we perceive the results for~\HypothesisRef{H17} as positive
because the issue of ``confusing safety with reliability'' raised
in~\cite[p.~7, Assumption 1]{Leveson2012} can at least not be
confirmed from the analysis of our responses.  In fact, we observe an
opposite tendency from our sample and assume this to be the effect of
those SPs having been trained on that issue.

\Finding{H17}{}{It is generally justified to not believe in the
  hypothesis ``safety is equivalent to reliability.''}
From the responses to \QuestionRef{H17_RQ_E2}-a, we derive that assured
reliability of a system does not reduce the need for safety
activities.  Consequently, these responses do not give rise to believe
in the hypothesis \emph{reliability implies safety}.  However, we
might sometimes expect to see agreement on the hypothesis \emph{safety
  implies reliability}~(\QuestionRef{H17_RQ_E2}-e).  Likewise, our
responses are ambiguous in that case.  The most reasonable
explanation for this ambiguity is that we missed to clearly explain
what such implications exactly mean when used as answer options.
Moreover, \HypothesisRef{H17} is not backed by redundant data.  The
data gathered from~\QuestionRef{H17_RQ_E2} makes it hard to draw a strong
conclusion.  To back a ``true extension'' of reliability---\ie safety
carries features essentially different from reliability or, even more,
safety is independent of reliability---we should have asked questions
like ``Does reliability imply safety?''  with an expected median of
``disagree''.

In conclusion, our data gives rise to the reasonable belief that
safety and safety activities are less dependent on issues of system
failures than on the more general issues of system accidents.

\Finding{H17.2}{}{From our responses, we cannot further characterise
  the relationship of safety and reliability.}

\paragraph{Questions \QuestionRef{H7_RQ_F1.a} and
  \QuestionRef{H18_RQ_F5.ab}}

56 respondents state that their role is clearly defined.  37 perceive
positive impacts on their activities, particularly, fostering clear
responsibilities, authority, and escalation routes.
  
30 respondents state that their role is not clearly defined.  15 of
them perceive negative impacts in form of unclear responsibilities,
limited authority, autonomy, and space for discretionary activity as
well as unclear or late decision making~(\Cref{sec:sum-H7_RQ_F1.a}).

\Finding{H7_RQ_F1.a}{}{The role of a SP is often not clearly defined
  and SPs experience negative impacts from this.}

24 participants experience a non-supportive, misunderstood, or
underrated safety culture.  As opposed to that, 20 respondents
perceive a supportive or highly-valued safety culture.  9 persons
provided an ambivalent picture of safety culture, stating that they
have gathered contrasting experiences~(\Cref{sec:sum-H18_RQ_F5.ab}).

\Finding{H18_RQ_F5.ab}{}{SPs perceive to a similar extent both,
      supportive and non-supportive notions of safety.}

\paragraph{Hypothesis \HypothesisRef{H20} is supported}

While responses to~\QuestionRef{H20_RQ_F2} 
support~\HypothesisRef{H20}, the frequent indication of ``medium
value'', particularly for~\QuestionRef{H16_RQ_F4}, suggests that some
SPs might either not be convinced of the role, their profession, or
even unsatisfied with their tasks and their job profile.
\Cref{sec:sum-H7_RQ_F1.a} provides some explanation for such a
dissatisfaction coming from an \emph{unclear} role definition and
\Cref{sec:sum-H18_RQ_F5.ab} delivers an explanation from a
\emph{non-supportive} safety culture.  However, for a solid
conclusion, this indication has to be investigated in more detail by
further studies. %

The perception of an SP's role and contribution by non-safety
co-workers slightly differs from how SPs perceive their own role.
This might not be too surprising because~\QuestionRef{H20_RQ_F2}
and~\QuestionRef{H16_RQ_F4} redundantly measure fragments of a
participant's self-perception.

\Finding{H20}{}{SPs seem to be self-confident about their
  contribution.}

\begin{table*}
  \centering
  \caption{Overview of main findings from hypothesis analysis}
  \label{tab:relat-exist-find}
  \footnotesize
  \begin{tabularx}{1.0\linewidth}{LLL}
    \toprule
    \textbf{RQ1:} Which means do SPs typically rely on? How helpful are those
    means to them?
    & \textbf{RQ2:} Which typical process factors have influence on
      SPs' decisions \& performance?
    & \textbf{RQ3:} How do SPs perceive and understand their role in the process or life cycle?
    \\\midrule
    \FindingRef{H3}: The responses suggest that safety mainly
    depends on expert opinion and project memory.
    \newline
    \FindingRef{H26}: Standards in the considered high automation domains
    seem to be inadequate.
    \newline
    \FindingRef{H11}: From our data, we are not able to provide a clear
    general picture about the adequacy of means.
    \newline
    \FindingRef{H11b}: Among informed respondents, formal methods are
    believed to be beneficial.
    & \FindingRef{H1}: Our data suggest that resources occasionally but
      not typically govern SPs' performance.
      \newline
      \FindingRef{H18}: Our responses suggest that safety is not typically a
      question of cost-benefit.
      \newline
      \FindingRef{H22-1}: A lack of collaboration or interaction
      downgrades the performance of safety activities.
      \newline
      \FindingRef{H22-2}: Interaction between safety and security
      practitioners rarely occurs in requirements and assurance
      activities.
      \newline
      \FindingRef{H22}: Collaboration of safety and security experts
      is clearly viewed as beneficial.
      \newline
      \FindingRef{H14}: Diverse expertise is perceived as beneficial for SPs.
      \newline
      \FindingRef{C2}: The improvement of skills towards new
      technologies is generally agreed among respondents.
    & \FindingRef{H17}: It is generally justified to not believe in the
      hypothesis ``safety is equivalent to reliability.''
      \newline
      \FindingRef{H17.2}: Our responses do not offer specific insights on the
      relationship between safety and reliability.
      \newline
      \FindingRef{H20}: SPs seem to be self-confident about their
      contribution.
      \newline
      \FindingRef{H7_RQ_F1.a}: Their role is often not clearly defined
      and SPs experience negative impacts from this.
      \newline
      \FindingRef{H18_RQ_F5.ab}: SPs perceive to a similar extent both,
      supportive and non-supportive notions of safety.
    \\\bottomrule
  \end{tabularx}
\end{table*}

\subsection{Relation to Existing Evidence}
\label{sec:relat-exist-find}

In \Cref{tab:relat-exist-find}, we summarise our findings and, below,
we compare them with findings from related studies.

\mycite[Graaf et al.]{Graaf2003} identify \emph{legacy
  incompatibility, lack of maturity, and additional complexity} of new
methods, languages, and tools as three major obstacles to the early or
timely adoption of such means.  Similar obstacles were observed in
software testing by \mycite[Kasurinen et al.]{Kasurinen2010}.  These
observations are \textbf{consistent with} \FindingRef{H3} that
knowledge from previous projects has the strongest influence.

\mycite[Martins and Gorschek]{Martins2016} observe a lack of evidence
for the usefulness and usability of new approaches from safety
research.  Their observation is \textbf{not in conflict with}
finding~\FindingRef{H11b}, because SPs can perceive usefulness of new
FMs independent of evidence.  The authors perceive a dominance of
conventional approaches in practice which is again \textbf{consistent
  with finding}~\FindingRef{H3}.
Furthermore, they observe a lack of studies that investigate how to
improve the communication process throughout the life cycle.
\FindingRef{H22} indicates that such studies would be of interest to
practitioners.

\mycite[Chen et al.]{Chen2018a} observe that assurance cases can
improve cross-disciplinary collaboration but are missing tool support
and experienced personnel.  We believe that a lack of research
transfer and training could explain the \textbf{contrast to}
finding~\FindingRef{C2}, given that assurance cases are seen as a new
method by SPs.

\mycite[Borg et al.]{Borg2016} and \mycite[De la Vara et
al.]{Vara2016} \textbf{clarify} finding~\FindingRef{H11} at least for the
specific case of change impact analysis in safety practice.

\mycite[McDermid and Rae]{McDermid2014} could find no satisfactory
explanation to their observation that systems got ``so safe despite
inadequate and inadequately applied techniques.'' However, their
assumption is \textbf{orthogonal to}
finding~\FindingRef{H26}, %
\textbf{contrasts} finding~\FindingRef{H11} and certainly emphasises
the need for further empirical research.  The lack of consensus on how
to combine case-based~\cite{Hatcliff2014} and
compliance-based~\cite{Ceccarelli2015} assurance underpins this lack
of clarity on the adequacy of means.

Finding~\FindingRef{H3} supports the observation of \mycite[Nair et
al.]{Nair2015} that expert judgements and checklists are among the most
frequently used references to assess safety arguments and
evidence~(see \Cref{fig:H3_RQ_E1}).
\FindingRef{H3} is also \textbf{shared by} \mycite[Rae and Alexander]{Rae2017a}
who conclude that critical aspects of safety analysis (\egs
identifying hazards, estimating risk probability and consequence
severity) often rely on expert opinion.
Moreover, \FindingRef{H3} \textbf{underpins} two out of \mycite[Wang's and
Wagner's]{Wang2018} top ten identified decision making pitfalls.

\mycite[Leveson]{Leveson2012} observes that safety is pervasively
confused (or assumed to correlate) with reliability.  The data for the
findings \FindingRef{H17} and \FindingRef{H17.2} \textbf{support} her
conclusion in parts, but the consensus of our responses suggests that
there is broad awareness that safety and reliability are first of all
two distinct properties of a system.
  
In summary, we found related supportive and contrasting evidence
regarding most findings for RQ1, RQ2, and RQ3.  

\subsection{General Feedback on the Survey}
\label{sec:gener-feedb-surv}

The last page of our questionnaire contains a text field to leave
general comments, \eg an overall opinion, on our survey.

One issue, our survey participants criticised, pertains to
\emph{the scope and the terminology} used in the questionnaire:

The respondents noted that the inquiry is general and does not
account for the diversity of safety practices in various industries.
Some questions rely on a particular interpretation of safety practice
leaving assumptions implicit and risking to get in conflict with other
views of system safety, \eg ``safety by introduction of controls''
versus ``safety assurance and assessment.''  Moreover, some of the
questions are hard to answer because of a lack of standardised
terminology across domains 
and because of missing topics, \eg legal safety requirements and
regulations,
human operators, socio-technical systems were not mentioned.

Although this is justified critique, we found it hard to arrive at a
terminology and at a level of detail suitable for all SPs while
keeping our construct lean~(\Cref{sec:hypoth-rese-quest}).  After
several iterations and an email-based focus group, we finalised the
questionnaire to be released.

When designing our questionnaire, we were driven by specific not
necessarily related findings from previous studies.  Moreover, we had
to prioritise and cut the question catalogue to stay within a maximum
duration of about 30 minutes, an amount of time we believe to be
affordable by the participants.

Except for~\QuestionRef{H26_RQ_B4} and~\QuestionRef{H11_RQ_B6}, the
acceptably low number~($<10\%$) of \emph{dnk}-responses indicates that
most respondents did not seem to struggle with answering most of the
questions.  However, frequent \emph{nand}-responses indicate
difficulties in deciding on the given answer options~(see,
\egs\QuestionRef{H11_RQ_B2}).

Another issue raised by our respondents deals with the \emph{survey
  method and design} we applied:

Some questions include bias, drive one to answer in a particular way
and solicit a specific support.
\textsc{Likert}-scales impose an abstraction with the risk to deny
more accurate answers such as ``I often highly agree and sometimes
I strongly disagree.''  Moreover, \textsc{Likert}-scales should be
substituted by open questions more appropriate for exploratory studies
where the construct is not known or (entirely) fixed beforehand.

On the one hand, we have gained good knowledge about the construct
from previous studies and, on the other, we provided several
possibilities to give open answers and, in fact, present results from
their qualitative analysis (\egs in the
\Cref{sec:sum-H18_RQ_F5,sec:sum-H3_RQ_E1,sec:sum-H1_RQ_G2,sec:sum-H18_RQ_F5.ab,sec:sum-H7_RQ_F1.a}).
More open questions reduce the risks of bias and constrained data
acquisition.  However, it is worth noting that, as opposed to
interviews, too many open answers in large-scale questionnaires can
also be demanding for the respondents and, thus, lead to a high number
of partial data points.

\subsection{Validity Procedure after Survey Execution}
\label{sec:threats-validity}

Here, we assess our survey design with respect to internal
and external validity as well as
reliability~\cite{Shull2008,Wohlin2012}.

\subsubsection{Internal Validity}
\label{sec:validity-internal-post}

To reduce internal threats to validity, we performed an
a-posteriori cross-validation with recommendations on
questionnaire-based surveys in the software and systems engineering
domain~\cite{Ciolkowski2003}.
\Cref{sec:gener-feedb-surv} discusses further arguments for internal
validity as a response to the general feedback on our survey.
Additionally, the everyday use of English among the majority of survey
participants~(\SubConstructRef{CQ_6_lang}) supports the accuracy of a
large fraction of the data points.

\subsubsection{External Validity} %
\label{sec:validity-external}

To which extent would the procedure in \Cref{sec:research-design}
lead to similar results with different samples?

Our sampling procedure is network-guided and, hence, \textbf{not
  uniformly random}~\cite{Haslam2009}.  However, on the one hand, from
\Cref{sec:sample}, our sample \textbf{varies over the scales of all
  classification criteria}~(\Cref{tab:classification}).  This
variation limits potential deficiencies of our sample resulting from
an overlap of the summer holiday season with our sampling period.  On
the other hand, regarding the notion of \emph{safety culture}, our
sample might be biased towards the more frequently occurring
backgrounds, domains, and geographical regions~(\Cref{sec:sample}).
However, the lack of evidence for \HypothesisRef{C6}~(\ies
practitioners differ from academics in their view of inadequacy of
means) reduces the extent to which the participation of researchers
biases the results towards a one-sided academic viewpoint.

According to \Cref{fig:CQ_8_role}, 19 out of 124 respondents stated
that they have been working on safety-related topics as a
\emph{researcher in academia}, \ie the role or responsibility profile
which we associate the least of all with genuine practical experience.
\textbf{Only 4 of them declared to be solely academic researchers.}  8
stated to be SPs, too; 7 have also done research in industry; 11 have
worked as software, systems, requirements, reliability, or health \&
safety practitioners in addition.  This again strengthens our belief
that our results are not biased towards and not significantly
influenced by a purely academic view.

We believe that, in comparison with focus groups and individual
interviews, on-line surveys can be a highly valuable instrument in
further investigations of this topic. %
There are two risks that can be mitigated by anonymous questionnaires:
\begin{enumerate}
\item In collaborations between academia and industry it is not
  unlikely that industrial participants in such projects are from the
  management, or senior engineers, or research engineers for several
  reasons not necessarily regularly connected to the
  operational teams.  Such collaborations bear the risk
  that the sample gets biased towards these roles.
  With an on-line survey advertised on multiple channels, we are
  convinced to have mitigated such a bias.
\item For legal reasons, safety activities can be quite critical to
  talk about personally and in an open way.  The authors' experience
  and impression is that in personal interviews, practitioners tend to
  avoid talking loosely about their organisations and, where
  aggravating, to moderately generalise.  Our impression from the
  respondents' occasionally quite open comments leads us to believe
  that the risk of this bias is lower in anonymous surveys such as our
  questionnaire.  Note that subjectivity has to be handled by other
  means in both questionnaires and interviews.
\end{enumerate}

Leveson~\cite[p.~211]{Leveson2012} states that FMEA, with its limited
applicability for safety analysis, is less frequently used as a hazard
analysis technique than FTA or ETA.  As opposed to Leveson's
observation, our respondents most often state that they work(ed) with
FMEA-based techniques in their safety
activities~(\cf\Cref{fig:CQ_5_meth}).  One reason for this discrepancy
could be that we only provided a small set of techniques as answer
options to check the criterion~\SubConstructRef{CQ_5_meth}, particularly,
ETA was not included.  Assuming that many respondents are reluctant to
add further techniques in the ``Other'' field, this might have led to
a bias towards the specified answer options.  Assuming that Leveson's
observation is drawn from US system safety cultures, this discrepancy
could also have arisen from the circumstance that our sample is biased
towards European safety cultures~(\cf\SubConstructRef{CQ_4_geo} in
\Cref{sec:sample}).  While this issue limits the external validity of
our exploratory study, we believe that the results for the questions
\QuestionRef{H26_RQ_B4}, \QuestionRef{H11_RQ_B2}, and
\QuestionRef{H11_RQ_B6} and the hypotheses \HypothesisRef{H26},
\HypothesisRef{H11}, \HypothesisRef{H11b}, \HypothesisRef{C7},
\HypothesisRef{C6}, and \HypothesisRef{C8} (relying on the constructs
\SubConstructRef{CQ_5_meth}, \SubConstructRef{Means}, and
\SubConstructRef{AdequacyOfStandardsMethods}) are not harmed by this
issue.

The independence of most of the questions allows a per-question
analysis.  Particularly, the 59 partial responses might not affect any
complete data points and thus were taken into consideration for the
questions for which they delivered responses~(\cf variation of $N$
values).  The relatively high number of registered views~(565) might
stem from users checking the questionnaire start page and concluding
that they do not belong to the target
group~(\Cref{sec:daten-collection}): Diverse preconceptions of safety,
diverse channel members, as well as short non-informative survey
advertisements might have played a role.  We believe, this issue has
not led to a significant loss of relevant respondents or a
participation of illegible respondents.

However, given that we expect the population of SPs to be 2 to 3
orders of magnitude larger than our sample~($N=91, N'=124$),
\textbf{confident general conclusions cannot be drawn}.  For this,
other sampling approaches such as the one employed by \mycite[Manotas
et al.]{Manotas2016} might be more appropriate, given proper
multilateral backing and preparation.  Their possibility to sample the
population with the \emph{support of global software companies} might
be more effective than our approach based on volunteer and
cluster-based sampling from several on-line discussion channels.

\subsubsection{Reliability} %
\label{sec:validity-reliability}

To which extent would a repetition of the procedure in
\Cref{sec:research-design} with the same sample lead to the same
results?

It is difficult to exactly repeat this survey in the short term because
our advertisements covered many of the relevant on-line channels and we
expect some of the respondents not willing to participate again within
short-term or at all.  This is a general problem for studies of this
kind.  Therefore, we suggest to
\begin{inparaenum}
\item provide incentives, 
\item pursue off-line channels as well, 
\item repeat the study in the long term, and
\item extend the sampling period. 
\end{inparaenum}
\Eg~\mycite[Mendez-Fernandez et al.]{DBLP:journals/ese/FernandezWKFMVC17} provide a longitudinal
design supporting repeatability and hence the determination of
reliability of the results.

\subsection{Lessons Learned}
\label{sec:personalnote}

Regarding the sample size~(\Cref{sec:sample}), we wished to get more
responses against the background of the effort we had in reaching out
to the population~(\Cref{sec:descr-data-points}).  From the Unipark
questionnaire view statistics, we saw that in some of the larger
discussion forums, users seemed to appear noticeably reluctant to
respond to our questionnaire.  The return rates estimated in
\Cref{sec:descr-data-points} can be considered low.  In few discussion
forums, our friendly, singular, and topic-related post of the
questionnaire was even penalised by deleting the post or by loosing
forum membership.  Unfortunately, important non-commercial panels such
as, \eg SoSciSurvey\footnote{See \url{https://www.soscisurvey.de}.}
or SurveyCircle~(\Cref{tab:adplatforms}) do not offer profiling
facilities to focus on engineering professionals.  In the case of no
budget for incentives and for paying commercial panels, these
circumstances make it very difficult for empirical (software)
researchers to approximate a representative sample.

\section{Conclusions and Future Work}
\label{sec:conclusions}

We designed and conducted a questionnaire-based cross-sectional
on-line survey of safety practitioners.  Our objective was to
investigate safety practice by asking practitioners
about means they rely on, process factors influencing their work, and
their role in the life cycle, and by checking several observations
stemming from previous research.

\subsection{Summary of Findings and Implications}
\label{sec:summary-findings}

Below, observations marked with $+$ represent our aspirations when
performing the study.  Observations marked with $-$ represent our
apprehensions.  Items labelled with $\bullet$ accommodate neutral
observations.

We collected evidence in \textbf{support} of several hypotheses
leading to the following observations:
\begin{itemize}
\item Our respondents confirm that safety decision making is mostly
  based on expert opinion and experience from previous
  projects\HypothesisRefP{H3}.
\item Safety practitioners think that for highly interconnected
  systems~(\egs systems of systems, connected transport systems),
  assurance of safety will have to rely on high assurance of IT
  security\HypothesisRefP{H8}.  Our experience suggests that the
  inverse relation is similarly strong.
\item[+] %
  They see a clear benefit in the interaction of safety and security
  activities\HypothesisRefP{H22}.  We like to support the agenda in
  \cite{Martins2016} and motivate research of strongly integrated
  safety-security approaches.
\item[+] %
  The survey participants believe that formal methods may have a
  positive impact on safety activities\HypothesisRefP{H11b}.
\item[--] %
  Currently applied standards and practised methods are believed to be
  largely inadequate to cope with the assurance of technologies~(\egs
  adaptive control, machine learning) used for high automation and
  autonomy in upcoming system applications\HypothesisRefP{H26}.
\end{itemize}
The last findings raise the question whether systems are
safe enough and why this would be the case~\cite{McDermid2014}?

Our analysis leads to further observations:
\begin{itemize}
\item Resources occasionally but not typically govern safety
  practitioners' decisions and performance\HypothesisRefP{H1}.
  The responses indicate that safety seems only rarely compromised by
  cost-benefit questions\HypothesisRefP{H18}.
\item Practitioners refrain from seeing safety as a special case of
  reliability\HypothesisRefP{H17}.  This stands in contrast with
  Leveson's former observation that safety is pervasively confused
  with reliability~\cite{Leveson2012}.
\item[--] Safety practitioners think that many of their non-safety
  co-workers' share at most medium appreciation of safety
  practitioners' contributions to the life cycle\HypothesisRefP{H16}.
\item[--] Respondents are indecisive on whether the conventional or
  ready-to-use methods they (could) apply scale
  sufficiently\HypothesisRefP{H11}.  
\end{itemize}
The last finding again motivates further analysis along the lines
of~\cite{McDermid2014}.  If we are left unsure about whether means
have become inadequate and, as found for safety RE in
\cite{Martins2016}, if conventional approaches are dominant and we
lack evidence for efficacy of novel research, how could safety
research help safety practitioners?
  
In summary, we share the impression that \emph{empirical research in
  system safety is still in an early stage}, on the one hand, offering
many opportunities to perform cross-disciplinary studies and, on the
other hand, bearing large risks of not exactly knowing to which extent
safety practitioners are applying state of the art and able to do
their best.  This is a severe issue to be discussed in software and
system safety research.

\subsection{Future Work}
\label{sec:futurework}

We seek to extend our analysis by revisiting findings from the
collected data set and not discussed in this work.
Furthermore, we are going to identify and evaluate further
hypotheses %
and ask more \emph{why}- and \emph{how}-questions.

Aspiring to the exploratory approach and grounded
theory~\cite[p.~298]{Shull2008}, we can further engage with our survey
participants using the \emph{focus group method}~\cite{Kontio2004},
request for comments on our findings, and ask them for approaches to
overcome the identified issues.
Additionally, we re-shape our construct and focus on a smaller set of
questions, \eg to investigate the applicability of formal methods.%
\footnote{The first author of this study has finished a
  follow-up survey on the \emph{use of formal
    methods}, available via
  \url{https://goo.gl/forms/FnKNQtTmI3A6BekM2}.}

Our research design can be extended towards the application of the
\emph{goal question metric} approach~\cite{Basili1994}: The results of
the hypotheses analysis promotes the definition of \emph{goals} of
safety activities, the \emph{survey questions} corresponding to the
hypotheses can be refined, and process and product \emph{metrics} be
derived from the refined questions, \eg as already suggested and
discussed in \cite{Murdoch2003,Luo2016}.  Our study object includes
SPs and, consequently, some of these metrics get measurable by
questionnaires.

Our setting as well as our findings coin a good starting point
for the design of a longitudinal study, offering possibilities to
identify and validate causal relationships among the measured
sub-constructs~(\Cref{tab:constructs}).

Inspired by previous work~\cite{Gleirscher2017a} and by
\cite{Nair2015}, it would be interesting to adapt our research design
to support investigations of phenomena such as \emph{confirmation
  bias} in practical safety arguments~\cite{Rae2017,Leveson2011}.

\balance

\subsubsection*{Acknowledgments}
\label{sec:ack}

The first author of this work is supported by the Deutsche
Forschungsgemeinschaft~(DFG, German Research Foundation) -- GL~915/1-1.
It is our pleasure to thank all survey participants for their valuable
responses, and several practitioners, researchers, and students for
acting as pilot run respondents and for providing us with initial
feedback.  Special thanks go to Mohammed Hussein and Dai Yang whose
analyses yielded important preliminary findings for initiating this
survey.  We are indebted to Martin Wildmoser for attending the final
interview for~\cite{Hussein2016} with friendly support of the
Validas~AG\footnote{See \url{http://www.validas.de}.} in Munich and to
further enthusiastic safety experts from various German industries for
participation in the interviews for~\cite{Yang2016}.
Technical University of Munich~(TUM) and University of York have been
excellent working environments.  I would like to thank Manfred Broy
for his senior advice and for providing the research infrastructure.
Daniel Mendez-Fernandez deserves cordial gratitude for giving us
Unipark advice and granting us access to this platform using the TUM
Informatics faculty license.
Finally, we would like to express our gratitude to the anonymous reviewers
for helpful comments leading to significant improvements.

\section*{References}
\label{sec:references}

\bibliographystyle{elsarticle-num} 
\bibliography{references}

\appendix

\section{Summary of All Responses}
\label{tab:datasummary}

For validation purposes, the following tables present data summaries
for all closed (q)uestions according to \Cref{tab:questions} and
questions for classification according to \Cref{tab:classification}.
The ``Option'' column refers to the parts~(if any) of multi-part
questions.  The ``NA's'' column signifies the number of invalid data
points for each~(part of a) question.  The checksum (including invalid
responses) of each row results in $N=152$ responses.  Rows with NA's
$ = 0$ result from parts (\ies answer categories) added after content
analysis of half-open questions~(\Cref{sec:openquestions}).  The
questions~\QuestionRef{H7_RQ_F1.a} and \QuestionRef{H18_RQ_F5.ab} are
open and, hence, not accompanied by a corresponding table.

\footnotesize
\begin{flushleft}
\begin{tabularx}{\linewidth}[t]{X|XXXXX|X}
\toprule
\multicolumn{2}{l}{\QuestionRef{H3_RQ_E1}} & \multicolumn{5}{c}{Value}
\\\midrule
Option / N & Very low & Low & Medium & High & Very high & NA's
\\
a~/~96 & 1 & 4 & 14 & 40 & 37 & 56
\\
b~/~96 & 1 & 3 & 25 & 36 & 31 & 56
\\
c~/~95 & 2 & 10 & 33 & 37 & 13 & 57
\\
d~/~95 & 1 & 1 & 13 & 53 & 27 & 57
\\
e~/~96 & 9 & 34 & 37 & 10 & 6 & 56
\\
f~/~96 & 1 & 8 & 47 & 32 & 8 & 56
\\
g~/~97 & 1 & 0 & 7 & 39 & 50 & 55
\\\bottomrule
\end{tabularx}
\end{flushleft}
\emph{Legend:} a.~Hazard list from previous projects, b.~Case (accident, incident) reports, c.~Inspection checklist, d.~Expert opinions, e.~Management recommendations, f.~Co-workers' recommendations, g.~Safety-related project experience

\vfill

\begin{flushleft}
\begin{tabularx}{\linewidth}[t]{X|XXXXX|X}
\toprule
\multicolumn{2}{l}{\QuestionRef{H1_RQ_G2}} & \multicolumn{5}{c}{Impact}
\\\midrule
Option / N & Do not know & No impact & Low impact & Medium impact & High impact & NA's
\\
a~/~98 & 5 & 5 & 12 & 39 & 37 & 54
\\
b~/~96 & 2 & 4 & 6 & 31 & 53 & 56
\\
c~/~95 & 6 & 2 & 5 & 36 & 46 & 57
\\
d~/~95 & 5 & 4 & 12 & 31 & 43 & 57
\\
e~/~97 & 4 & 3 & 9 & 25 & 56 & 55
\\
f~/~97 & 1 & 4 & 23 & 35 & 34 & 55
\\
g~/~98 & 2 & 2 & 11 & 27 & 56 & 54
\\\bottomrule
\end{tabularx}
\end{flushleft}
\emph{Legend:} a.~Budget cuts, b.~Late or unclear choice of safety concepts, c.~Postponed safety decisions, d.~Schedule pressure, e.~Erroneous hazard analyses, f.~Vague safety standards, g.~Inexperienced safety engineers

\vfill

\begin{flushleft}
\begin{tabularx}{\linewidth}[t]{X|XXXX|X}
\toprule
\multicolumn{2}{l}{\QuestionRef{H1_RQ_G3}} & \multicolumn{4}{c}{Frequency}
\\\midrule
Option / N & Often & Rarely / Occasionally & Never & Do not know & NA's
\\
--~/~99 & 36 & 48 & 9 & 6 & 53
\\\bottomrule
\end{tabularx}
\end{flushleft}

\vfill

\begin{flushleft}
\begin{tabularx}{\linewidth}[t]{X|XXXXX|X}
\toprule
\multicolumn{2}{l}{\QuestionRef{H26_RQ_B4}} & \multicolumn{5}{c}{Adequacy}
\\\midrule
Option / N & Do not know & Not adequate & Slightly adequate & Adequate & Very adequate & NA's
\\
a~/~100 & 32 & 31 & 26 & 11 & 0 & 52
\\
b~/~101 & 22 & 30 & 33 & 13 & 3 & 51
\\
c~/~101 & 26 & 31 & 27 & 16 & 1 & 51
\\
d~/~101 & 22 & 48 & 23 & 8 & 0 & 51
\\
e~/~100 & 36 & 7 & 25 & 27 & 5 & 52
\\
f~/~101 & 29 & 10 & 26 & 30 & 6 & 51
\\
g~/~99 & 32 & 28 & 21 & 17 & 1 & 53
\\\bottomrule
\end{tabularx}
\end{flushleft}
\emph{Legend:} a.~Self-adaptive systems, b.~Highly automated and autonomous driving, c.~Distributed networked systems, d.~AI/ML-based applications, e.~Medical and healthcare applications, f.~Highly automated air traffic control, g.~Consumer or commercial drones

\vfill

\begin{flushleft}
\begin{tabularx}{\linewidth}[t]{X|XXXXXX|X}
\toprule
\multicolumn{2}{l}{\QuestionRef{H11_RQ_B2}} & \multicolumn{6}{c}{Agreement}
\\\midrule
Option / N & Do not know & Strongly disagree & Disagree & Neither agree nor disagree & Agree & Strongly agree & NA's
\\
--~/~102 & 5 & 3 & 28 & 18 & 37 & 11 & 50
\\\bottomrule
\end{tabularx}
\end{flushleft}

\vfill

\begin{flushleft}
\begin{tabularx}{\linewidth}[t]{X|XXXXX|X}
\toprule
\multicolumn{2}{l}{\QuestionRef{H11_RQ_B6}} & \multicolumn{5}{c}{Impact}
\\\midrule
Option / N & Do not know & No impact & Low impact & Medium impact & High impact & NA's
\\
--~/~62 & 4 & 3 & 15 & 23 & 17 & 90
\\\bottomrule
\end{tabularx}
\end{flushleft}

\vfill

\begin{flushleft}
\begin{tabularx}{\linewidth}[t]{X|XXXXXX|X}
\toprule
\multicolumn{2}{l}{\QuestionRef{H6_RQ_C1}} & \multicolumn{6}{c}{Agreement}
\\\midrule
Option / N & Do not know & Strongly disagree & Disagree & Neither agree nor disagree & Agree & Strongly agree & NA's
\\
a~/~96 & 1 & 1 & 3 & 5 & 37 & 49 & 56
\\
b~/~95 & 9 & 1 & 14 & 17 & 36 & 18 & 57
\\
c~/~96 & 6 & 1 & 4 & 13 & 53 & 19 & 56
\\
d~/~95 & 12 & 6 & 18 & 21 & 32 & 6 & 57
\\\bottomrule
\end{tabularx}
\end{flushleft}
\emph{Legend:} a.~Adapt skills to new technologies, b.~Study state-of-the-art safety principles, c.~Juniors learn from seniors, d.~Juniors learn from accident reports

\vfill

\begin{flushleft}
\begin{tabularx}{\linewidth}[t]{X|XXXXXX|X}
\toprule
\multicolumn{2}{l}{\QuestionRef{H22_RQ_D1}} & \multicolumn{6}{c}{Agreement}
\\\midrule
Option / N & Do not know & Strongly disagree & Disagree & Neither agree nor disagree & Agree & Strongly agree & NA's
\\
a~/~94 & 3 & 5 & 13 & 14 & 35 & 24 & 58
\\
b~/~95 & 3 & 10 & 30 & 23 & 19 & 10 & 57
\\
c~/~93 & 4 & 4 & 11 & 15 & 36 & 23 & 59
\\
d~/~93 & 6 & 8 & 22 & 28 & 20 & 9 & 59
\\
e~/~93 & 5 & 3 & 10 & 9 & 42 & 24 & 59
\\
f~/~94 & 11 & 1 & 9 & 13 & 43 & 17 & 58
\\
g~/~92 & 11 & 0 & 12 & 16 & 41 & 12 & 60
\\
h~/~94 & 4 & 1 & 4 & 12 & 38 & 35 & 58
\\
i~/~94 & 5 & 4 & 3 & 12 & 49 & 21 & 58
\\
j~/~94 & 3 & 0 & 3 & 5 & 41 & 42 & 58
\\\bottomrule
\end{tabularx}
\end{flushleft}
\emph{Legend:} a.~Security is prerequisite for safety, b.~Safety is prerequisite for security, c.~SPs depend on security practitioners, d.~Security practitioners depend on SPs, e.~Safety assurance requires security assurcance, f.~Rare interaction in requirements stage, g.~Rare interaction in assurance stage, h.~Lack of collaboration is hazardous, i.~Lack of collaboration is inefficient, j.~Involvement in RE improves safety

\vfill

\begin{flushleft}
\begin{tabularx}{\linewidth}[t]{X|XX|X}
\toprule
\multicolumn{2}{l}{\QuestionRef{H18_RQ_F5}} & \multicolumn{2}{c}{Multiple Choice}
\\\midrule
Option / N & Checked & Unchecked & NA's
\\
a~/~95 & 30 & 65 & 57
\\
b~/~95 & 27 & 68 & 57
\\
c~/~95 & 50 & 45 & 57
\\
d~/~95 & 30 & 65 & 57
\\
e~/~95 & 21 & 74 & 57
\\\bottomrule
\end{tabularx}
\end{flushleft}
\emph{Legend:} a.~A cost factor, b.~A beneficial factor, c.~A necessity independent of cost, d.~A tedious mandated task, e.~A secondary issue

\vfill

\begin{flushleft}
\begin{tabularx}{\linewidth}[t]{X|XXXXXX|X}
\toprule
\multicolumn{2}{l}{\QuestionRef{H24_RQ_E3}} & \multicolumn{6}{c}{Agreement}
\\\midrule
Option / N & Do not know & Strongly disagree & Disagree & Neither agree nor disagree & Agree & Strongly agree & NA's
\\
a~/~97 & 1 & 1 & 1 & 7 & 44 & 43 & 55
\\
b~/~97 & 3 & 2 & 3 & 8 & 40 & 41 & 55
\\
c~/~97 & 2 & 2 & 2 & 16 & 44 & 31 & 55
\\
d~/~97 & 2 & 2 & 1 & 7 & 41 & 44 & 55
\\\bottomrule
\end{tabularx}
\end{flushleft}
\emph{Legend:} a.~Safety is given high priority, b.~Management highly values safety, c.~SPs have declared authority, d.~Safety process is defined

\vfill

\begin{flushleft}
\begin{tabularx}{\linewidth}[t]{X|XXXXXX|X}
\toprule
\multicolumn{2}{l}{\QuestionRef{H17_RQ_E2}} & \multicolumn{6}{c}{Agreement}
\\\midrule
Option / N & Do not know & Strongly disagree & Disagree & Neither agree nor disagree & Agree & Strongly agree & NA's
\\
a~/~97 & 2 & 57 & 28 & 3 & 5 & 2 & 55
\\
b~/~96 & 1 & 11 & 18 & 12 & 37 & 17 & 56
\\
c~/~96 & 1 & 4 & 6 & 15 & 44 & 26 & 56
\\
d~/~97 & 2 & 17 & 18 & 21 & 29 & 10 & 55
\\
e~/~96 & 3 & 25 & 25 & 15 & 19 & 9 & 56
\\\bottomrule
\end{tabularx}
\end{flushleft}
\emph{Legend:} a.~Lack of failures reduces need for safety, b.~Accidents drive need for safety, c.~Accidents help SPs argue for safety, d.~Lack of accidents reduces need for safety, e.~Safety implies reliability

\vfill

\begin{flushleft}
\begin{tabularx}{\linewidth}[t]{X|XXXXX|X}
\toprule
\multicolumn{2}{l}{\QuestionRef{H20_RQ_F2}} & \multicolumn{5}{c}{Value}
\\\midrule
Option / N & Very low & Low & Medium & High & Very high & NA's
\\
--~/~95 & 2 & 6 & 25 & 48 & 14 & 57
\\\bottomrule
\end{tabularx}
\end{flushleft}

\vfill

\begin{flushleft}
\begin{tabularx}{\linewidth}[t]{X|XXXXX|X}
\toprule
\multicolumn{2}{l}{\QuestionRef{H16_RQ_F4}} & \multicolumn{5}{c}{Value}
\\\midrule
Option / N & Very low & Low & Medium & High & Very high & NA's
\\
--~/~95 & 4 & 19 & 37 & 28 & 7 & 57
\\\bottomrule
\end{tabularx}
\end{flushleft}

\vfill

\begin{flushleft}
\begin{tabularx}{\linewidth}[t]{X|XXXXXX|X}
\toprule
\multicolumn{2}{l}{\QuestionRef{H14_RQ_C2}} & \multicolumn{6}{c}{Agreement}
\\\midrule
Option / N & Do not know & Strongly disagree & Disagree & Neither agree nor disagree & Agree & Strongly agree & NA's
\\
a~/~95 & 3 & 2 & 6 & 11 & 41 & 32 & 57
\\
b~/~95 & 3 & 1 & 4 & 13 & 47 & 27 & 57
\\
c~/~96 & 16 & 2 & 2 & 27 & 29 & 20 & 56
\\\bottomrule
\end{tabularx}
\end{flushleft}
\emph{Legend:} a.~Senior SPs outperform junior SPs, b.~Previous projects experience is beneficial, c.~Adversarial thinking improves hazard analysis

\vfill

\begin{flushleft}
\begin{tabularx}{\linewidth}[t]{X|XX|X}
\toprule
\multicolumn{2}{l}{\SubConstructRef{CQ_1_edu}} & \multicolumn{2}{c}{Multiple Choice (Classification)}
\\\midrule
Option / N & Checked & Unchecked & NA's
\\
a~/~124 & 49 & 75 & 28
\\
b~/~124 & 13 & 111 & 28
\\
e~/~124 & 23 & 101 & 28
\\
f~/~124 & 31 & 93 & 28
\\
g~/~124 & 22 & 102 & 28
\\
h~/~124 & 12 & 112 & 28
\\
k~/~124 & 6 & 118 & 28
\\\bottomrule
\end{tabularx}
\end{flushleft}
\emph{Legend:} a.~Computer Science, b.~Systems Engineering, e.~Safety Science, f.~Electrical and Electronics Engineering, g.~Mechanical and Aerospace Engineering, h.~Physics and Mathematics, k.~Other Discipline

\vfill

\begin{flushleft}
\begin{tabularx}{\linewidth}[t]{X|XX|X}
\toprule
\multicolumn{2}{l}{\SubConstructRef{CQ_2_dom}} & \multicolumn{2}{c}{Multiple Choice (Classification)}
\\\midrule
Option / N & Checked & Unchecked & NA's
\\
a~/~124 & 51 & 73 & 28
\\
b~/~124 & 54 & 70 & 28
\\
c~/~124 & 16 & 108 & 28
\\
d~/~124 & 19 & 105 & 28
\\
e~/~124 & 30 & 94 & 28
\\
f~/~124 & 17 & 107 & 28
\\
g~/~124 & 24 & 100 & 28
\\
h~/~124 & 9 & 115 & 28
\\
j~/~124 & 25 & 99 & 28
\\
l~/~124 & 15 & 109 & 28
\\
m~/~124 & 8 & 116 & 28
\\
o~/~152 & 15 & 137 & 0
\\
p~/~152 & 6 & 146 & 0
\\\bottomrule
\end{tabularx}
\end{flushleft}
\emph{Legend:} a.~Automotive and Transport Systems, b.~Aerospace Industry, c.~IT Infrastructure and Networking, d.~Power and Nuclear Industry, e.~Industrial Processes and Plant Automation, f.~Electronic Devices and Appliances, g.~Healthcare Systems, h.~Construction and Building Automation, j.~Industrial Machinery, l.~Naval Systems, m.~Other Domain, o.~Railway and Cablecar Systems, p.~Military and Defense Systems

\vfill

\begin{flushleft}
\begin{tabularx}{\linewidth}[t]{X|XXXXX|X}
\toprule
\multicolumn{2}{l}{\SubConstructRef{CQ_3_exp}} & \multicolumn{5}{c}{Single Choice (Years Of Experience In Levels)}
\\\midrule
Option / N & $<$ 3 & 3 - 7 & 8 - 15 & 16 - 25 & $>$ 25 & NA's
\\
--~/~119 & 27 & 27 & 27 & 18 & 20 & 33
\\\bottomrule
\end{tabularx}
\end{flushleft}

\vfill

\begin{flushleft}
\begin{tabularx}{\linewidth}[t]{X|XX|X}
\toprule
\multicolumn{2}{l}{\SubConstructRef{CQ_5_std}} & \multicolumn{2}{c}{Multiple Choice (Classification)}
\\\midrule
Option / N & Checked & Unchecked & NA's
\\
a~/~124 & 70 & 54 & 28
\\
b~/~124 & 19 & 105 & 28
\\
c~/~124 & 47 & 77 & 28
\\
e~/~124 & 3 & 121 & 28
\\
f~/~124 & 47 & 77 & 28
\\
h~/~124 & 14 & 110 & 28
\\
i~/~124 & 8 & 116 & 28
\\
k~/~152 & 3 & 149 & 0
\\
l~/~152 & 3 & 149 & 0
\\
m~/~152 & 12 & 140 & 0
\\
n~/~152 & 4 & 148 & 0
\\
o~/~152 & 14 & 138 & 0
\\\bottomrule
\end{tabularx}
\end{flushleft}
\emph{Legend:} a.~Generic, b.~Machinery, c.~Automotive and Transport, e.~Agriculture, f.~Aerospace and Avionics, h.~Not Familiar, i.~Other Standard, k.~Nuclear and Other Energy, l.~Medical Devices, m.~Railway, n.~Methodology and Tooling, o.~Military and Defense

\vfill

\begin{flushleft}
\begin{tabularx}{\linewidth}[t]{X|XX|X}
\toprule
\multicolumn{2}{l}{\SubConstructRef{CQ_5_meth}} & \multicolumn{2}{c}{Multiple Choice (Classification)}
\\\midrule
Option / N & Checked & Unchecked & NA's
\\
a~/~124 & 90 & 34 & 28
\\
b~/~124 & 78 & 46 & 28
\\
c~/~124 & 16 & 108 & 28
\\
d~/~124 & 59 & 65 & 28
\\
e~/~123 & 4 & 119 & 29
\\
f~/~124 & 15 & 109 & 28
\\
g~/~124 & 15 & 109 & 28
\\
i~/~152 & 9 & 143 & 0
\\
j~/~152 & 3 & 149 & 0
\\
k~/~152 & 5 & 147 & 0
\\
l~/~152 & 3 & 149 & 0
\\
m~/~152 & 5 & 147 & 0
\\
n~/~152 & 4 & 148 & 0
\\
o~/~152 & 6 & 146 & 0
\\
p~/~152 & 3 & 149 & 0
\\\bottomrule
\end{tabularx}
\end{flushleft}
\emph{Legend:} a.~Failure Mode Effects, b.~Fault Trees, c.~STAMP-based Methods, d.~Hazard Operability, e.~Security Threats, f.~Not Familiar, g.~Other Concept/Method, i.~Functional and Design Risk, j.~Probabilistic Risk, k.~Assurance Cases, l.~Root Causes, m.~Event Trees, n.~Automated VandV, o.~Dependencies and Interactions, p.~Bidirectional Methods

\vfill

\begin{flushleft}
\begin{tabularx}{\linewidth}[t]{X|XX|X}
\toprule
\multicolumn{2}{l}{\SubConstructRef{CQ_6_lang}} & \multicolumn{2}{c}{Multiple Choice (Classification)}
\\\midrule
Option / N & Checked & Unchecked & NA's
\\
a~/~124 & 54 & 70 & 28
\\
b~/~124 & 38 & 86 & 28
\\
c~/~124 & 5 & 119 & 28
\\
d~/~124 & 6 & 118 & 28
\\
e~/~124 & 2 & 122 & 28
\\
g~/~152 & 3 & 149 & 0
\\
h~/~152 & 3 & 149 & 0
\\
i~/~152 & 14 & 138 & 0
\\\bottomrule
\end{tabularx}
\end{flushleft}
\emph{Legend:} a.~English, b.~German, c.~Italian, d.~French, e.~Chinese, g.~Portuguese, h.~Swedish, i.~Other language

\vfill

\begin{flushleft}
\begin{tabularx}{\linewidth}[t]{X|XX|X}
\toprule
\multicolumn{2}{l}{\SubConstructRef{CQ_7_buslang}} & \multicolumn{2}{c}{Multiple Choice (Classification)}
\\\midrule
Option / N & Checked & Unchecked & NA's
\\
a~/~124 & 113 & 11 & 28
\\
b~/~124 & 41 & 83 & 28
\\
c~/~124 & 5 & 119 & 28
\\
d~/~124 & 10 & 114 & 28
\\
e~/~124 & 12 & 112 & 28
\\\bottomrule
\end{tabularx}
\end{flushleft}
\emph{Legend:} a.~English, b.~German, c.~Italian, d.~French, e.~Other language

\vfill

\begin{flushleft}
\begin{tabularx}{\linewidth}[t]{X|XX|X}
\toprule
\multicolumn{2}{l}{\SubConstructRef{CQ_8_role}} & \multicolumn{2}{c}{Multiple Choice (Classification)}
\\\midrule
Option / N & Checked & Unchecked & NA's
\\
a~/~124 & 62 & 62 & 28
\\
e~/~124 & 29 & 95 & 28
\\
f~/~124 & 9 & 115 & 28
\\
g~/~124 & 39 & 85 & 28
\\
h~/~124 & 23 & 101 & 28
\\
j~/~124 & 11 & 113 & 28
\\
k~/~124 & 3 & 121 & 28
\\
m~/~66 & 12 & 54 & 86
\\
n~/~124 & 4 & 120 & 28
\\
q~/~152 & 3 & 149 & 0
\\\bottomrule
\end{tabularx}
\end{flushleft}
\emph{Legend:} a.~Safety Practitioner, e.~Software Practitioner, f.~Electrical or Electronics Practitioner, g.~Systems Practitioner, h.~Requirements Practitioner, j.~Health and Safety Practitioner, k.~IT Security Practitioner, m.~Reliability Practitioner, n.~Other Role, q.~V and V Practitioner

\vfill

\begin{flushleft}
\begin{tabularx}{\linewidth}[t]{X|XX|X}
\toprule
\multicolumn{2}{l}{\SubConstructRef{CQ_8_role}} & \multicolumn{2}{c}{Multiple Choice (Classification)}
\\\midrule
Option / N & Checked & Unchecked & NA's
\\
b~/~124 & 26 & 98 & 28
\\
c~/~124 & 19 & 105 & 28
\\
d~/~124 & 5 & 119 & 28
\\
l~/~76 & 13 & 63 & 76
\\
p~/~152 & 12 & 140 & 0
\\\bottomrule
\end{tabularx}
\end{flushleft}
\emph{Legend:} b.~Researcher in Industry, c.~Researcher in Academia, d.~Assistant, Trainee, or Junior, l.~Practitioner with PhD degree, p.~Consultant / Assessor

\vfill

 \normalsize

\section{List of Abbreviations}
\label{sec:list-abbreviations}

See \Cref{tab:abbrev}.

\begin{table}[h]
  \centering
  \footnotesize
  \caption{List of abbreviations used in this article
    \label{tab:abbrev}}
  \begin{tabularx}{1.0\linewidth}{lX}
    \toprule
    AC & Acceptance Criterion \\
    CAST & Causal Analysis using System Theory \\
    CCA & Common Cause Analysis \\
    CIA & Change Impact Analysis \\
    CMA & Common Mode Analysis \\
    DV & Dependent Variable \\
    ETA & Event Tree Analysis \\
    FFA & Functional Failure Analysis \\
    FHA & Functional Hazard Analysis \\
    FM & Formal Method \\
    FMEA & Failure Mode Effects Analysis \\ 
    FMECA & Failure Mode, Effects, and Criticality Analysis \\
    FMEDA & Failure Mode, Effects, and Diagnostic Analysis \\
    FRAM & Functional Resonance Analysis Method \\
    FTA & Fault Tree Analysis \\
    GSN & Goal Structuring Notation \\
    HRA & Health Risk Assessment \\
    IV & Independent Variable \\
    MC & Multiple Choice \\
    NA & Not Available \\
    OSHA & Operation \& Support Hazard Analysis\\
    PHA & Preliminary Hazard Analysis \\
    PHL & Preliminary Hazard List \\
    RCA & Root Cause Analysis \\
    RE & Requirements Engineering \\
    RQ & Research Question \\
    SACM & Structured Assurance Case Meta-Model \\
    SCRA & Supply Chain Risk Assessment \\
    SE & Software Engineering\\ 
    SHA & System Hazard Analysis \\
    SHARD & Software Hazard Analysis and Resolution in Design \\
    SP & Safety Practitioner \\
    SPP & Safety Practice and its Practitioners \\
    STAMP & System-Theoretic Accident Model \& Processes \\
    STPA & System-Theoretic Process Analysis \\
    STRIDE & Spoofing, Tampering, Repudiation, Information disclosure,
             Denial of service, Elevation of privilege \\
    SSHA & System Safety (or Sub-System) Hazard Analysis \\
    WBA & Why-Because Analysis \\
    ZHA & Zonal Hazard Analysis \\
    ZSA & Zonal Safety Analysis \\ %
    \bottomrule
  \end{tabularx}
\end{table}

\end{document}